\documentclass[11pt,prd,onecolumn,amsmath,amssymb,aps,floats,floatfix,nofootinbib]{revtex4-2}
\usepackage[colorlinks=true,urlcolor=blue,anchorcolor=blue,citecolor=blue,filecolor=blue,linkcolor=blue,menucolor=blue,linktocpage=true]{hyperref} 

\usepackage[inline]{enumitem}
\usepackage[multidot]{grffile}  
\usepackage{dcolumn}
\usepackage{bm}
\usepackage{amsmath}
\usepackage{amsfonts}
\usepackage{amssymb}
\usepackage{color}
\usepackage[table]{xcolor}
\usepackage{float}
\usepackage{latexsym}
\usepackage{slashed} 
\usepackage{pstricks}
\usepackage{indentfirst}
\usepackage{mathrsfs}
\usepackage{multirow}
\usepackage{epsfig,psfrag}
\usepackage{graphicx}
\usepackage{enumitem}
\usepackage{geometry,amssymb,yfonts}
\usepackage{yhmath}
\usepackage{anysize}
\usepackage{subfigure}
\usepackage{mathtools}
\usepackage{setspace} 
\usepackage[utf8]{inputenc} 
\usepackage[scientific-notation=true]{siunitx} 

\usepackage{url}
\usepackage{makecell}

\graphicspath{{fig/}}

\allowdisplaybreaks 

\begin{document}

\title{Dark matter phenomenology and phase transition dynamics of the next to minimal composite Higgs model with dilaton}

\author{Borui Zhang$^{a,c}$}%
\author{Zhao Zhang$^{a}$}
\author{Chengfeng Cai$^{b}$}%
\email[Corresponding author. ]{caichf3@mail.sysu.edu.cn}
\author{Hong-Hao Zhang$^{a}$}
\email[Corresponding author. ]{zhh98@mail.sysu.edu.cn}
\affiliation{$^a$School of Physics, Sun Yat-Sen University, Guangzhou 510275, China}
\affiliation{$^b$School of Science, Sun Yat-Sen University, Shenzhen 518107, China}
\affiliation{$^c$Department of Physics, Tsinghua University, Beijing 100084, China}

\bigskip
	
	
\begin{abstract}
In this paper, we conduct a comprehensive study of the Next-to-Minimal Composite Higgs Model (NMCHM) extended with a dilaton field $\chi$ (denoted as NMCHM$_\chi$). A pseudo-Nambu-Goldstone boson (pNGB) $\eta$, resulting from the SO(6)$\to$SO(5) breaking, serves as a dark matter (DM) candidate. 
The inclusion of the dilaton field is helpful for evading the stringent constraints from dark matter direct detection, as it allows for an accidental cancellation between the amplitudes of DM-nucleon scattering, an outcome of the mixing between the dilaton and Higgs fields. The presence of the dilaton field also enriches the phase transition patterns in the early universe. We identify two types of phase transitions: (i) a 1-step phase transition, where the chiral symmetry and electroweak symmetry breaking (EWSB) occur simultaneously, and (ii) a 2-step phase transition, where the chiral symmetry breaking transition takes place first, followed by a second phase transition corresponding to EWSB. 
Since the first-order phase transitions can be strong due to supercooling in our model, we also examine the stochastic background of gravitational waves generated by these phase transitions. We find that these gravitational waves hold promise for detection in future space-based gravitational wave experiments, such as LISA, Taiji, BBO, and DECIGO.

\end{abstract}
	

\maketitle
\section{Introduction}\label{sec:intro}
The Composite Higgs Model (CHM) is one of the most compelling models beyond the Standard Model (SM) for addressing the hierarchy problem~\cite{Georgi:1984af,agashe_minimal_2005,Agashe:2005dk,cacciapaglia_fundamental_2014,csaki_tasi_2018,marzocca_general_2012,panico_composite_2016,daza_composite_2019}. In this paradigm, the Higgs boson is not viewed as a fundamental particle but emerges as a pseudo-Nambu-Goldstone boson (pNGB) corresponding to the spontaneous breaking of an approximate global symmetry.
The Minimal Composite Higgs Model (MCHM)~\cite{agashe_minimal_2005,Agashe:2005dk}, which is the minimal realistic CHM that possesses a custodial symmetry, is built on the symmetry breaking pattern $\mathrm{SO(5)}\to\mathrm{SO(4)}$. However, this minimal realization does not introduce any new degrees of freedom that could serve as a dark matter candidate.

A natural extension to include a dark matter candidate is to consider the Next-to-Minimal Composite Higgs Model (NMCHM), which is based on the symmetry breaking pattern $\mathrm{SO(6)}\to\mathrm{SO(5)}$~\cite{gripaios_beyond_2009}. This pattern of symmetry breaking implies the existence of five pNGBs, four of which correspond to the Higgs fields, while the remaining one corresponds to a new scalar boson $\eta$, potentially serving as the Weakly Interacting Massive Particle (WIMP) dark matter candidate. Note that the $Z_2$ symmetry associated with $\eta\to-\eta$ can be violated by the Wess-Zumino-Witten (WZW) terms in the $\mathrm{SO(6)}\to\mathrm{SO(5)}$ model. Therefore, we will consider an $\mathrm{O(6)}\to\mathrm{O(5)}$ model~\cite{frigerio_composite_2012,marzocca_composite_2014,xu_identifying_2023}, which assumes that the $Z_2$ symmetry is respected by the whole strong sector.
Moreover, if the NMCHM is equipped with an approximate conformal symmetry, for instance, when it is built on the 5D duality framework~\cite{Arkani-Hamed:2000ijo,Rattazzi:2000hs}, both the scale invariance and the chiral symmetry can be broken simultaneously during a confinement phase transition (PT). The spontaneous breaking of this approximate scale invariance also results in a pNGB, known as a dilaton $\chi$ (its dual counterpart in 5D theory is the radion~\cite{Arkani-Hamed:2000ijo,Rattazzi:2000hs,goldberger_modulus_1999}).

In this work, we will conduct a comprehensive study of the NMCHM extended with $\chi$, which we refer to as NMCHM$_\chi$. We also make the assumption that the conformal symmetry breaking scale of NMCHM$_\chi$, which is determined by the vacuum expectation value (VEV) of the dilaton, coincides with the confinement scale. Nonetheless, this condition is not generally required.
A potential of the dilaton can be generated due to the explicit breaking of conformal symmetry, and this will govern both the vacuum expectation value (VEV) and the mass of the dilaton. If the effect of explicit breaking is small, the theory would remain near a fixed point over a large scale range from the ultraviolet (UV) scale to the confinement scale, exhibiting a walking behavior~\footnote{While we refer to the behavior here as 'walking' for visualization purposes, it is in reality a tuning model with fixed points set to be real, which differs from the complex ones found in the true walking model. This signifies the emergence of a tachyon in the 5D dual theory~\cite{pomarol_holographic_2019,baratella_supercooled_2019}.}. 

In the NMCHM$_\chi$, the dilaton can influence dark matter (DM) phenomenology in two ways. Firstly, the dilaton field can mix with the Higgs field, potentially leading to a suppression of the DM-nucleon scattering amplitude if an accidental cancellation occurs. This could help the DM evade stringent constraints from direct detection. Secondly, if $m_\eta>m_\chi$, the DM candidate $\eta$ could open an annihilation window, $\eta+\eta\to \chi+\chi$. This would enhance the annihilation cross section, impacting both indirect detection and the relic density.

It has been proposed that the MCHM, when combined with a dilaton field, can achieve a strong first-order electroweak phase transition, particularly in the context of a supercooled phase transition~\cite{bruggisser_electroweak_2018,bruggisser_status_2022,bian_electroweak_2019,chung_125_2013,croon_tasi_2023,patel_stepping_2013,mazumdar_cosmic_2019,inoue_two-step_2016,curtin_thermal_2016}. 
We can anticipate that this would also occur in the NMCHM$_\chi$. As we know, the strong first-order electroweak phase transition is not only essential for electroweak baryogenesis~\cite{cohen_spontaneous_1991,cohen_progress_1993,trodden_electroweak_1999,cline_baryogenesis_2006,petropoulos_baryogenesis_2003,White:2016nbo,mazumdar_cosmic_2019,braconi_bubble_nodate,espinosa_electroweak_2012,de_curtis_composite_2019,chala_unified_2016}, but it also provides a source of stochastic background of gravitational waves. These gravitational wave signals are promising to be detected in future experiments, such as LiSA, TianQin, Taiji, BBO, and DECIGO~\cite{Audley:2017drz,Mei:2020lrl,Guo:2018npi,Cutler:2005qq,Kudoh:2005as}.

This paper is organized as follows. In Sect.~\ref{sec:model}, we discuss the basic construction of the NMCHM$_\chi$, which includes the effective Lagrangian and the effective potential. In Sect.~\ref{sec:DM}, we analyze the DM phenomenology, taking into account constraints from direct detection, indirect detection, and relic density. In Sect.~\ref{sec:PT}, we investigate the phase transition dynamics of the NMCHM$_\chi$ in conjunction with the results from the DM phenomenology. In Sect.~\ref{sec:GW}, we discuss the gravitational waves produced by the FOPTs. Finally, we present our conclusion in Sect.~\ref{sec:conclusion}.

\section{Model construction}\label{sec:model}
\subsection{Effective theory of dilaton}

In the Next-to-Minimal Composite Higgs Model (NMCHM), an approximate conformal symmetry is typically assumed. The strong dynamics in the infrared region (IR) lead to a condensate of techni-quarks, resulting in the spontaneous breaking of both chiral symmetries and conformal symmetry at the IR scale $f$. The Goldstone boson corresponding to the breaking of scale invariance is referred to as the dilaton, $\chi$. If the conformal invariance is explicitly broken by quantum effects, a potential for the dilaton can develop, causing the dilaton field's excitation to become massive. We will follow the derivations and conventions constructing the dilaton potential as presented in Ref.~\cite{bruggisser_electroweak_2018,bruggisser_status_2022}, focusing exclusively on the case of the meson-like dilaton.

Given that both the Higgs boson and the dilaton originate from the same underlying strongly-coupled sector, the vacuum expectation value (VEV) of the dilaton, $v_\chi$, is equal to the global symmetry breaking scale $f$. We assume that the main source of the violation of conformal invariance arises from the ultraviolet (UV) theory. Consequently, the breaking effects infiltrate the effective theory through the renormalization group equation (RGE) as follows:
\begin{eqnarray}
    \frac{\partial\mathrm{log}\epsilon(\chi)}{\partial\mathrm{log}\chi}=\gamma_\epsilon+c^{(1)}\epsilon(\chi),
\end{eqnarray}
where $c^{(1)}=c_\epsilon/g_\chi^2$, with $c_\epsilon$ assumed to be tiny to generate a flat potential.
Thus, the effective potential of the dilaton with a non-zero VEV can be expressed as follows~\cite{bruggisser_electroweak_2018,bruggisser_status_2022}:
\begin{eqnarray}
    V(\chi)=C[N,\chi]\chi^4= \left[c_\chi g_\chi^2-\epsilon(\chi)\right]\chi^4
\end{eqnarray}
where $c_\chi$ is an $\mathcal{O}(1)$ coefficient, and $g_\chi=4\pi/\sqrt{N}$, as per the large-$N$ expansion description~\cite{Panico:2015jxa}.

\subsection{The next-to-minimal composite Higgs model}
For this study, we use a model where all the SM fermions fall under the $\bm{6+6}$ representation of SO(6) as a benchmark (fermions in higher dimensional representations have been examined in Ref.~\cite{bian_electroweak_2019,xie_electroweak_2020}).
The Goldstone bosons, which correspond to global SO(6) symmetry breaking, are encoded in the $\Sigma$ field. This field is defined by rotating the vacuum using a Goldstone matrix, denoted as $\Sigma(x)\equiv e ^{i \Pi/v_\chi}$. Typically, it is expressed using the following patterns of parametrization:
	\begin{eqnarray}
		\Sigma&=&\mathrm{sin}\frac{\pi}{v_\chi}\left(\frac{\pi^{\hat{1}}}{\pi},\frac{\pi^{\hat{2}}}{\pi},\frac{\pi^{\hat{3}}}{\pi},\frac{\pi^{\hat{4}}}{\pi},\frac{\pi^{\hat{5}}}{\pi},\mathrm{cot}\frac{\pi}{v_\chi}\right)\notag\\
		&=&\frac{1}{v_\chi}\left(h_1,h_2,h_3,h_4,\eta,\sqrt{v_\chi^2-h^2-\eta^2}\right).
	\end{eqnarray}
By choosing the unitary gauge, these parametrizations are reduced to:
\begin{eqnarray}
	\Sigma&=&\mathrm{sin}\frac{\pi}{v_\chi}\left(0,0,\frac{\pi^{\hat{3}}}{\pi},0,\frac{\pi^{\hat{5}}}{\pi},\mathrm{cot}\frac{\pi}{v_\chi}\right)\notag\\
	&=&\frac{1}{v_\chi}\left(0,0,h,0,\eta,\sqrt{v_\chi^2-h^2-\eta^2}\right)\notag.
\end{eqnarray}
However, we will use the expression in the second line as it is more conducive for discussing phenomenologies. Note that in this parametrization, $h$ and $\eta$ are independent with $\chi$.

The leading order of the chiral Lagrangian, derived from the Callan-Coleman-Wess-Zumino (CCWZ) construction, can be expressed as follows:
\begin{eqnarray}
    \mathcal{L}_{chiral}^{(2)}&=&\frac{f^2}{2}(D_\mu\Sigma)^T(D^\mu\Sigma)\notag\\
    &=&\frac{1}{2}\partial_{\mu}h\partial^{\mu}h+\frac{1}{2}\partial_{\mu}\eta\partial^{\mu}\eta+\frac{1}{2}\frac{(h\partial_{\mu}h+\eta\partial_{\mu}\eta)^2}{v_\chi-h^2-\eta^2}\notag\\
    &&+\frac{h^2}{v^2}\left(M^2_WW^+_\mu W^{-\mu}+\frac{M^2_Z}{2}Z_\mu Z^{\mu}\right).
\end{eqnarray}
The effective Lagrangian of the fermion sector can be constructed either by spurions, which restore the broken symmetries, or by directly integrating out the heavy resonances. Finally, we obtain the following:
\begin{eqnarray}
    \mathcal{L}_f &=& i\bar{t}_{L} \slashed{\partial} \left[\Pi^q_0-\frac{\Pi^q_1}{2}\left(\frac{h}{v_\chi}\right)^2\right]t_L+i\bar{t}_{R} \slashed{\partial} \left\{\Pi^t_0-\Pi^t_1\left[1-\left(\frac{h}{v_\chi}\right)^2-\left(\frac{\eta}{v_\chi}\right)^2\right]\right\} t_R\notag\\
&&-\frac{m_{t0}}{\sqrt{1-\left(\frac{v}{v_\chi}\right)^2}}\sqrt{1-\left(\frac{h}{v_\chi}\right)^2-\left(\frac{\eta}{v_\chi}\right)^2}\frac{h}{v}\bar{t}_Lt_R+h.c.+...
\end{eqnarray}
where the ellipses denote the other flavors of quarks, which are also under the $\bm{6+6}$ representation of SO(6).
The form factors $\Pi_0^q,~\Pi_1^q,~\Pi_0^t,\text{~and~}\Pi_1^t$ have a detailed construction that can be found in Refs.~\cite{marzocca_general_2012,panico_composite_2016,daza_composite_2019,bian_electroweak_2019,marzocca_composite_2014,gripaios_beyond_2009}.

\subsection{Dilaton extension of the NMCHM}

We can incorporate the dilaton field into our model by considering it as a spurion that compensates for the scale transformation of the effective Lagrangian for the NMCHM. Alternatively, we can promote the global symmetry breaking scale $f$ to a dynamic field $\chi$.
The dilaton extension of the NMCHM (denoted as NMCHM$_\chi$) results in the following effective Lagrangian:
\begin{eqnarray}\label{eq:effLagchi}
\mathcal{L}_{eff}&=&\frac{1}{2}\left(\frac{\chi}{v_\chi}\right)^2\partial_{\mu}h\partial^{\mu}h+\frac{1}{2}\left(\frac{\chi}{v_\chi}\right)^2\partial_{\mu}\eta\partial^{\mu}\eta+\frac{1}{2}\left(\frac{\chi}{v_\chi}\right)^2\frac{(h\partial_{\mu}h+\eta\partial_{\mu}\eta)^2}{v_\chi-h^2-\eta^2}+\frac{1}{2}\partial_{\mu}\chi\partial^{\mu}\chi\notag\\
&&+\frac{h^2}{v^2}\left(\frac{\chi}{v_\chi}\right)^2\left(M^2_WW^+_\mu W^{-\mu}+\frac{M^2_Z}{2}Z_\mu Z^{\mu}\right)\notag\\
&&+i\bar{t}_{L}\partial\mkern -9.5 mu /\left(\Pi^q_0-\frac{\Pi^q_1}{2}\left(\frac{h}{v_\chi}\right)^2\right)t_L+i\bar{t}_{R}\partial\mkern -9.5 mu /\left(\Pi^t_0-\Pi^t_1\left(1-\left(\frac{h}{v_\chi}\right)^2-\left(\frac{\eta}{v_\chi}\right)^2\right)\right)t_R+...\notag\\
&&-\frac{m_{t0}}{\sqrt{1-\left(\frac{v}{v_\chi}\right)^2}}\sqrt{1-\left(\frac{h}{v_\chi}\right)^2-\left(\frac{\eta}{v_\chi}\right)^2}\frac{h}{v}\frac{\chi}{v_\chi}\bar{t}_Lt_R+h.c.-V_{eff}(h,\eta,\chi)\notag\\
&&+\frac{\alpha_s}{8\pi}\left(b^3_{IR}-b^3_{UV}\right)\mathrm{log}\left(\frac{\chi}{v_\chi}\right)G^a_{\mu\nu}G^{a\mu\nu}+\frac{\alpha_{em}}{8\pi}\left(b^{em}_{IR}-b^{em}_{UV}\right)\mathrm{log}\left(\frac{\chi}{v_\chi}\right)F_{\mu\nu}F^{\mu\nu}.
\end{eqnarray}
The effective potential term becomes
\begin{eqnarray}
	V_{eff}(h,\eta,\chi)=\left(\frac{\chi}{v_\chi}\right)^4\left(\frac{\mu_h^2}{2}h^2+\frac{\mu_\eta^2}{2}\eta^2+\frac{\lambda_h}{4}h^4+\frac{\lambda_\eta}{4}\eta^4+\frac{\lambda_{h\eta}}{2}h^2\eta^2\right)+c_\chi g_\chi^2\chi^4-\epsilon(\chi)\chi^4.
\end{eqnarray}
If the IR contribution predominates, the effective potential is dictated by the effective Lagrangian resulting from integrating out the heavy resonances.Consequently, the coefficients in the potential are interrelated, as they can all be represented in terms of integrals of the form factors.~\cite{bian_electroweak_2019}.
The Weinberg sum rules are also taken into account to eliminate the divergence~\cite{marzocca_general_2012}.
Note that the last line of Eq.~\eqref{eq:effLagchi} demonstrates the interactions between the dilaton and the SM gauge bosons, facilitated by trace anomalies~\cite{chacko_effective_2013}. The parameter $b_{UV}$ receives contributions from both the strongly-coupled sector and elementary fields, whereas $b_{IR}$ receives additional contributions from light degrees of freedom that emerge after the spontaneous breaking of SO(6) symmetry. The values of these parameters are model-specific.

To render the kinetic terms of physical fields canonical, we can choose the parameterization of the fluctuations around the vacuum as follows:
\begin{eqnarray}\label{eq:physiei}
	h=v+\sqrt{1-\xi}\hat{h},\quad \eta=\eta,\quad \chi=v_\chi+\hat{\chi},
\end{eqnarray}
where $\xi\equiv v^2/v_\chi^2$ quantifies the extent of the vacuum misalignment angle.
Finally, the complete effective Lagrangian is presented in Appendix~\ref{A}.

\section{Dark matter phenomenology}\label{sec:DM}

It is straightforward to verify that the Lagrangian Eq.~\eqref{eq:fullEffL} presented in Appendix~\ref{A} remains invariant under a $Z_2$ transformation, $\eta\to-\eta$.
We assume that this symmetry is maintained throughout the composite sector, and higher derivative terms involving Wess-Zumino-Witten terms are absent. Under these conditions, $\eta$ is protected by the $Z_2$ symmetry and thus can serve as a viable dark matter (DM) candidate.

\begin{figure}[htp!!]
	\centering
	\subfigure{
		\begin{minipage}[t]{0.5\linewidth}
			\centering
			\includegraphics[scale=0.7]{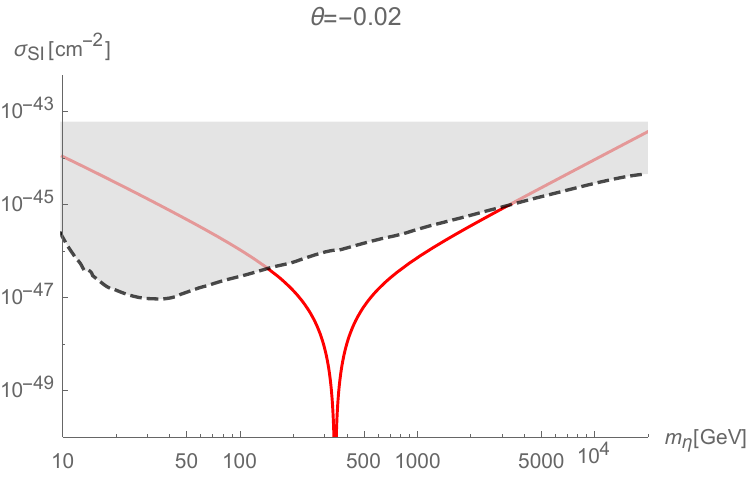}
	\end{minipage}}%
	\subfigure{
		\begin{minipage}[t]{0.5\linewidth}
			\centering
			\includegraphics[scale=0.7]{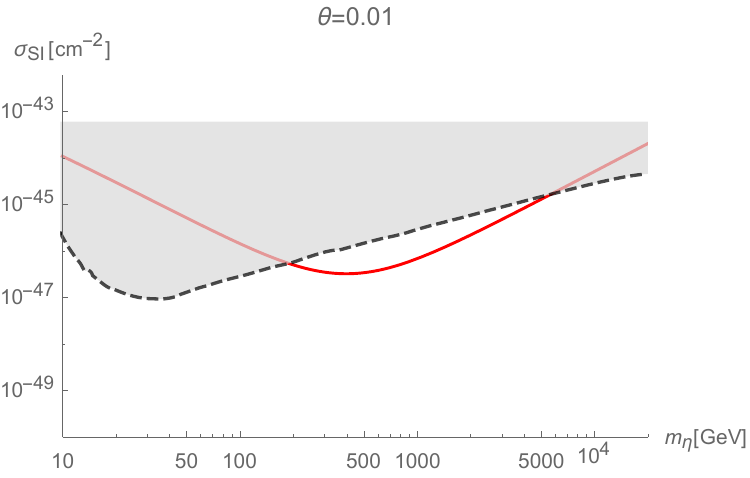}
	\end{minipage}}	
	\caption{
		The DM-nucleon cross section with different mixing angle $\theta$. The gray shaded area is excluded by the LUX-ZEPLIN (LZ) bound and the red solid lines denote the theoretical calculation of the spin independent scattering cross section. Different mixing angle leads to different exclusion range of DM mass, with $\theta=-0.02$ for the left panel and $\theta=0.01$ for the right panel. The accidental cancellation occurs only if $\theta$ is negative.
		}\label{fig:sigmaSI_theta}
\end{figure}

Direct detection experiments for DM can impose stringent constraints on the parameter space of WIMP DM. In the following discussion, we will use the upper bound of the DM-nucleon cross-section adopted from the LUX-ZEPLIN (LZ) experiment~\cite{aalbers_first_2023}.
In the context of our model, there are basically three kinds of interaction vertices related to the direct detection: the Higgs portal $\eta\mbox{-}\eta\mbox{-}h$, the dilaton portal $\eta\mbox{-}\eta\mbox{-}\chi$, and the contact interactions with fermions $\eta\mbox{-}\eta\mbox{-}f\mbox{-}\bar{f}$.
The first two interactions contribute to the DM-nucleon scattering through t-channel processes mediated by Higgs and dilaton. 
The contact interactions between the DM and quarks are relatively suppressed as they are induced by higher dimensional operators. The effective Lagrangian for these interactions is formulated as follows:
\begin{eqnarray}
  \mathcal{L}_{direct}&=&\frac{\xi}{\sqrt{1-\xi}}\frac{\eta}{v}\partial_{\mu}\hat{h}\partial^\mu\eta-\lambda_{h\eta}v\sqrt{1-\xi}\eta^2\hat{h}\notag\\
  &&+\sqrt{\xi}\frac{\hat{\chi}}{v}(\partial_{\mu}\eta)^2-2\frac{m^2_{\eta}}{v_\chi}\eta^2\hat{\chi}\notag\\
  &&+\frac{\xi}{2(1-\xi)v^2}\eta^2\bar{\psi}_f\psi_f.
\end{eqnarray}
Since the incoming and outgoing $\eta$ are on-shell state during the DM-nucleon scattering process, we can use integration by part and equation of motion to rewrite the derivative terms. Finally, the portal interactions can be expressed as 
\begin{eqnarray}
	\mathcal{L}_{\eta\eta\hat{h}}&\simeq&-\lambda_{h\eta}v\sqrt{1-\xi}\eta^2\hat{h}\notag\\
	\mathcal{L}_{\eta\eta\hat{\chi}}&\simeq& -\sqrt{\xi}\frac{m^2_\eta}{v}\hat{\chi}\eta^2-\frac{2m_{\eta}^2}{v_\chi}\hat{\chi}\eta^2=-3\frac{m_{\eta}^2}{v_\chi}\hat{\chi}\eta^2.
\end{eqnarray}
On the other hand, the Higgs field and dilaton field are mixing, and thus we should diagonalize their mixing mass matrix to figure out the mass eigenstates $(h_p,\chi_p)$.
\begin{eqnarray}
	h_p=\mathrm{cos}\theta \hat{h}+\mathrm{sin}\theta \hat{\chi},\quad \chi_p=\mathrm{cos}\theta \hat{\chi}-\mathrm{sin}\theta \hat{h}.
\end{eqnarray}
where $\theta$ is the mixing angle. $h_p$ is regarded as the SM-like Higgs boson whose mass is about $126$~GeV.
If we consider the physical mass of Higgs and dilaton as input parameters, the mass matrix elements can be written in terms of $\theta$, physical mass $m_{hp}$, and $m_{\chi p}$ in the following way
\begin{eqnarray}
	m_{h}^2=c_\theta^2 m^2_{h_p}+s_\theta^2 m^2_{\chi_p},\quad m_{\chi}^2=c_\theta^2 m^2_{\chi_p}+s_\theta^2 m^2_{h_p},\quad m^2_{h\chi}=s_\theta c_\theta(m^2_{h_p}-m^2_{\chi_p}).
\end{eqnarray}
The mass mixing term is induced by CFT violation and shift symmetry violation which can be parametrized as~\cite{bruggisser_dilaton_2023}:
\begin{eqnarray}
	m^2_{h\chi}\propto g_{CFT}^2v_\chi^2\lambda_\psi^2(\gamma_{\text{violation}}).
\end{eqnarray}
where $\gamma_{\text{violation}}$ is an undetermined parameter depending on UV theory.
In practice, we will use the the mixing angle $\theta$ as an input instead of the mixing mass-squared.
The spin-independent DM-nucleon scattering cross section $\sigma_\mathrm{SI}$ can be computed as:
\begin{eqnarray}\label{eq:sigma_SI}
	\sigma_\mathrm{SI}&=&\frac{\mu^2m_N^2}{4\pi v^2m_\eta^2}\left[\frac{c_\theta a_{h_p}}{m^2_{h_p}}-\frac{s_\theta a_{\chi_p}}{m^2_{\chi_p}}+\omega_\chi\left(\frac{c_\theta a_{\chi_p}}{m^2_{\chi_p}}+\frac{s_\theta a_{h_p}}{m^2_{h_p}}\right)+\frac{\xi}{2(1-\xi)v}\right]^2F_N^2\notag\\
	&=&\frac{\mu^2m_N^2}{4\pi v^2m_\eta^2}\left[\frac{(c_\theta+s_\theta\omega_\chi) a_{h_p}}{m^2_{h_p}}+\frac{(c_\theta\omega_\chi -s_\theta)a_{\chi_p}}{m^2_{\chi_p}}+\frac{\xi}{2(1-\xi)v}\right]^2F_N^2,
\end{eqnarray}
where $\omega_\chi=\sqrt{\xi}(1+\gamma_\psi)$, $\mu=\frac{m_Nm_\eta}{m_N+m_\eta}$ is the reduced mass, and $F_N\sim 0.3$ is the hardron matrix element~\cite{xu_identifying_2023}.
The effective coupling coefficients between $(h_p,\chi_p)$ and DM $\eta$ are encoded in
\begin{eqnarray}
	a_{h_p}=3\frac{m_\eta^2}{v_\chi}s_\theta+\lambda_{h\eta}v\sqrt{1-\xi}c_\theta,\quad a_{\chi_p}=3\frac{m_\eta^2}{v_\chi}c_\theta-\lambda_{h\eta}v\sqrt{1-\xi}s_\theta.
\end{eqnarray}
Due the mass mixing between the $h$ and $\hat{\chi}$, the cross-section is different from the ordinary NMCHM without dilaton.
In NMCHM$_\chi$ model, the scattering amplitude could happen to be suppressed due to an accidental cancellation between the diagrams induced by $h_p$ and $\chi_p$ mediation~\cite{chao_gravitational_2017}. In such situation, the DM-nucleon scattering cross section can circumvent the stringent constraint from direct detection data (see Fig.~\ref{fig:sigmaSI_theta}).

\begin{figure}
	\centering
	\subfigure{
		\begin{minipage}[t]{0.5\linewidth}
			\centering
			\includegraphics[width=\linewidth]{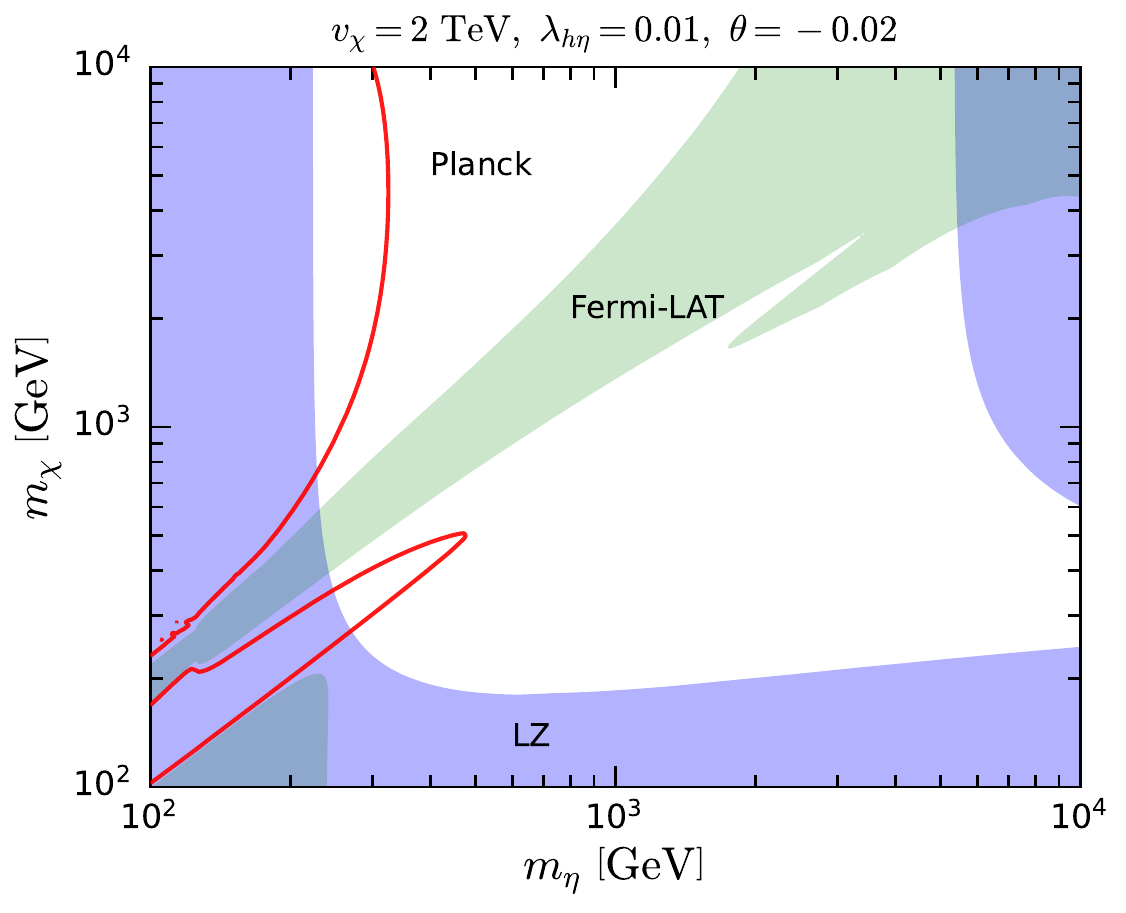}
	\end{minipage}}%
	\subfigure{
		\begin{minipage}[t]{0.5\linewidth}
			\centering
			\includegraphics[width=\linewidth]{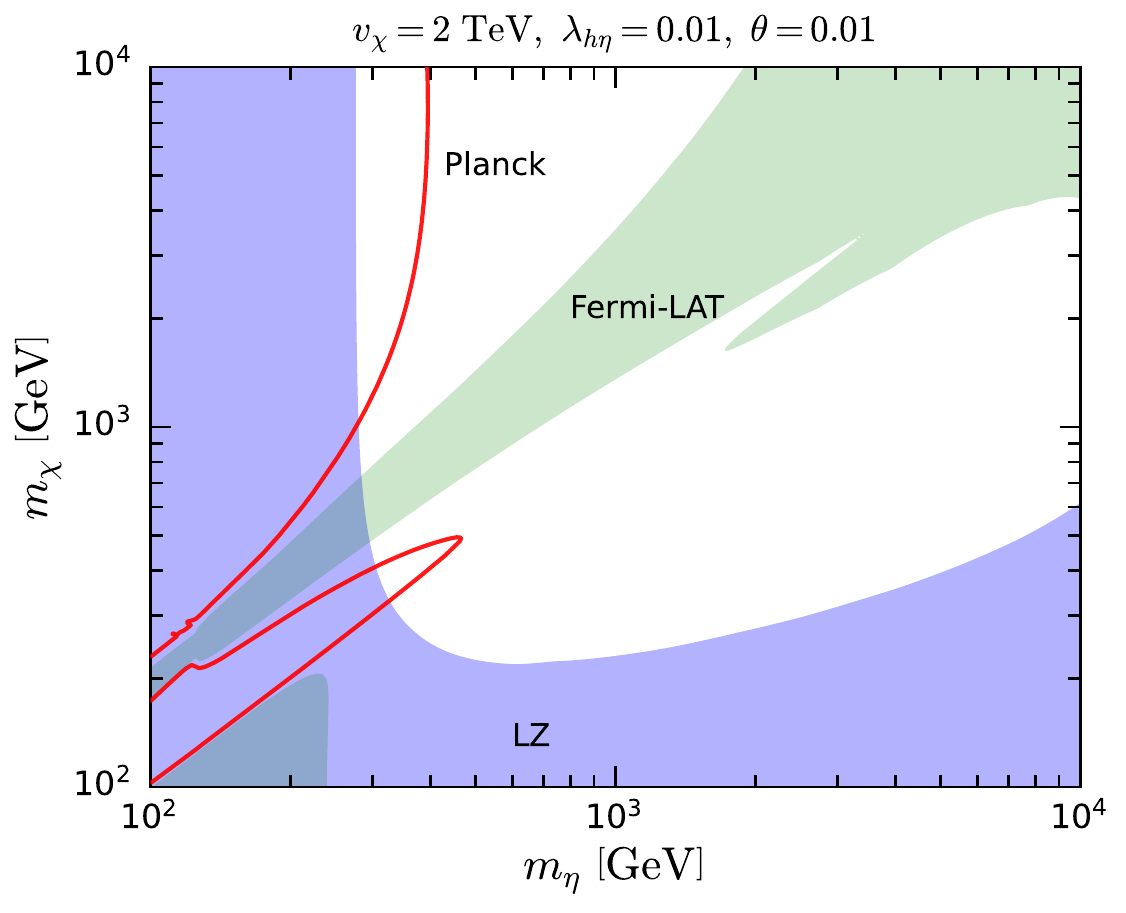}
	\end{minipage}}	
	\caption{
		 The exclusion regions with different mixing angle $\theta=-0.02$ on the left and $\theta=0.01$ on the right. The blue and green regions are excluded by the LZ experiment~\cite{aalbers_first_2023} and the Fermi-MAGIC indirect detection respectively~\cite{MAGIC:2016xys}. The red solid lines denote the DM relic density $\Omega_\mathrm{DM}h^2=0.12$ observed by Planck experiment~\cite{planck_collaboration_planck_2020}. 
		}\label{fig:DM_theta}
\end{figure}

In addition to direct detection, indirect detection bound and observed relic density of DM also strictly constrain on the parameter space of WIMPs. Both of these constraints are sensitive to the annihilation of cold dark matter, and we find the dominant annihilation channels are $\eta+\eta\to h_p+h_p$ and $\eta+\eta\to \chi_p+\chi_p$ if $m_\eta>m_{\chi_p}$. For the indirect detection constraint, we use the bound implied by the gamma-ray data from Major Atmospheric Gamma-ray Imaging Cherenkov (MAGIC) telescopes and the Fermi Large Area Telescope (LAT)~\cite{MAGIC:2016xys}. 
For the relic density, we adopt result $\Omega_{\textrm{DM}}=0.12$ from the Planck experiment~\cite{planck_collaboration_planck_2020}.
The numerical package \texttt{microOMEGAs}~\cite{belanger_micromegas41_2015} is used to compute the DM annihilation cross section, relic density, and DM-nucleon scattering cross section.
In Fig.~\ref{fig:DM_theta}, we show all these constraints in $m_\eta$-$m_\chi$ plane for two benchmark models with different $\theta$. In both cases, there are available parameter region surviving from all the DM constraints and the mass of DM $\eta$ is about $300$ GeV$\sim$ $600$ GeV.

\begin{figure}[htp!]
\vspace{-0.8cm}
\begin{center}
\fbox{\footnotesize $v_\chi =2\text{~TeV}$} \\
\hspace*{-0.65cm} 
\begin{minipage}{0.43\linewidth}
\begin{center}
	\includegraphics[width=\linewidth]{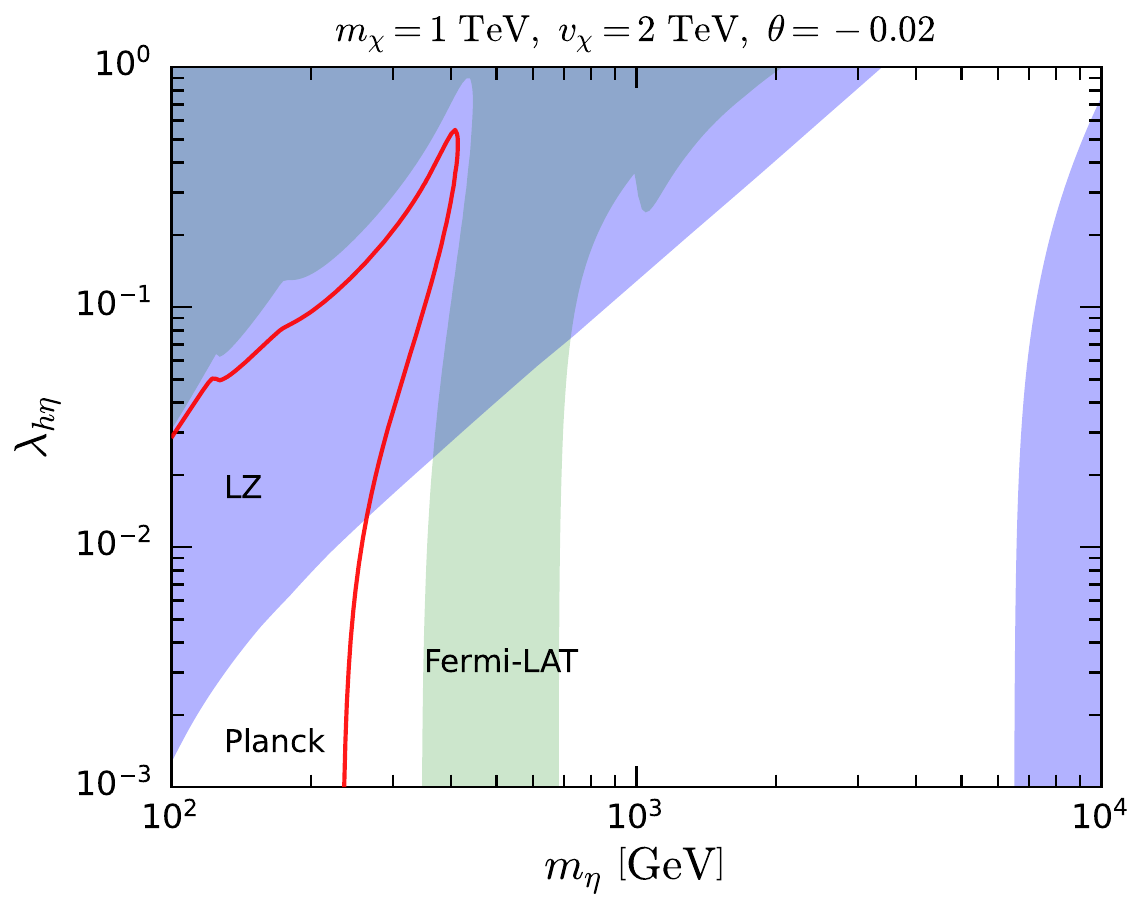}
\end{center}
\end{minipage}
\begin{minipage}{0.43\linewidth}
\begin{center}
	\includegraphics[width=\linewidth]{DM_vchi_2.0E+03,lamheta_1.0E-02,theta_-2.0E-02.pdf}
\end{center}
\end{minipage}
\\[0.1cm]
\fbox{\footnotesize $v_\chi =2.5\text{~TeV}$} \\
\hspace*{-0.65cm} 
\begin{minipage}{0.43\linewidth}
\begin{center}
	\includegraphics[width=\linewidth]{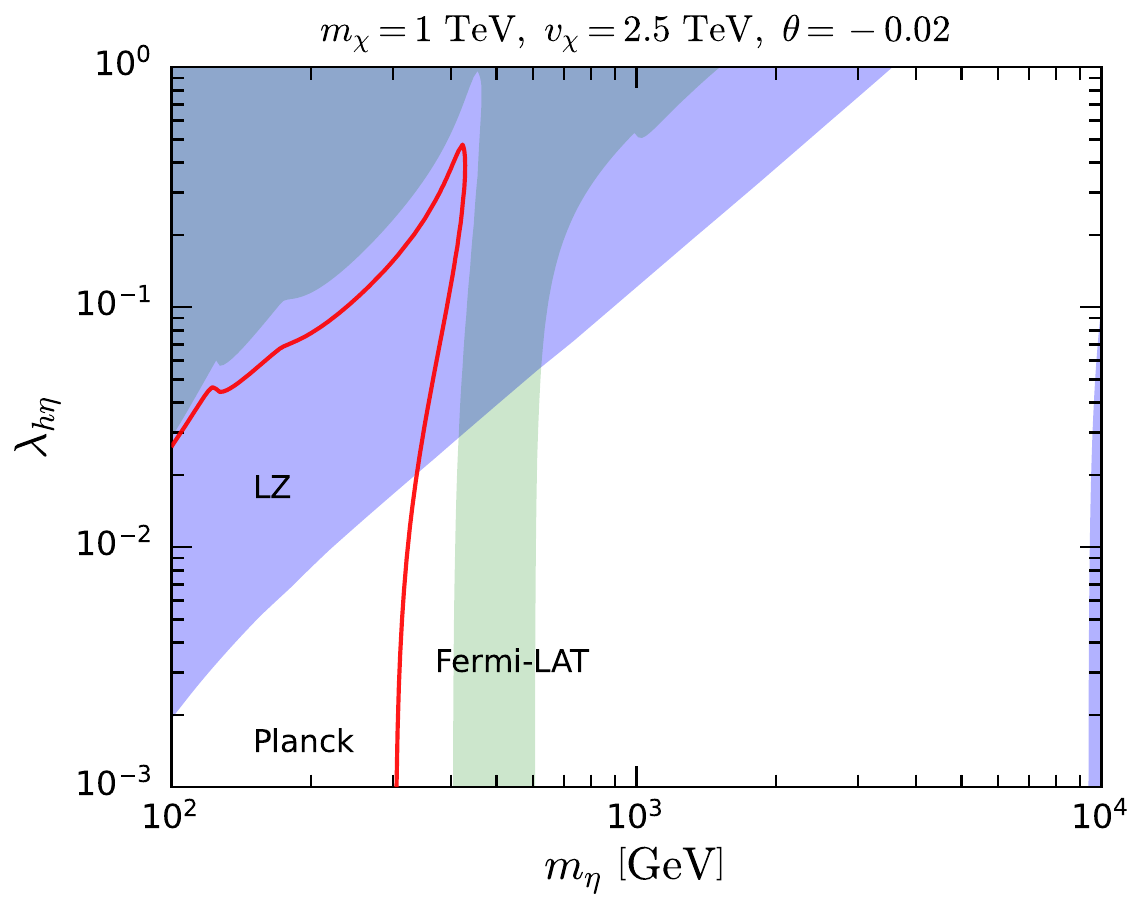}
\end{center}
\end{minipage}
\begin{minipage}{0.43\linewidth}
\begin{center}
	\includegraphics[width=\linewidth]{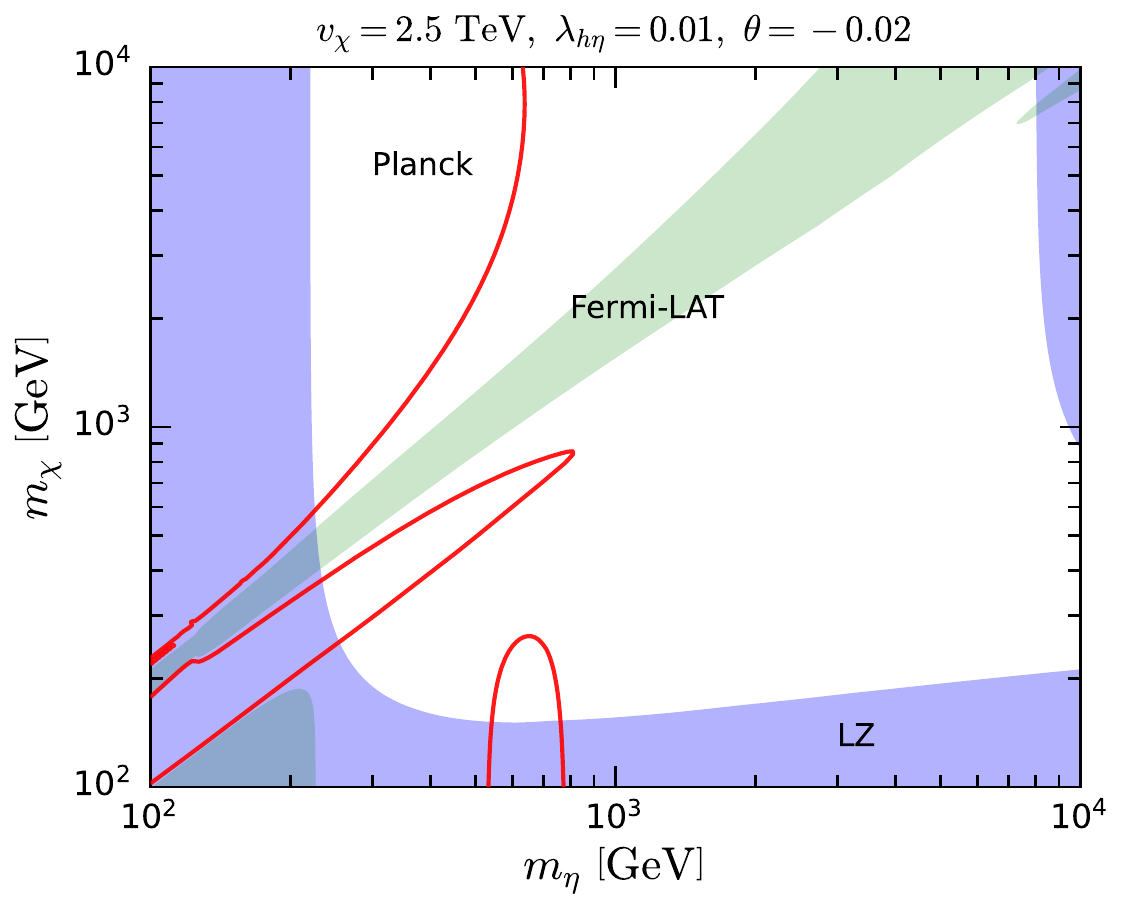}
\end{center}
\end{minipage}
\\[0.1cm]
\fbox{\footnotesize $v_\chi =3\text{~TeV}$} \\
\hspace*{-0.65cm} 
\begin{minipage}{0.43\linewidth}
\begin{center}
	\includegraphics[width=\linewidth]{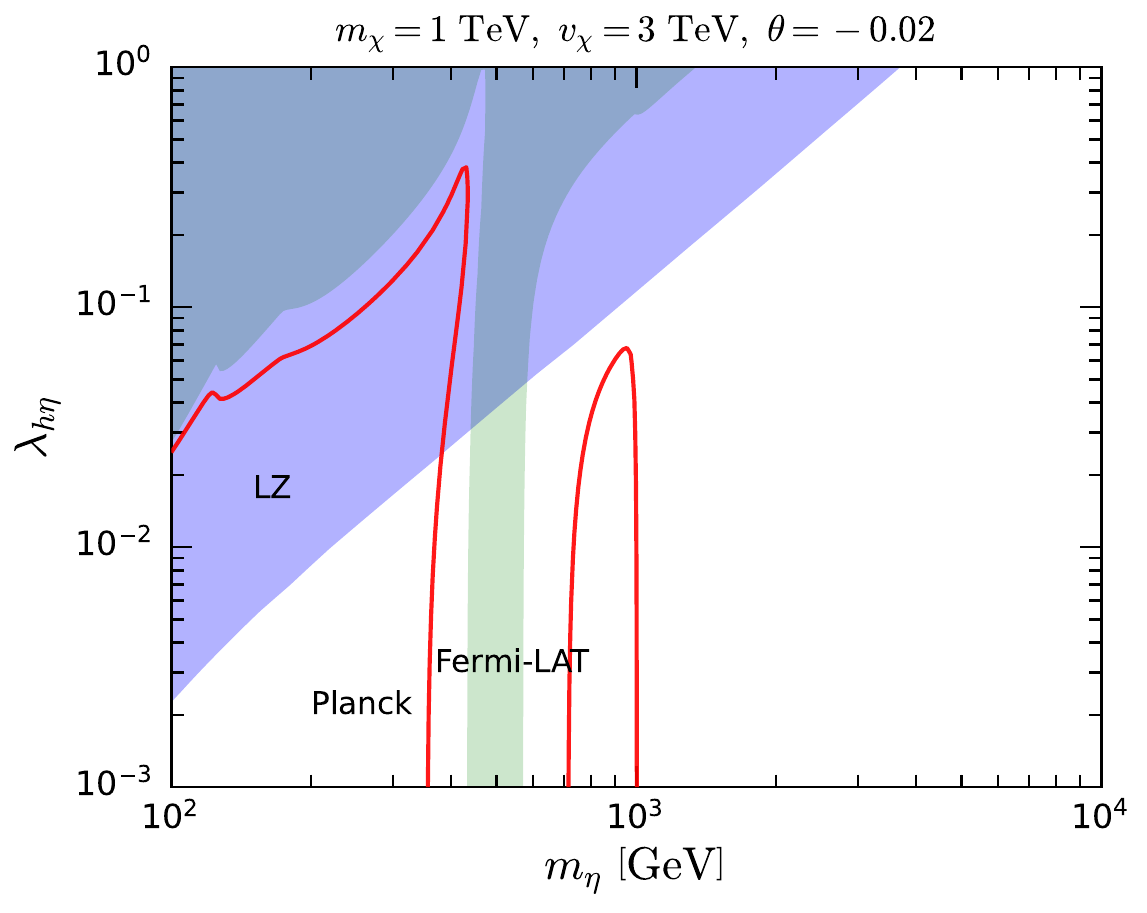}
\end{center}
\end{minipage}
\begin{minipage}{0.43\linewidth}
\begin{center}
	\includegraphics[width=\linewidth]{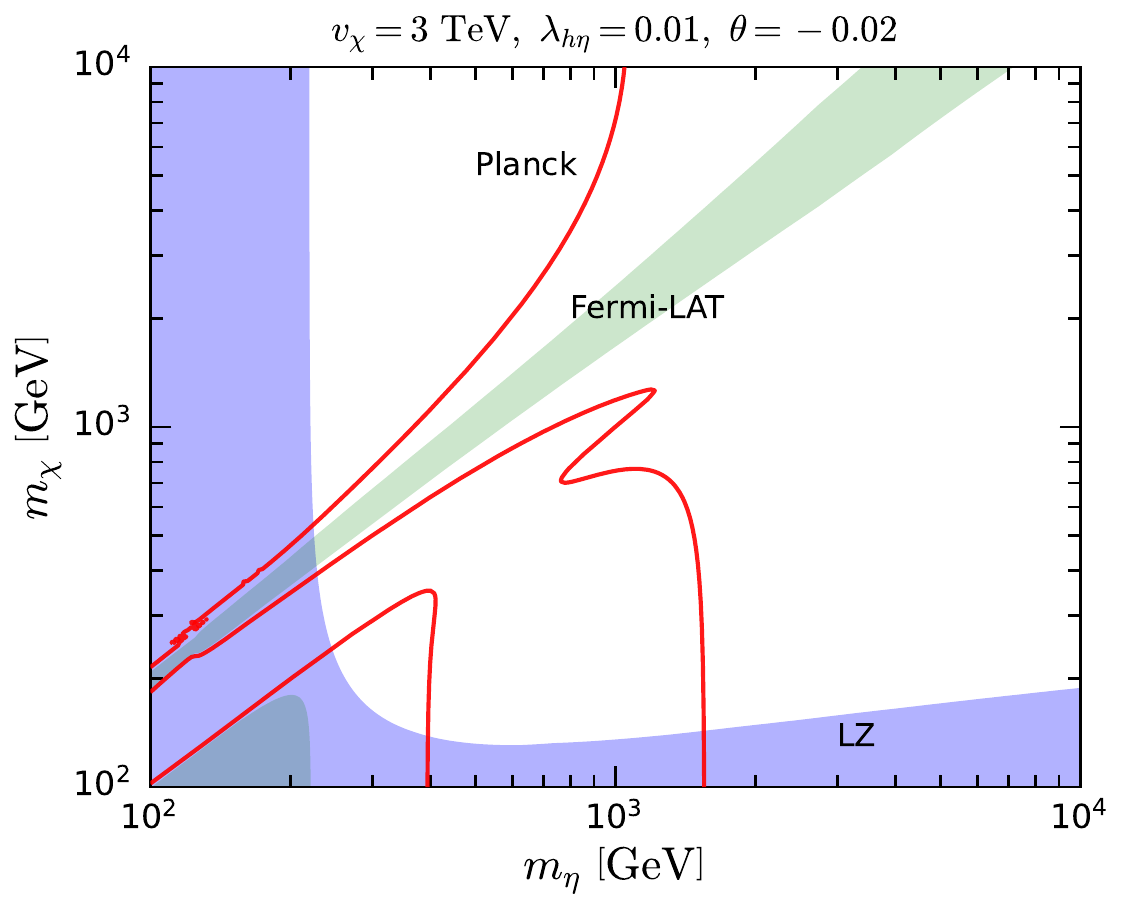}
\end{center}
\end{minipage}
\end{center}
\vspace*{-0.2cm}
\caption{The parameter constraint from DM phenomenology. The legend is as same as the illstration in Fig.~\ref{fig:DM_theta}. On the left panel we scan the $m_\eta$-$\lambda_{h\eta}$ plane with fixed dilaton mass $m_\chi=1~\text{TeV}$ while on the right panel we scan $m_\eta$-$m_{\chi}$ plane fixed Higgs-DM coupling $\lambda_{h\eta}=0.01$.
}\label{fig:CombinedPlot}
\end{figure}

In Fig.~\ref{fig:CombinedPlot}, we fix the mixing angle $\theta=-0.02$ and scan the $m_\eta$-$\lambda_{h\eta}$ plane (left hand side) and $m_\eta$-$m_{\chi}$ plane (right hande side) with different dilaton VEV $v_\chi=2~\text{TeV},~2.5~\text{TeV},~3~\text{TeV}$.
For the panels in the left, the regions around $m_\eta=500~\text{GeV}$ are excluded by the indirect detection bound. It is caused by the resonant effect due to the energy of incoming states reaches the mass pole at $m_\eta$. Since the annihilation cross section changes rapidly around this resonant region, the mass preferred by relic density is restricted in a a narrow interval. 
Note that there is another red contour of relic density appears in the region of $700~\textrm{GeV}<m_\eta<1$~TeV in the case of $v_\chi=3\text{~TeV}$. The nearly vertical line located at the value $700~\textrm{GeV}$ can be understood as the result of the resonance, while the vertical line located at the value $1~\textrm{TeV}$ is due to the opening of $\eta+\eta\to\chi_p+\chi_p$ channel.
Since the coupling of DM and dilaton is proportional to $m_\eta^2$, the annihilation probability of $\eta$ turns out to be larger with the increase of $m_\eta$, and thus reduce the dark matter relic density.
As a result, the relic density in the large DM mass region is too small to meet the observed data.

For the panels in the right, which show the constraints on the $m_\eta$-$m_\chi$ plane, we can also find that a region with $m_\eta/m_\chi\approx2$ is excluded by the indirect detection due to the resonant effect. In cases of $v_\chi=2.5$~TeV and $3$~TeV, there is another contour of $\Omega_{\textrm{DM}}h^2=0.12$ near the region $m_\eta\sim700$~GeV. The reason is that the $\eta\eta\to\chi\chi$ process dominates the DM annihilation in this region, and there are interference among the $s,t,u$-channels and the contact interaction channel diagrams.
In the limit of $\theta\to0$, the annihilation cross section can be approximated by
\begin{eqnarray}
  \sigma v\propto \frac{1}{ m_\eta^2} \left[-\frac{1}{ \left(4 m_\eta^2-m_\chi^2\right)}\lambda^{(3)}_{\chi}\left(\frac{6m_\eta^2}{v_\chi}\right) -\frac{2}{2 m_\eta^2-m_\chi^2} \left(\frac{4m_\eta^2 }{v_\chi}\right)^2+\frac{14 m_\eta^2}{v_\chi^2}\right]^2.
\end{eqnarray}
The minimum of the annihilation cross section is found to locate at $m_\eta\approx562~\text{GeV}$ for $v_\chi=2~\text{TeV}$, at $m_\eta\approx639~\text{GeV}$ for $v_\chi=2.5~\text{TeV}$, and at $m_\eta\approx708~\text{GeV}$ for $v_\chi=3~\text{TeV}$.
We can see that these $m_\eta$ can match the median point between the intersection points of the red contours and the $m_\eta$ axis.

\section{Phase transition dynamics of NMCHM$_\chi$}\label{sec:PT}
\subsection{The effective potential}
Before we start exploring the features and the patterns of PTs, it is imperative to have a thorough understanding of the effective potential~\cite{laine_basics_2016,linde_phase_1979,Linde:1980tt,sher_electroweak_nodate,quiros_finite_1999,hindmarsh_phase_2021}.
Typically, the effective potential of the Next-to-Minimal Composite Higgs Model with Dilaton (NMCHM$_\chi$), which encompasses the Higgs boson $h$, the dark matter $\eta$, and the dilaton $\chi$ is comprised of the following components:
\begin{eqnarray}\label{Veff0}
    V_\mathrm{eff}(h,\eta,\chi) = V_\mathrm{tree}(h,\eta,\chi)+V_\mathrm{1-loop}(h,\eta,\chi)+V_\mathrm{CT}(h,\eta,\chi)+V_\mathrm{T}(h,\eta,\chi).
\end{eqnarray}
In the Composite Higgs Model, the potential of pNGBs originates from the explicit breaking of global symmetry. This is already at one-loop order and is renormalized by the Weinberg sum rules, as discussed in Section~\ref{sec:model}.
Consequently, there is no necessity to incorporate an additional zero-temperature one-loop correction to the potential. The daisy resummation is also omitted due to its negligible impact on the PT dynamics. The first three terms of Eq.\eqref{Veff0}, denoted as $V(h,\eta,\chi)$, possess a specific form:
\begin{eqnarray}
	V(h,\eta,\chi)
	=\frac{\chi^{2}}{v_{\chi}^{2}}\left(\frac{1}{2}\mu_{h}^{2}h^{2}+\frac{1}{2}\mu_{\eta}^{2}\eta^{2}\right)+\frac{1}{4}\lambda_{h}h^{4}+\frac{1}{4}\lambda_{\eta}\eta^{4}+\frac{1}{2}\lambda_{h\eta}h^{2}\eta^{2}+V_0(\chi),
\end{eqnarray}
where $v_{\chi}$ is the VEV of $\chi$ at zero temperature and 
\begin{eqnarray}
V_{0}(\chi)=c_{\chi}g_{\chi}^{2}\chi^{4}-\epsilon(\chi)\chi^{4}.
\end{eqnarray}

To compute the finite temperature effective potential and circumvent the singularity at the origin $\chi=0$, we introduce the canonical field variables $\chi_1$, $\chi_2$, and $\chi_3$ as replacements for $h$, $\eta$, and $\chi$. The scalar fields are redefined such that $\chi_{1} \equiv h$, $\chi_{2} \equiv \eta$, and $\chi_{3} \equiv \sqrt{\chi^{2}-\chi_{1}^{2}-\chi_{2}^{2}}$. 
Then the zero-temperature potential in terms of the new field variables are given by
\begin{eqnarray}
	V(\chi_{1},\chi_{2},\chi_{3}) &=(\frac{\mu_{h}^{2}}{2v_{\chi}^2}+\frac{\lambda_{h}}{4}+c_{\chi}g_{\chi}^{2})\chi_{1}^{4}+(\frac{\mu_{\eta}^{2}}{2v_{\chi}^2}+\frac{\lambda_{\eta}}{4}+c_{\chi}g_{\chi}^{2})\chi_{2}^{4}+c_{\chi}g_{\chi}^{2}\chi_{3}^{4}\notag\\
	&\quad +(\frac{\mu^{2}_{\eta}+\mu^{2}_{h}}{2v_{\chi}^{2}}+\frac{\lambda_{h\eta}}{2}+2c_{\chi}g_{\chi}^{2})\chi_{1}^{2}\chi_{2}^{2}+(\frac{\mu_{\eta}^{2}}{2v_{\chi}^{2}}+2c_{\chi}g_{\chi}^{2})\chi_{2}^{2}\chi_{3}^{2}\notag\\
	&\quad +(\frac{\mu_{h}^{2}}{2v_{\chi}^{2}}+2c_{\chi}g_{\chi}^{2})\chi_{1}^{2}\chi_{3}^{2} -\epsilon(\chi)(\chi_{1}^{2}+\chi_{2}^{2}+\chi_{3}^{2})^{2}.
\end{eqnarray}

The finite-temperature terms are vital in accomplishing the first-order phase transition. These terms induce a dip at the origin of the scalar field coordinates, thereby forming a barrier between the origin $(\chi_1,\chi_2,\chi_3)=(0,0,0)$ (deconfined phase) and the EW vacuum $(\chi_1,\chi_2,\chi_3)\approx(v,0,v_\chi)$ (confined phase). Following the strategy outlined in Ref.~\cite{bruggisser_electroweak_2018}, we assume that the theory in the deconfined phase is characterized by a free energy given by $F\simeq -cN^2T^4$, where $c=\pi^2/8$ corresponds to the $\mathcal{N}=4 ~\mathrm{SU}(N)$ super-Yang-Mills theory. As the temperature decreases, the vacuum configuration $(\chi_1,\chi_2,\chi_3)$ undergoes a transition from the origin to the EW vacuum or an intermediate state due to quantum tunneling effect. In the confined phase ($\chi\gtrsim T/g_\chi$), most of confined states have masses larger than temperature and their thermal corrections can be neglected.
In the low-energy effective theory, only the light degrees of freedom need to be considered. These include the SM gauge bosons $W$ and $Z$, the SM quark $q$, the Goldstone bosons arising from SSB, the dilation $\chi$, and the CFT resonances from the composite sector. The free energy for the confined phase can be expressed as
\begin{eqnarray}
F\simeq V(\chi_1,\chi_2,\chi_3)+ V_{T}(\chi_1,\chi_2,\chi_3).
\end{eqnarray}
The second term represents the finite temperature potential~\cite{quiros_finite_1999}:
\begin{eqnarray}\label{eq:VT}
	 V_\mathrm{T}(\chi_1,\chi_2,\chi_3)=\sum_{i=\text{bosons}}\frac{n_i T^4}{2\pi^2}J_\mathrm{b}\left[\frac{m_i^2(\chi_1,\chi_2,\chi_3)}{T^2}\right]-\sum_{i=\text{fermions}}\frac{n_i T^4}{2\pi^2}J_\mathrm{f}\left[\frac{m_i^2(\chi_1,\chi_2,\chi_3)}{T^2}\right],
\end{eqnarray}
where the expressions for the effective masses $m_i^2$ of each particle species are detailed in appendix~\ref{appdB}. The degrees of freedom for each types of particles are given by
\begin{eqnarray}
 n_{W}=6,\quad n_{Z}=3,\quad n_{t}=12,\quad n_{\Pi}=3,\quad n_{\chi_1}=n_{\chi_2}=n_{\chi_1}=1,\quad \sum n_\mathrm{CFT}=\frac{45}{4}N^2.
\end{eqnarray}
Note that we have simplified the contributions of CFT resonances by assuming that all of them are bosonic states, as suggested by Ref.~\cite{bruggisser_electroweak_2018}. The value of $n_\mathrm{CFT}$ is chosen to reproduce the free energy of the deconfined phase in the $\chi\to0$ limit. This assumption defines an interpolation of the effective potential between the interval $0<\chi\lesssim T/g_\chi$. 
The functions $J_{\mathrm{b(f)}}$ in Eq.\eqref{eq:VT} are standard and defined as follows:
\begin{eqnarray}
	J_\mathrm{b}[x]=\int^{\infty}_0dk k^2\mathrm{log}\left[1-e^{-\sqrt{k^2+x}}\right],\quad J_\mathrm{f}[x]=\int^{\infty}_0dk k^2\mathrm{log}\left[1+e^{-\sqrt{k^2+x}}\right].
\end{eqnarray}

\subsection{Phase transition and bubbles nucleation}

To gain an intuitive understanding of the first-order phase transition, we can simplify the analysis by neglecting the $\chi_1$ and $\chi_2$ directions for the moment. This approximation is motivated by the large number of degrees of freedom in the CFT sector whose masses are $\sim g_\chi \chi\approx g_\chi \chi_3$. Consequently, the potential barrier is predominantly determined by $\chi_3$.

As temperature decreases, the free energy of the symmetry-preserving vacuum ($\chi_3 = 0$) increases and equals that of the symmetry-breaking vacuum ($\chi_3 = v_\chi$) at a critical temperature $T_c$. Consequently, the symmetry-preserving vacuum becomes a false vacuum, prone to quantum tunneling to the true vacuum ($\chi_3 = v_\chi$). This process leads to the random formation of true vacuum bubbles throughout the universe.

To estimate the critical temperature $T_c$, we approximate the free energy of the symmetry-breaking vacuum by considering only the zero-temperature terms, since the $J_\mathrm{b(f)}(m/T)$ terms are exponentially suppressed in the limit $m/T\gg1$. This yields the free energy:
\begin{eqnarray}\label{eq:V0min}
	F(\chi_3=v_\chi)\approx V_{\chi}^\mathrm{min}\simeq \frac{\gamma_\epsilon}{4}c_{\chi}g^2_\chi v_{\chi}^4=-\frac{1}{16}m_\chi^2v_\chi^2,
\end{eqnarray}
which agrees with the result obtained in Ref.~\cite{bruggisser_electroweak_2018}. For the symmetry-preserving vacuum, we approximate the free energy as $F(0)\simeq -cN^2T^4$. Solving for $T_c$, we find
\begin{eqnarray}
    -\frac{\pi^2}{8}N^2T_c^4=-\frac{1}{16}m_\chi^2v_\chi^2\Rightarrow T_c=\frac{1}{(2\pi^2)^{\frac{1}{4}}}\left(\frac{m_\chi v_\chi}{N}\right)^{\frac{1}{2}}.
\end{eqnarray}

At the critical temperature $T_c$, the tunneling probability is still too low to produce bubbles, so the phase transition does not occur yet. The universe remains in the false vacuum state until a bubble is able to be produced within the Hubble volume per Hubble time. The temperature corresponding to this is called the nucleation temperature $T_n$. The computation of the nucleation temperature $T_n$ is detailed in Appendix~\ref{appdE}. It is worth noting that, in the NMCHM$_\chi$ model, $T_n$ can be much lower than the critical temperature, a situation known as supercooling, especially when $N$ is large and the dilaton mass $m_\chi$ is small~\cite{bruggisser_electroweak_2018}. However, to ensure dilaton potential to be bounded from below, it requires
\begin{eqnarray}\label{eq:mchi_lowbound}
    c_\epsilon<-\frac{\gamma_{\epsilon}}{c_\chi}
    \Rightarrow m_\chi^2 N&>4(4\pi)^2c_\chi^2v_\chi^2,
\end{eqnarray}
which sets a lower bound of $m_\chi$ for a given $N$.

Supercooling occurs in this model because the effective potential for the dilaton has a very wide barrier between $\chi=0$ and $\chi\sim v_\chi$, which is common in nearly conformally symmetric models~\cite{von_harling_qcd-induced_2018}. This barrier significantly suppresses the tunneling rate. However, it is possible that bubble nucleation never occurs because the false vacuum decay rate always stays below the Hubble rate. In such cases, the supercooled state would end through the growth of quantum fluctuations~\cite{bea_spinodal_2021}. This type of situation is beyond the scope of our research, and we only consider the case of phase transitions that can be accomplished by vacuum decay.

When the $\chi_1,\chi_2$ directions are included, the overall phase transition behavior from the deconfined phase to the confined phase remains essentially unchanged. However, the EW symmetry can be preserved during the confinement transition, followed by a second phase transition that spontaneously breaks the EW symmetry. In the $\chi>T/g_\chi$ region, the minimum of the potential along the $\chi_3$ direction is almost $T$-independent, and thus we can treat $\chi$ as a constant when we determine the second step phase transition. The concrete expression of the free energy with respect to $\chi_1,~\chi_2$ and $\chi$ is given in Appendix~\ref{appdC}.
From Eq.\eqref{eq:chithermalpotential} and \eqref{eq:chithermalpotentialcoe}, we can obtain the critical temperature $\tilde{T}_\mathrm{c}^h$ at which $h=0$ is no longer the local minimum.
\begin{eqnarray}
    \frac{\chi^2}{v^2_\chi}\mu_h^2+c_h \tilde{T}_\mathrm{c}^h=0\Rightarrow \tilde{T}_\mathrm{c}^h=\sqrt{\frac{-12\mu_h^2}{\left(\frac{6m_W^2}{v^2}+\frac{3m_Z^2}{v^2}+6\lambda_h+\lambda_{h\eta}+\frac{\mu^2_h}{v_\chi^2}+\frac{6m_t^2}{v^2(1-\frac{v^2}{v_\chi^2})}\right)}}\approx 140 \text{~GeV},
\end{eqnarray}
where we have made an approximation $\chi\approx v_\chi$ in the second equality. 
Note that our numerical computation, which will be presented in the next subsection, indicates that a phase with $\chi_2\neq0$ does not exist in the parameter space of interest.

The transition rate from the false vacuum to the true vacuum (the bubble nucleation rate) per unit time per unit volume is given by:
\begin{eqnarray}\label{eq:nucleationrate}
    \Gamma \sim A T^4 e^{-S_\mathrm{E}},
\end{eqnarray}
where $S_\mathrm{E}$ is the Euclidean action of the bounce solution.
Assuming a spherical solution, the equation of motion (EOM) becomes:
\begin{eqnarray}
\frac{d^2 \vec{\phi}}{d r^2} + \frac{d-1}{r} \frac{d\vec{\phi}}{d r} = \frac{\partial V(\vec{\phi},T)}{\partial \vec{\phi}},\qquad \vec{\phi}(\infty)=\vec{\phi}_\mathrm{false},\quad \frac{d \vec{\phi}}{d{r}}\bigg|_{{r}=0}=0.
\end{eqnarray}
where the $d=3$ or $4$ depending on the dimension of the spherical solution considered.
In principle, $S_E$ is determined by 
\begin{eqnarray}
S_\mathrm{E}(T)=\text{min}\left[\frac{S_3(T)}{T},S_4(T)\right]
\end{eqnarray}
where:
\begin{eqnarray}
    \frac{S_3}{T} &=&\frac{4\pi}{T} \int^{\infty}_0 dr r^2 \left[\frac{1}{2} \frac{d \vec{\phi}}{d r} \cdot \frac{d \vec{\phi}}{d r} + V_\mathrm{eff}(\vec{\phi}(r),T)\right],\\
    S_4&=&2\pi^2\int^{\infty}_0d r r^3 \left[\frac{1}{2} \frac{d \vec{\phi}}{d r} \cdot \frac{d \vec{\phi}}{d r} + V_\mathrm{eff}(\vec{\phi}(r),T)\right].
\end{eqnarray}

Note that in the case of supercooling, since the evolution of the universe is dominated by vacuum energy instead of radiation energy, the nucleation condition in our model is modified to:
\begin{eqnarray}\label{eq:Tn}
    S_\mathrm{E} \left(T_{\text{n}} \right) \simeq  131.98 - 4 \ln \left( \frac{m_\chi}{1~\text{TeV}}\right) - 4
  \ln \left(\frac{v_\chi}{2.5~\text{TeV}} \right) + 4 \ln \left(\frac{T_{\text{n}}}{100~\text{GeV}}\right) - \ln \left(
  \frac{\tilde{\beta}_\text{n}}{100} \right),
\end{eqnarray}
where $\tilde{\beta}$ is the inverse of the duration of phase transition.
More details can be found in the Appendix~\ref{appdE}.

\subsection{Numerical results}

The most relevant model parameters related to the phase transition in our model are the color number $N$, the dilaton VEV $v_\chi$ and the dilaton mass $m_\chi$.
In our numerical analysis, we scan the parameter space of
\begin{eqnarray}
N=3\text{-}10, v_\chi=2~\mathrm{TeV}, 2.5~\mathrm{TeV}, 3~\mathrm{TeV}, m_\chi \in [10^2,10^4]~\mathrm{GeV}
\end{eqnarray}
to explore the impact of these parameters on the phase transition. Other parameters are fixed by
\begin{eqnarray}
\quad v_h=246~\text{GeV},\quad c_{\chi}=0.5,\quad c_{\epsilon}=0.001,\quad m_{\eta}=1~\text{TeV},\quad \lambda_{\eta}=0.05,\quad\lambda_{h\eta}=0.01.
\end{eqnarray}

\begin{figure}[t!]
	\centering
	\subfigure{
		\begin{minipage}[t]{0.5\linewidth}
			\centering
			\includegraphics[scale=0.55]{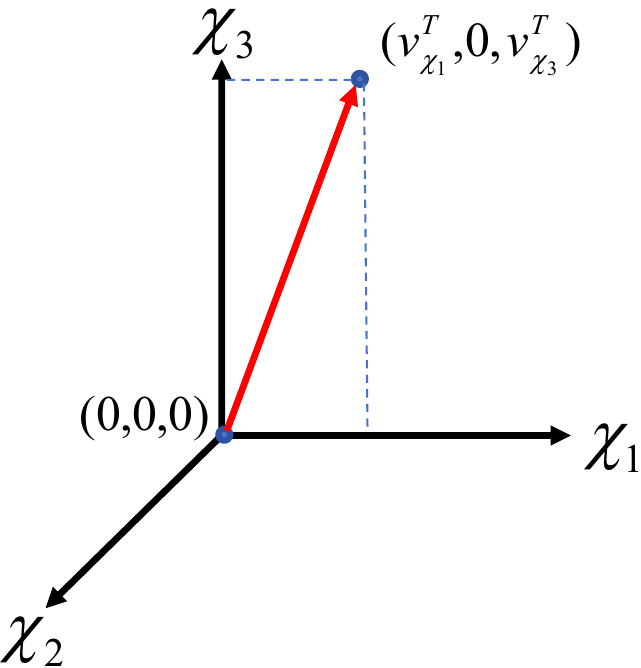}
	\end{minipage}}%
	\subfigure{
		\begin{minipage}[t]{0.5\linewidth}
			\centering
			\includegraphics[scale=0.55]{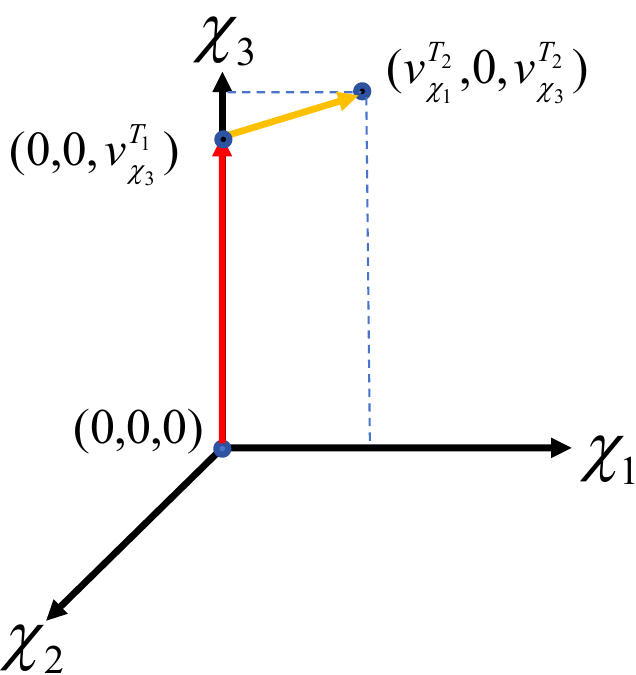}
	\end{minipage}}	
	\caption{The PT patterns with 1-step PT on the left and 2-step PT on the right.}\label{fig:PT_patt}
\end{figure}

\begin{figure}[htp!!]
	\centering
	\subfigure{
		\begin{minipage}[t]{0.48\linewidth}
			\centering
			\includegraphics[width=\linewidth]{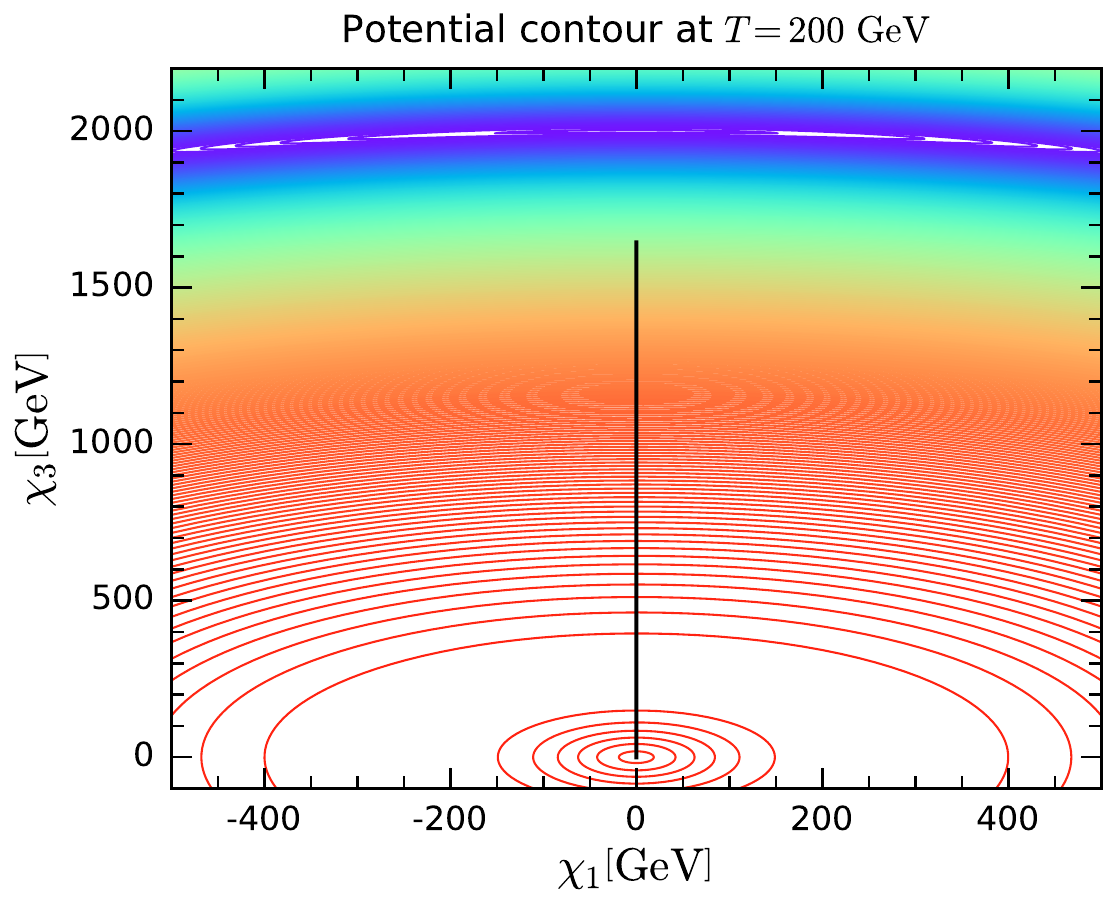}
	\end{minipage}}%
	\subfigure{
		\begin{minipage}[t]{0.5\linewidth}
			\centering
			\includegraphics[width=\linewidth]{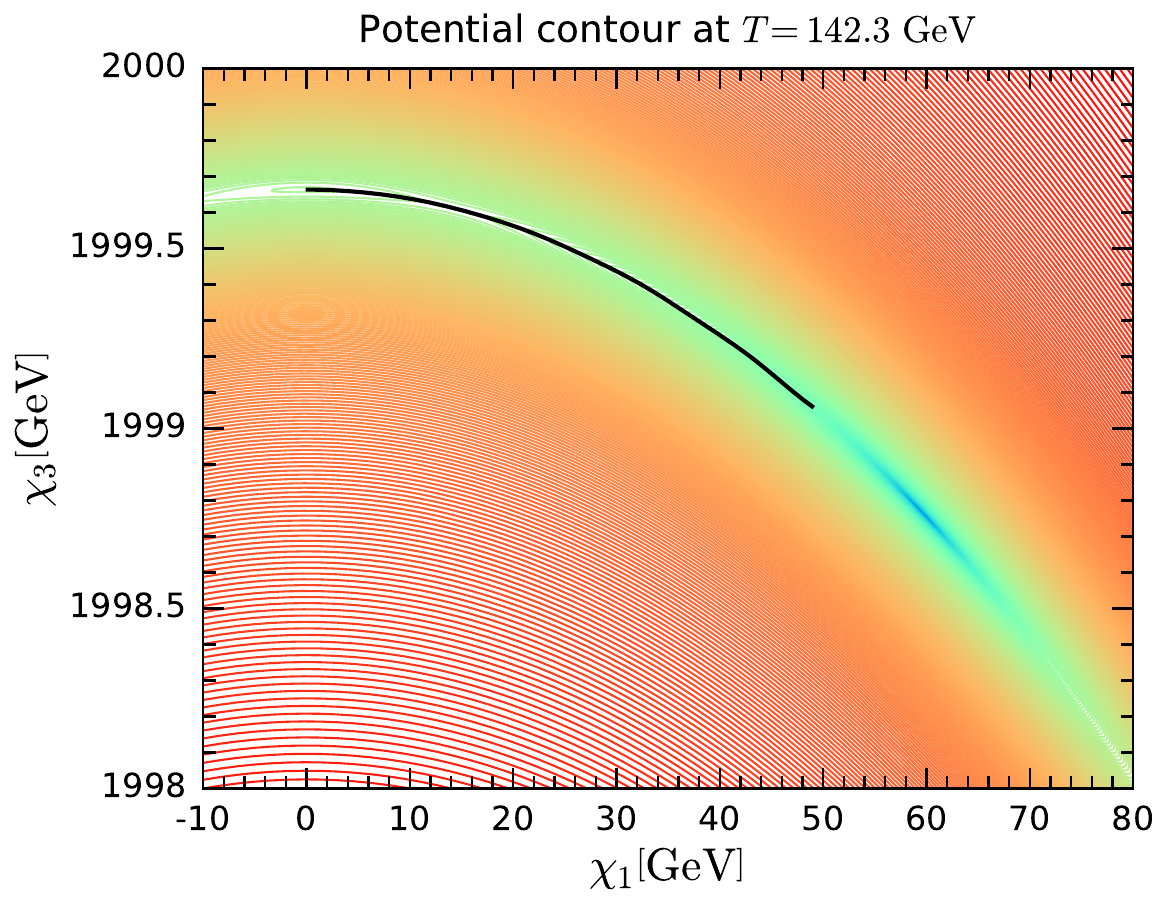}
	\end{minipage}}	
	\caption{
		The 2-step FOPT potential contours at $\chi_1$-$\chi_3$ plane with $N=8, v_\chi=2~\mathrm{TeV},m_\chi=3~\mathrm{TeV}$. The red contours corresponds to large potential value, while the blue contours corresponds to small potential value. The black curves represent the the tunneling paths, $(0,0,0)\rightarrow(0,0,v_{\chi_3}^{T_1})$ (left panel) and $(0,0,v_{\chi_3}^{T_1})\rightarrow(v_{\chi_1}^{T_2},0,v_{\chi_3}^{T_2})$(right panel)}\label{fig:V_contour}
\end{figure}

As discussed in the previous subsection, the dilaton ($\chi_3$)  direction dominates the phase transition in our model. Moreover, the phase transition of confinement is expected to be first order since the finite-temperature effective potential always exhibits a barrier between the deconfined phase and the confined phase. 
Our numerical calculations, performed using the \texttt{CosmoTransitions} package~\cite{wainwright_cosmotransitions_2012}, confirms that there are two distinct patterns of phase transitions: 1-step FOPTs and the 2-step FOPTs. Fig.~\ref{fig:PT_patt} shows schematic pictures of these two patterns.
Fig.~\ref{fig:V_contour} shows an example of 2-step phase transition trajectories and potential contours in the $\chi_1$-$\chi_3$ plane.
The black curve in the left panel corresponds to the tunneling path of first step, while the black curve in the right panel corresponds to the tunneling path of second step.
Our calculations also show that the confinement transition (either the 1-step transition or the first step of the 2-step phase transition) is supercooled, meaning that $T_n\ll T_c$~\cite{baratella_supercooled_2019,von_harling_qcd-induced_2018,nardini_confining_2007,agashe_cosmological_2020}.

\begin{figure}[t!]
	\centering
	\subfigure{
		\begin{minipage}[t]{0.5\linewidth}
			\centering
			\includegraphics[width=\linewidth]{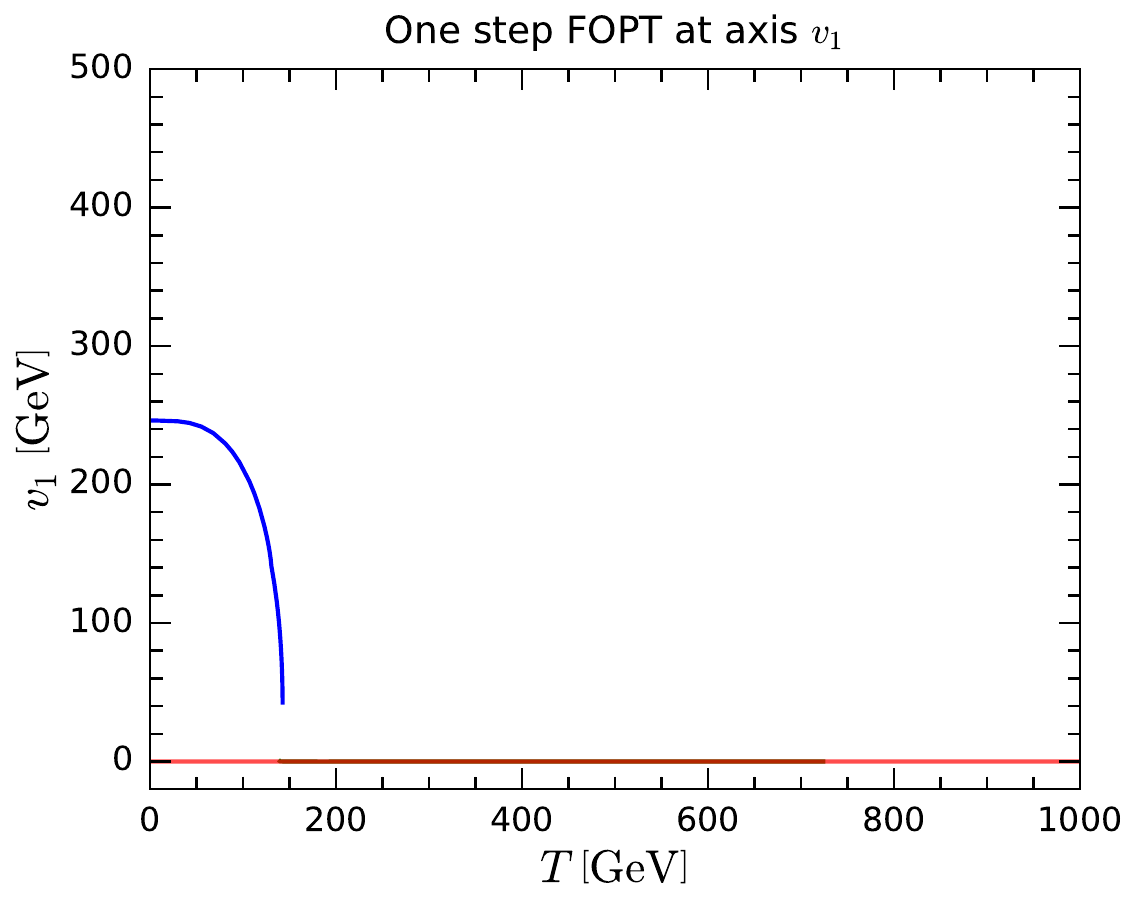}
	\end{minipage}}%
	\subfigure{
		\begin{minipage}[t]{0.5\linewidth}
			\centering
			\includegraphics[width=\linewidth]{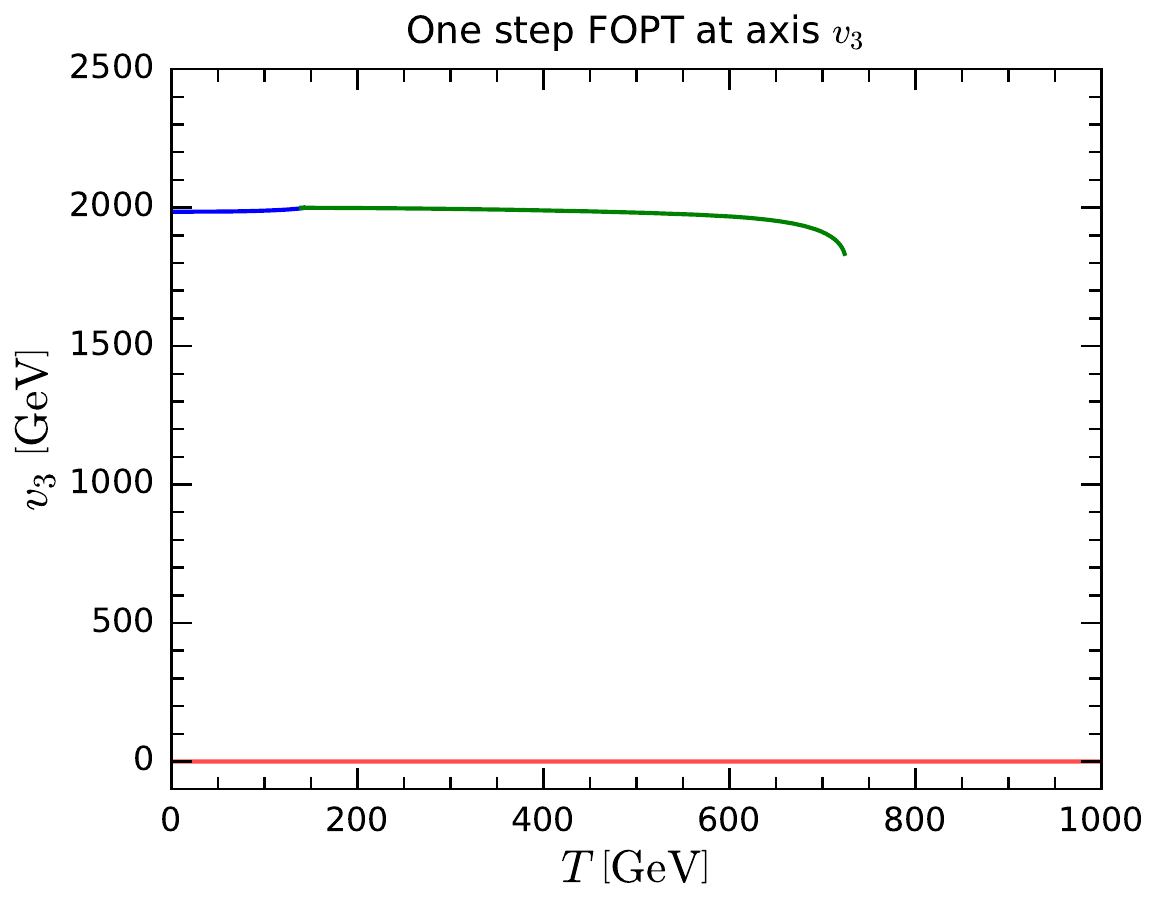}
	\end{minipage}}	
	\caption{
		Evolution of the minima in $\chi_1$ and $\chi_3$ directions for the 2-step FOPT. The red, green, and blue curves represent phases $(0,0,0)$, $(0,0,v_{\chi_3}^T)$, and $(v_{\chi_1}^T,0,v_{\chi_3}^T)$, respectively.
		}\label{fig:phases}
\end{figure}

In the case of 1-step phase transition, due to the low tunneling rate in the dilaton direction, $(0,0,v_{\chi})$ is no longer a minimum before the tunneling towards $(v_h,0,v_{\chi})$ directly. Therefore the phase transition from $(0,0,0)$ to $(0,0,v_{\chi})$ would never occur.
In Fig.~\ref{fig:phases}, we plot the evolution of $\chi_1$ and $\chi_3$ with temperature  in the case of a 1-step phase transition, for fixed values of $N=8, v_\chi=2~\mathrm{TeV}$, and $m_\chi=1.5~\mathrm{TeV}$.

We also find that the most sensitive parameter to the phase transition pattern is the dilaton mass $m_\chi$. The critical dilaton mass that distinguishes two phase transition patterns can be figured out analytically using the action approximation methods~\cite{Duncan:1992ai}. The details of derivation can be found in Appendix~\ref{appdD}.
We compare the numerical results for the critical dilaton mass and corresponding nucleation temperatures with the analytical approximation results in Table.~\ref{tab:critical_mass}.
\begin{table}[htp]
	\centering
	\begin{tabular}{ccccc}
		\hline\hline
        $N$& ~$T_\text{n}^{(\text{num.})}~\text{[GeV]}$~& ~$T_\text{n}^{(\text{ana.})}~\text{[GeV]}$~&~$m_{\text{crit.}}^{(\text{num.})}~\text{[GeV]}~$ & ~$m_{\text{crit.}}^{(\text{ana.})}~\text{[GeV]}$ \\
		\hline
		4 & 152.793&140.000  &1291.549 &945.374  \\
		7 & 145.273& 140.000 &1707.353 &1553.890  \\
		10& 146.317  &140.000  &2056.512     &2129.810   \\\hline
	\end{tabular}
	\caption{The comparison between numerical and analytical approximation results of critical dilaton masses which dividing two PT patterns. We have fixed $v_\chi=2~\text{TeV}$, $m_\eta=1 \text{~TeV}$ and $\lambda_{h\eta}=0.01$. $m_{\text{crit}}^{(\text{num})}$ and $m_{\text{crit}}^{(\text{theo})}$ are the numerical and analytical approximations of critical dilaton masses while $T_\text{n}^{(\text{num})}$ and $T_\text{n}^{(\text{theo})}$ are the nucleation temperatures corresponding to them, respectively.}\label{tab:critical_mass}
\end{table}

\begin{figure}[t!]
	\centering
	\subfigure{
		\begin{minipage}[t]{0.5\linewidth}
			\centering
			\includegraphics[width=\linewidth]{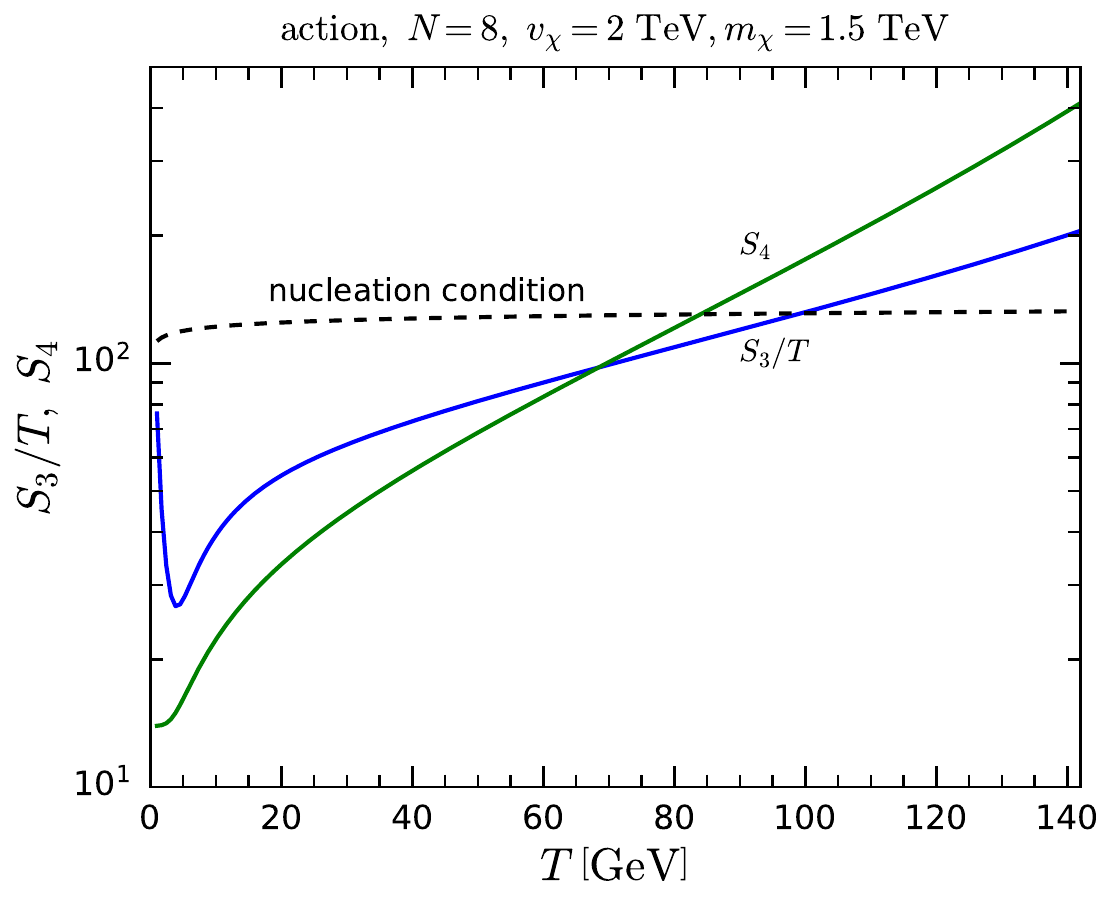}
	\end{minipage}}%
	\subfigure{
		\begin{minipage}[t]{0.5\linewidth}
			\centering
			\includegraphics[width=\linewidth]{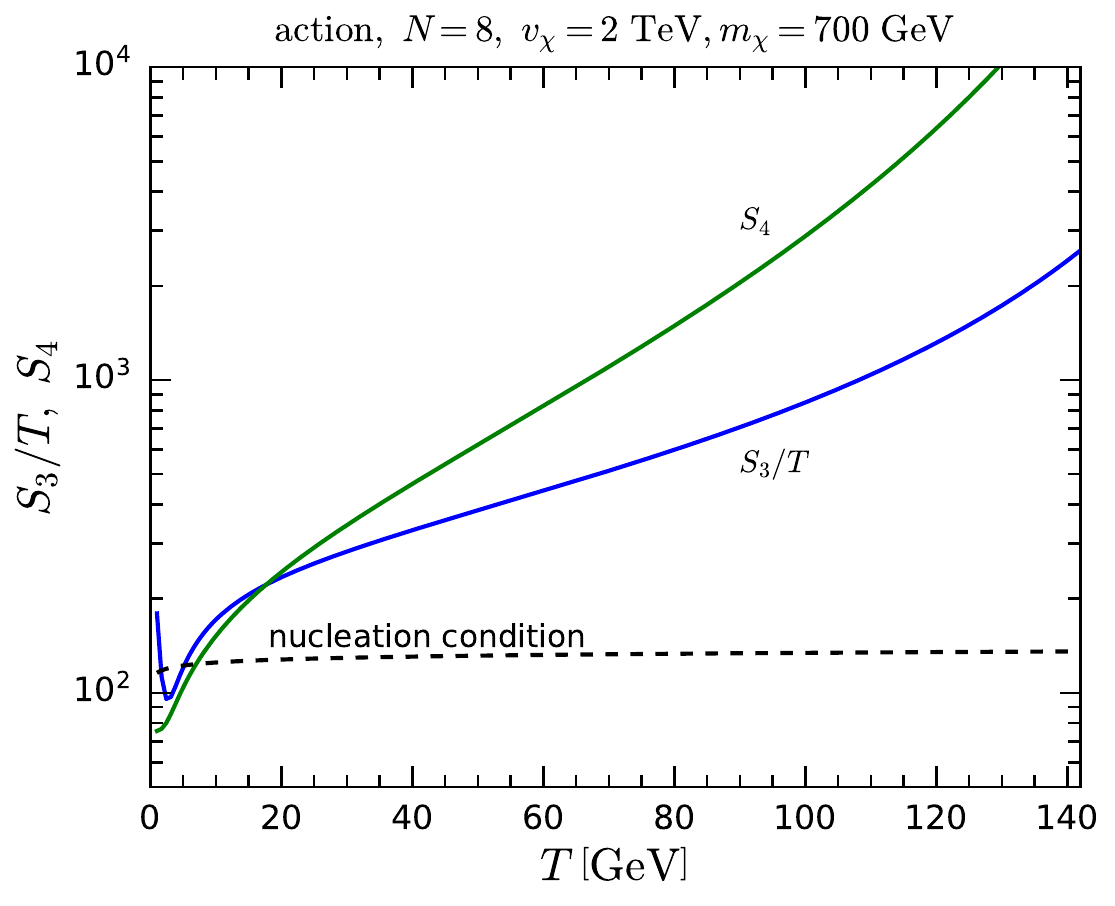}
	\end{minipage}}	
	\caption{
		Evolution of $S_4$ and $S_3/T$ for fixed values of $N=8,v_\chi=2~\mathrm{TeV}$. The blue and green lines correspond to $S_3/T$ and $S_4$, respectively. The black dashed line represents the nucleation condition.
		}\label{fig:action}
\end{figure}

Note that in the parameter space we considered, the quantum tunneling process can overwhelm the thermal fluctuation when $N\geq7$ and dilaton mass is small, leading to low temperature nucleation. In this case, the Euclidean action $S_E$ is determined by 4-d action $S_4$ for the bounce solution.
In Fig.~\ref{fig:action}, we compare the 3-d action $S_3/T$ (blue line) with the 4-d action $S_4$ (green line) for fixed values of  $N=8,v_\chi=2~\mathrm{TeV},m_\chi=1.5~\mathrm{TeV}~(\textrm{left panel}),700~\mathrm{GeV}~(\textrm{right panel})$.
The black dashed line represents the nucleation condition given by Eq.~\eqref{eq:Tn}.
The nucleation temperatures for these two types of actions, denoted as $T_{\mathrm{n}_3}$ and $T_{\mathrm{n}_4}$, are determined by the intersecting points of the dashed line with the blue ($S_3/T$) and green ($S_4$) lines.
We can see that for large dilaton mass (left panel), $T_{\mathrm{n}_3}>T_{\mathrm{n}_4}$, while for small dilaton mass (right panel), $T_{\mathrm{n}_3}<T_{\mathrm{n}_4}$.

\begin{figure}[t!] 
\vspace{-0.8cm}
\begin{center}
\fbox{\footnotesize $N=4$} \\
\hspace*{-0.65cm} 
\begin{minipage}{0.32\linewidth}
\begin{center}
	\includegraphics[width=\linewidth]{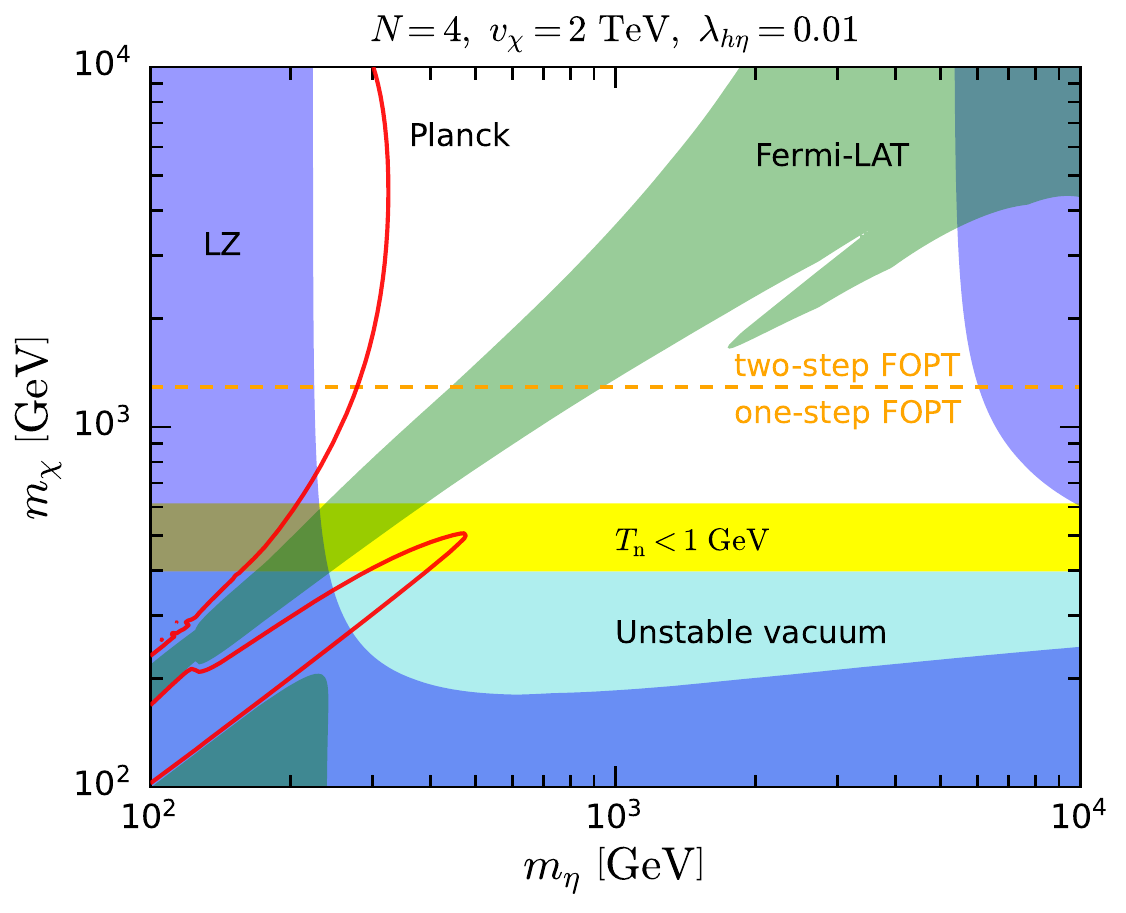}
\end{center}
\end{minipage}
\begin{minipage}{0.32\linewidth}
\begin{center}
	\includegraphics[width=\linewidth]{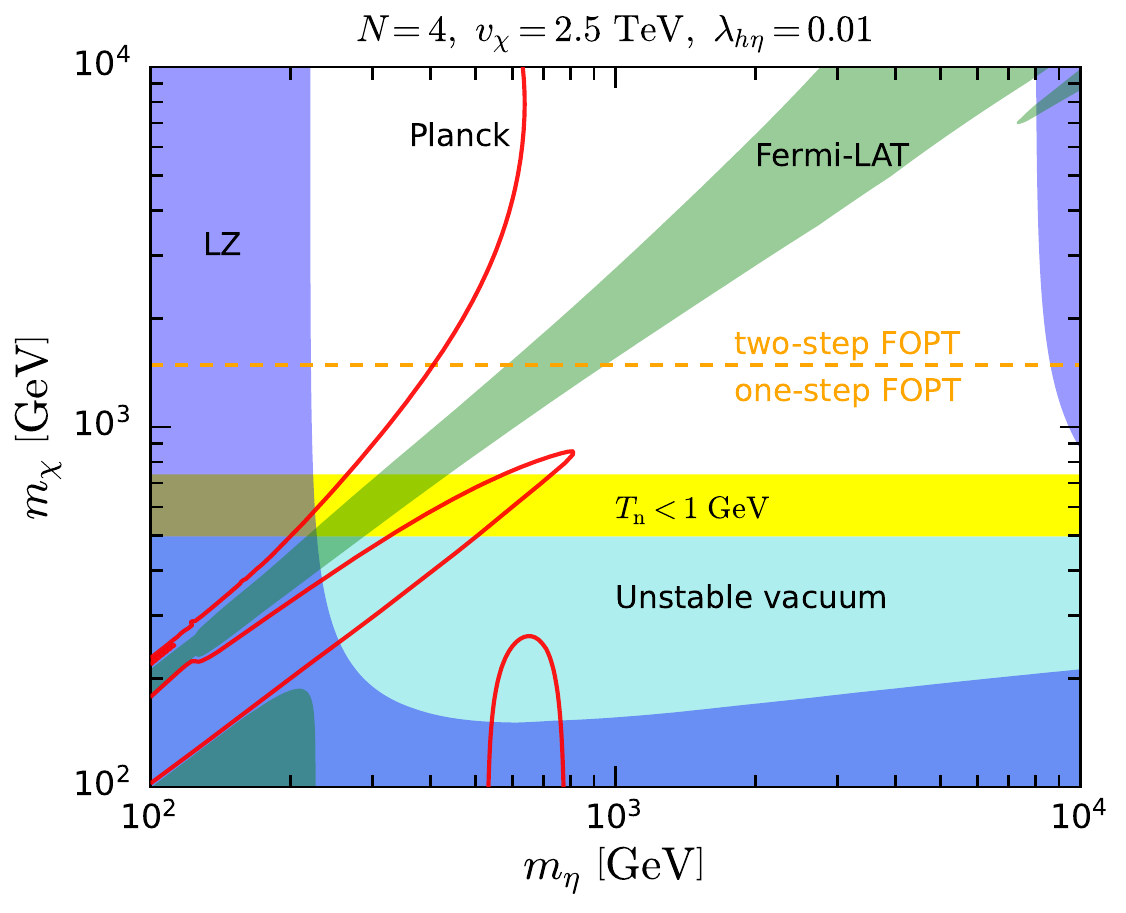}
\end{center}
\end{minipage}
\begin{minipage}{0.32\linewidth}
\begin{center}
	\includegraphics[width=\linewidth]{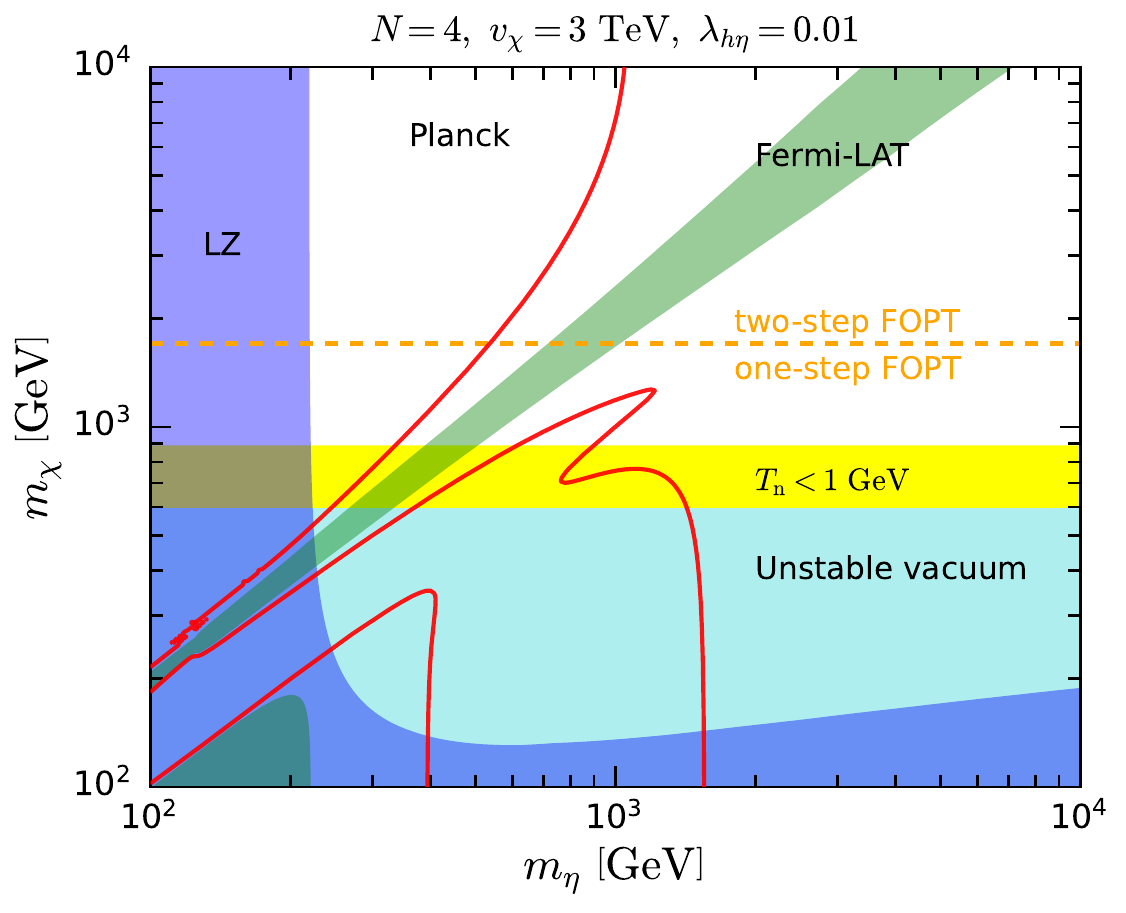}
\end{center}
\end{minipage}
\\[0.1cm]
\fbox{\footnotesize $N=7$} \\
\hspace*{-0.65cm} 
\begin{minipage}{0.32\linewidth}
\begin{center}
	\includegraphics[width=\linewidth]{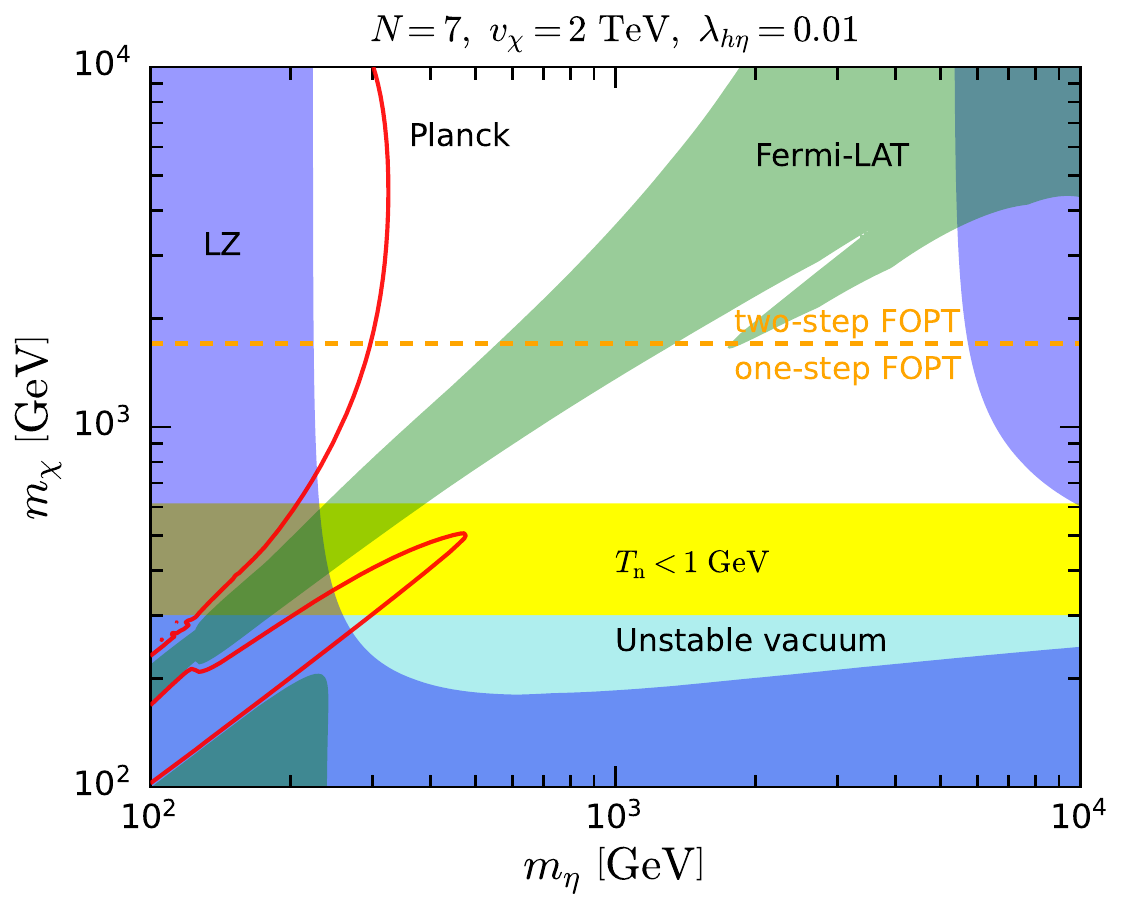}
\end{center}
\end{minipage}
\begin{minipage}{0.32\linewidth}
\begin{center}
	\includegraphics[width=\linewidth]{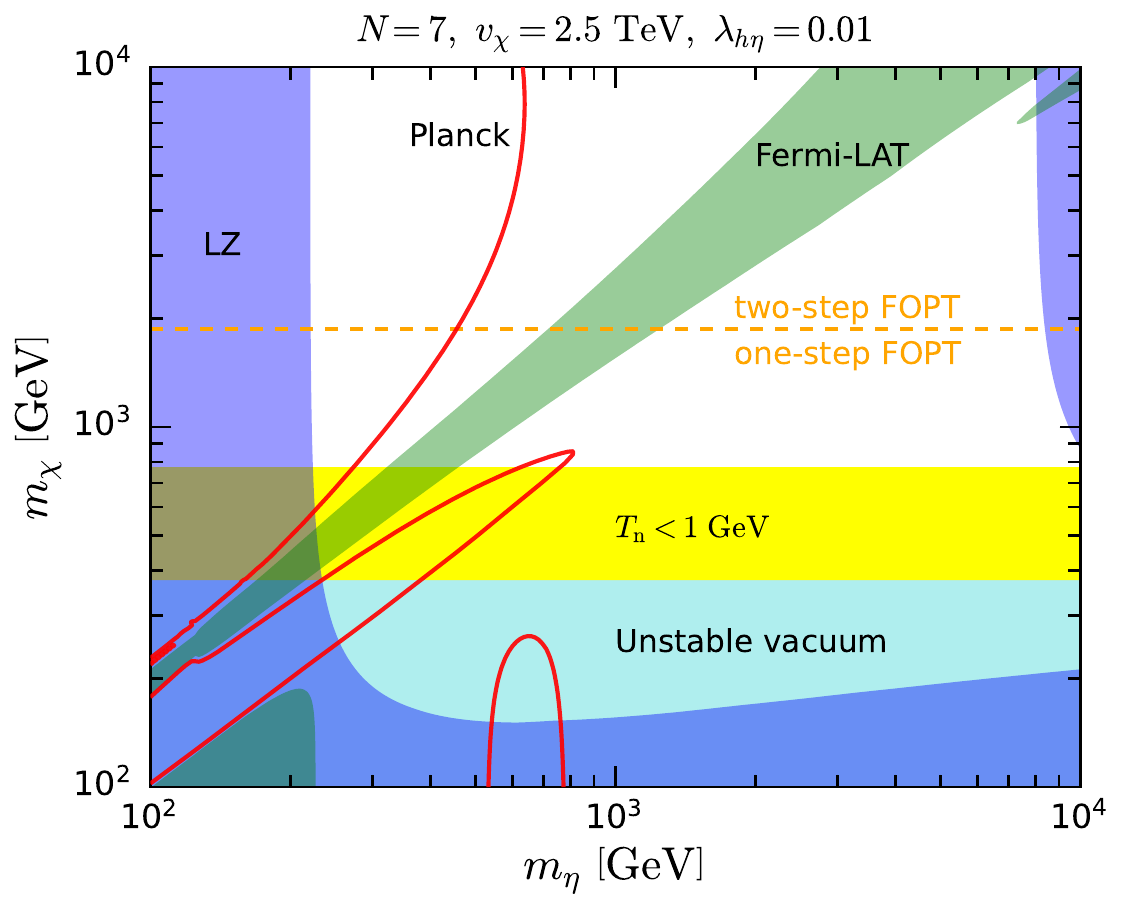}
\end{center}
\end{minipage}
\begin{minipage}{0.32\linewidth}
\begin{center}
	\includegraphics[width=\linewidth]{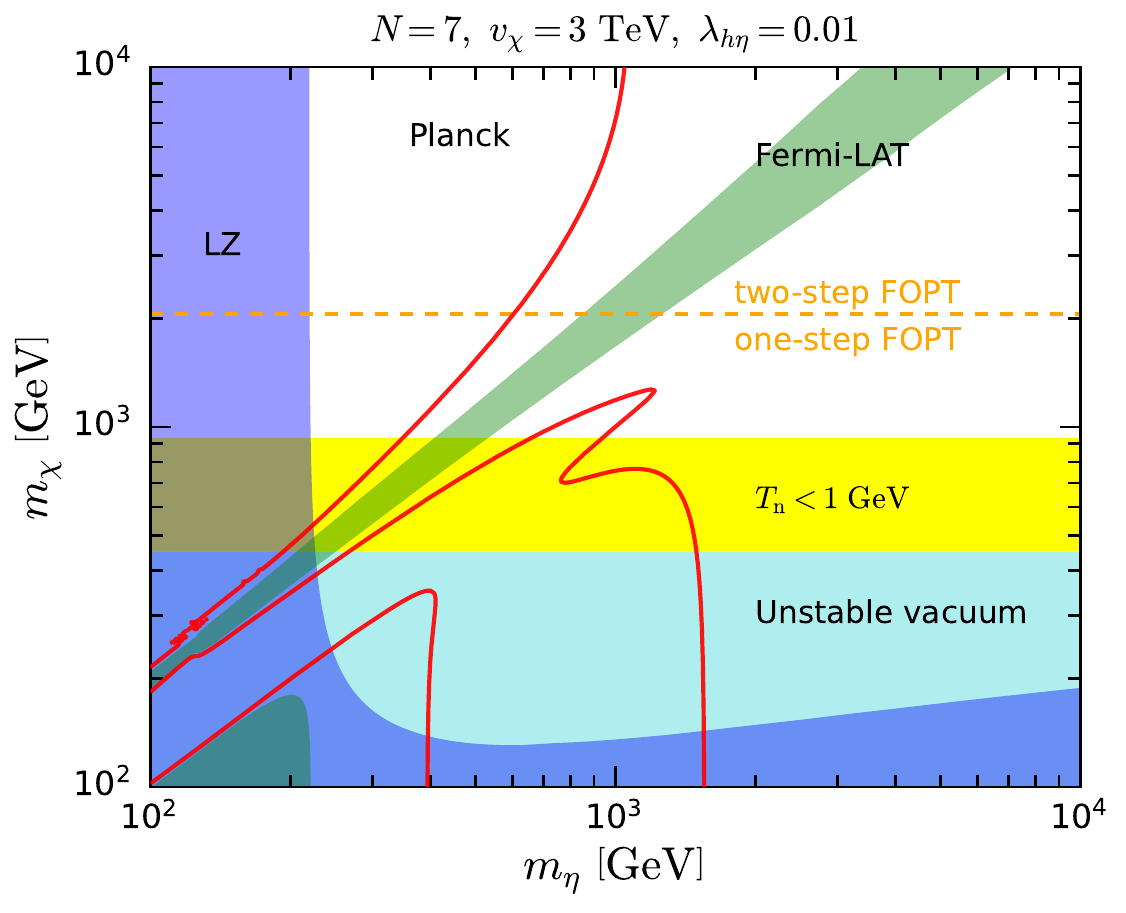}
\end{center}
\end{minipage}
\\[0.1cm]
\fbox{\footnotesize $N=10$} \\
\hspace*{-0.65cm} 
\begin{minipage}{0.32\linewidth}
\begin{center}
	\includegraphics[width=\linewidth]{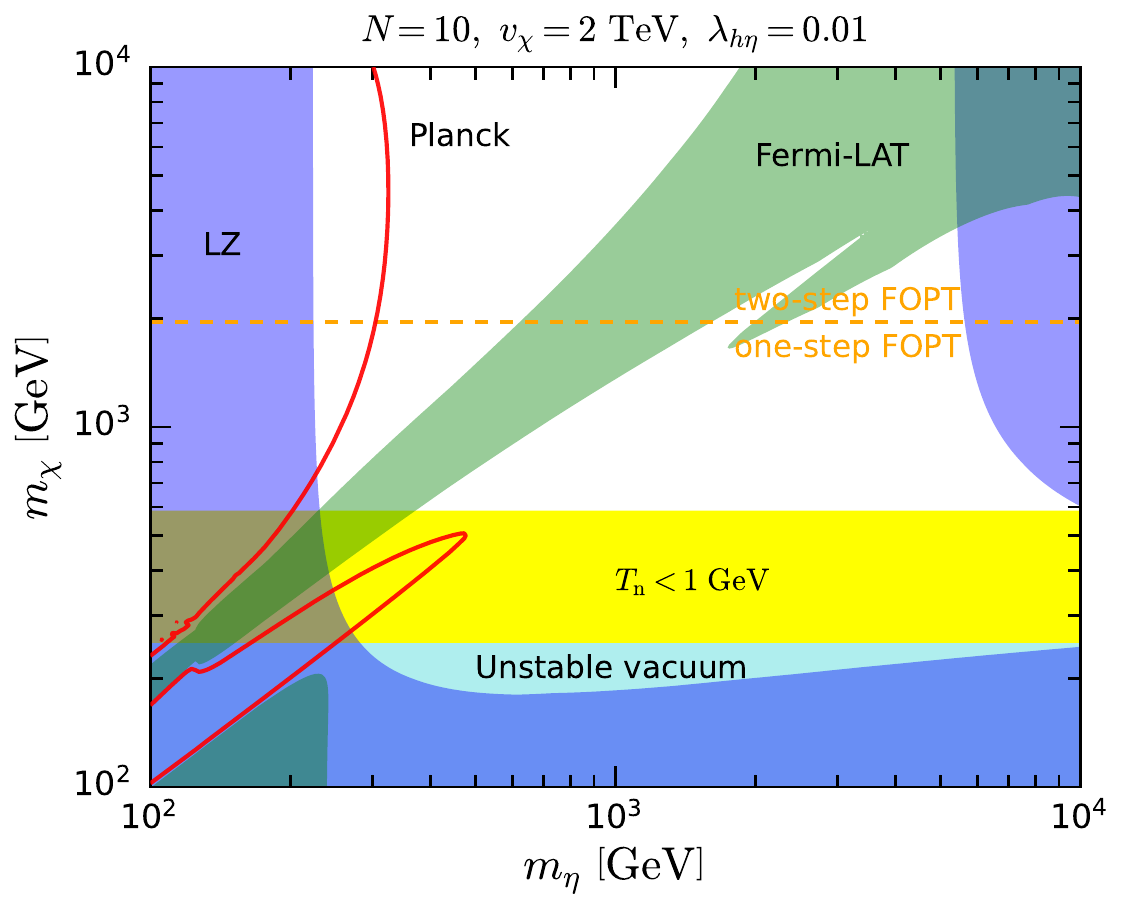}
\end{center}
\end{minipage}
\begin{minipage}{0.32\linewidth}
\begin{center}
	\includegraphics[width=\linewidth]{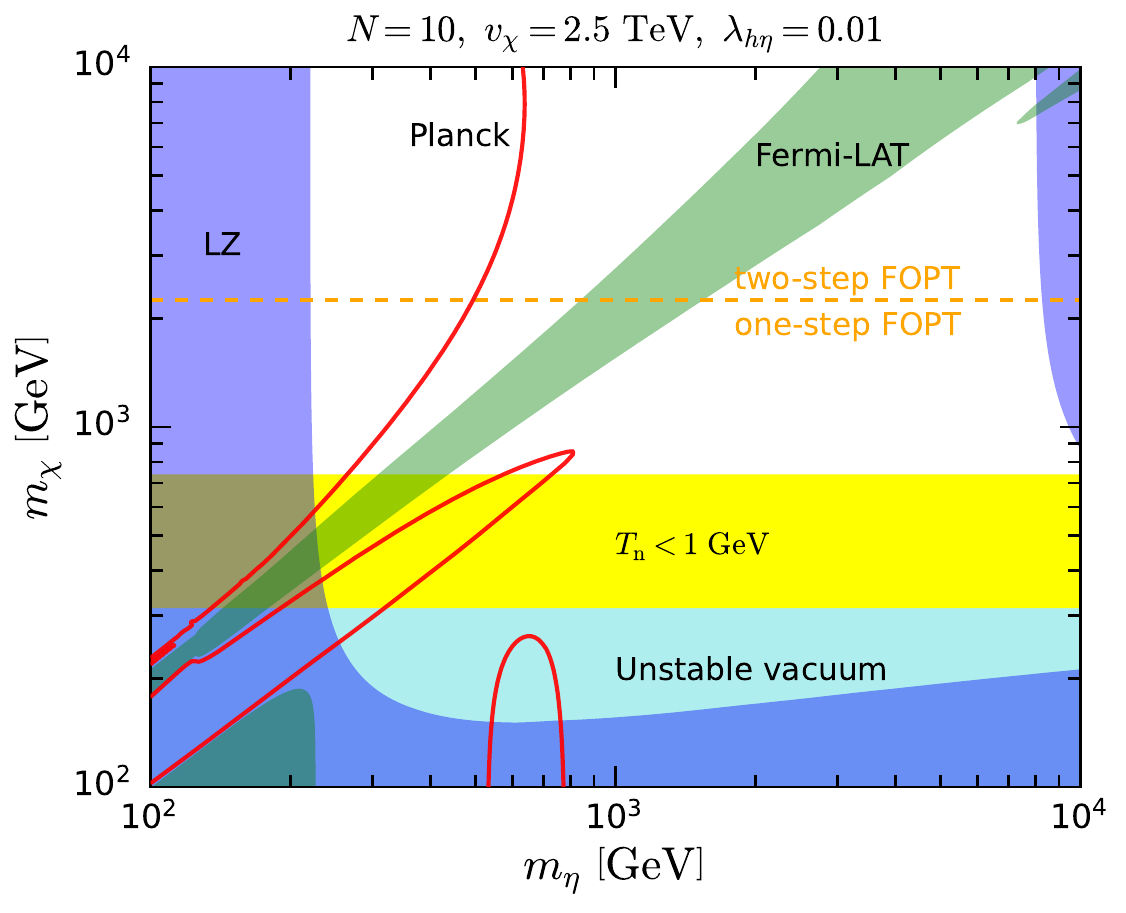}
\end{center}
\end{minipage}
\begin{minipage}{0.32\linewidth}
\begin{center}
	\includegraphics[width=\linewidth]{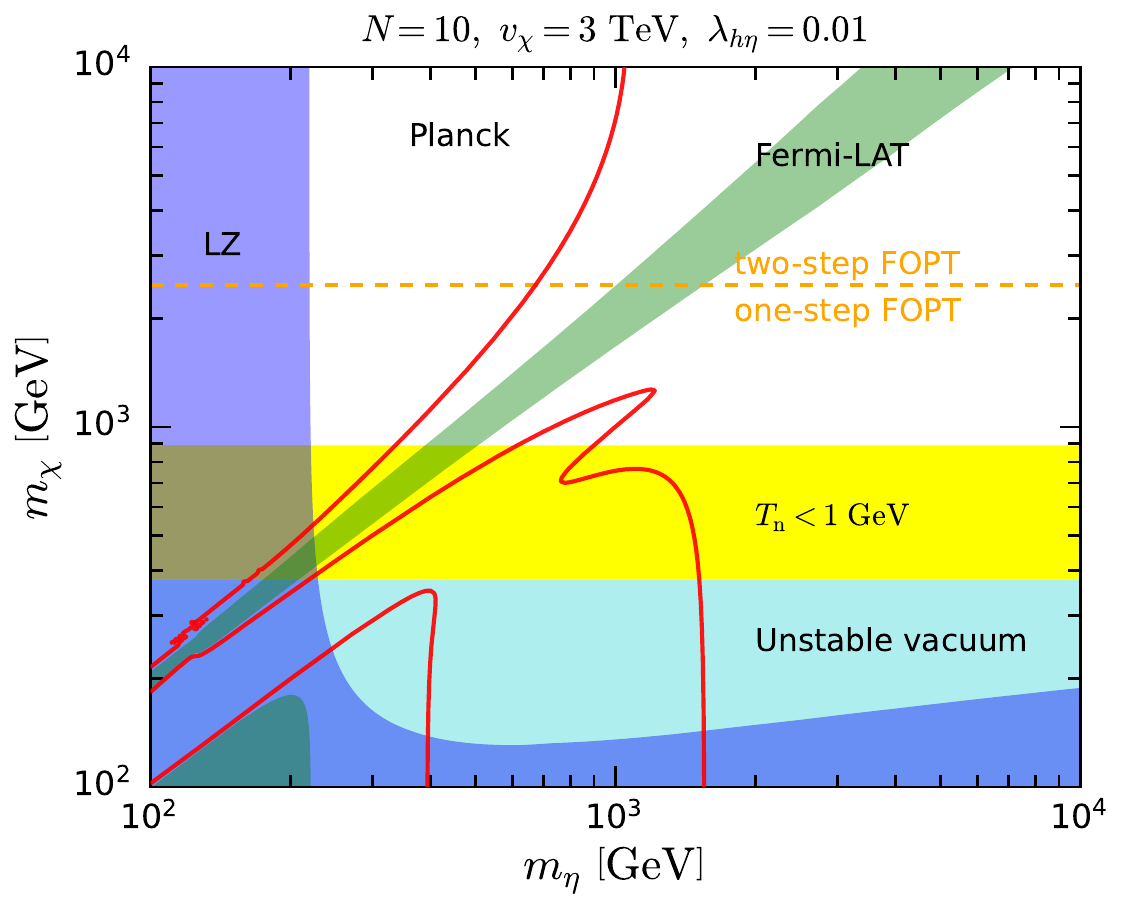}
\end{center}
\end{minipage}
\end{center}
\vspace*{-0.2cm}
\caption{The PT patterns and DM phenomenology in $m_\eta$-$m_\chi$ plane for $N=4,7,10$ and $v_\chi=2~\mathrm{TeV},2.5~\mathrm{TeV},3~\mathrm{TeV}$.}
\label{fig:CombinedPlot2}
\end{figure}

In order to conduct a comprehensive analysis of the parameter space, we scan the $m_\eta$-$m_\chi$ parameter space and display the constraints from the DM phenomenology along with the phase transition patterns in Fig.~\ref{fig:CombinedPlot2}. We present three rows of panels corresponding to $N=4,~7,~10$, and in each row, we feature three panels corresponding to  $v_\chi=2$~TeV, $2.5$~TeV, and $3$~TeV.
These plots confirm that from $10~\mathrm{GeV}$ to $10~\mathrm{TeV}$, the DM mass is almost irrelevant to the phase transition.
The region shaded in cyan indicates a potential unbounded from below (refer to the condition provided in Eq.~\eqref{eq:mchi_lowbound}).
We also exclude the region corresponding to $T_n<1~\mathrm{GeV}$ (represented in yellow) to prevent any significant impact from strong FOPTs on the BBN.
The orange dashed line demarcates the boundary between 1-step FOPTs and 2-step FOPTs. 
As previously discussed, a smaller dilaton mass results in a lower tunneling rate, making 1-step phase transition more likely. Conversely, a larger dilaton mass leads to a higher tunneling rate, favoring a 2-step phase transition. 
In Tab.~\ref{benchmark}, we provide some benchmark points for these two types of phase transition patterns.

In conclusion, Fig.~\ref{fig:CombinedPlot2} shows that our model can simultaneously provide a viable DM candidate and achieve strong FOPTs at the same time. The DM mass is constrained within a range of $200~\mathrm{GeV}$ to $1~\mathrm{TeV}$.

\begin{table}[htp]
	\centering
        \setlength\tabcolsep{0.6em}
	\renewcommand{\arraystretch}{1}
	\begin{tabular}{ccccccc}
		\hline\hline
        &$N$& $m_\chi~\text{[GeV]}$
		& $v_\chi~\text{[TeV]}$ & \text{pattern}&$T_\mathrm{c}~[\mathrm{GeV}]$& $T_\mathrm{n}~[\mathrm{GeV}]$ \\
		\hline
		$\bm{\mathrm{BP}_1}\quad$ & 4 & 800 & 2.5 & 1-step  & 297  & 7.78 \\
		$\bm{\mathrm{BP}_2}\quad$ & 4 & 2000 &  2.5 & 2-step & 521 &  256  \\
		$\bm{\mathrm{BP}_3}\quad$ & 4 & 8000 &  2.5 & 2-step & 1070 & 922 \\
		$\bm{\mathrm{BP}_4}\quad$ & 7 & 800 &  2.5 & 1-step & 238 & 5.00  \\
        $\bm{\mathrm{BP}_5}\quad$ & 7 & 2000 &  2.5 & 2-step  & 397 & 160 \\
		$\bm{\mathrm{BP}_6}\quad$ & 7 & 8000 &  2.5 & 2-step  & 813  & 685 \\
        $\bm{\mathrm{BP}_7}\quad$ & 10 & 800 &  2.5 & 1-step  & 203 & 5.14 \\
        $\bm{\mathrm{BP}_8}\quad$ & 10 & 2000 &  2.5 & 1-step  & 334 & 114 \\
        $\bm{\mathrm{BP}_9}\quad$ & 10 & 8000 &  2.5 & 2-step  & 685 & 568 \\
        $\bm{\mathrm{BP}_{10}}\quad$ & 7 & 2000 &  2.0 & 2-step  & 357 & 186 \\
        $\bm{\mathrm{BP}_{11}}\quad$ & 7 & 2000 &  3.0 & 1-step  & 434 & 127 \\
        \hline
	\end{tabular}
	\caption{PT benchmark points. $T_\mathrm{c}$ and $T_\mathrm{n}$ are the critical and nucleation temperature of the first step FOPTs (dilaton domination) respectively.}\label{benchmark}
    \label{tb.2}
\end{table}

Note that the mass and coupling strength of the DM candidate $\eta$ in our model suggest that $\eta$ falls within the WIMPs DM paradigm.
However, supercooling might influence the production of DM and potentially skew our calculations of the DM relic density. 
More specifically, $\eta$ is a pNGB which originates from the confinement of the techni-quarks in the composite sector.
Before the confinement phase transition occurs, the universe contains only deconfined techni-quarks in thermal equilibrium with the SM particles, not $\eta$. Therefore, it is crucial to compare the DM freeze-out temperature with the nucleation temperature. The freeze-out temperature of WIMPs DM can be estimated as follows:
\begin{eqnarray}
    T_\mathrm{FO} \simeq \frac{m_\eta}{25}.
\end{eqnarray}
If $T_n$ is lower than $T_\mathrm{FO}$, then the reheating after the supercooled phase transition should be taken into account~\cite{baratella_supercooled_2019}. 
During the strong FOPT, the latent heat between the deconfined and confined phases is released, with the majority of the energy being transferred into the plasma, resulting in reheating the universe. 
At the end of the PT, the free energy of the deconfined phase, $\mathcal{F}_\mathrm{dec.}\simeq -T_\mathrm{n}^4$, is negligible compared to the free energy of the confined phase, $\mathcal{F}_\mathrm{con.}$.Consequently, the latent heat between the two phases can be approximated by the free energy of the deconfined phase at the critical temperature $T_c$ of the phase transition. Assuming that the latent heat is entirely transferred to the radiation plasma, we can estimate the reheating temperature as follows:
\begin{eqnarray}
    \mathcal{F}_\mathrm{dec}-\mathcal{F}_\mathrm{con} \simeq \mathcal{F}_\mathrm{rh}\quad\Rightarrow\quad \frac{\pi^2}{8} N^2 T_\mathrm{c}^4\simeq \frac{\pi^2}{90}g_{\ast}T_\mathrm{rh}^4\quad\Rightarrow\quad T_\mathrm{rh}\simeq \left(\frac{45}{4g_\ast}\right)^{\frac{1}{4}}\sqrt{N}T_\mathrm{c},
\end{eqnarray}
where $g_\ast$ is the effective degrees of freedom after reheating.
If $T_\mathrm{FO}<T_\mathrm{rh}$, then $\eta$ will thermalize after reheating, and the DM production scenario is similar to that of the traditional WIMPs DM. 
However, if $T_\mathrm{FO}>T_\mathrm{rh}$, then $\eta$ will instantaneously freeze-out as soon as they are produced by the oscillating inflaton field ($\chi$).
Our numerical calculations indicate that, within the parameter space of interest, the critical temperature is typically around several hundred GeV, even in scenarios where supercooling occurs. 
Therefore, the situation of $T_\mathrm{FO}>T_\mathrm{rh}\sim T_\mathrm{c}$ only arises when $m_\eta\gtrsim \mathcal{O}(10)~\text{TeV}$. Given that our main focus is on a scale where $m_\eta,v_\chi<10~\text{TeV}$, the traditional WIMPs paradigm is consistently applicable.

\section{Gravitational waves}\label{sec:GW}

In the previous section, we discussed the PT patterns of our model and found that strong FOPTs exist.
The occurrence of these strong FOPTs in the early universe can generate a stochastic background of gravitational waves (GWs).
The most important PT parameters that characterize the GWs are the ratio of vacuum energy density to radiation energy density $\alpha$, the inverse of PT duration in unit of Hubble time $\tilde{\beta}$, and the bubble wall velocity $\xi_\mathrm{w}$:
\begin{eqnarray}
    \alpha\equiv \frac{\rho_\mathrm{vac}}{\rho_\mathrm{rad}}, \qquad \tilde{\beta} \equiv -\frac{1}{H} \frac{dS}{dt}=T \frac{dS}{dT}.
\end{eqnarray}
We will focus on the GWs produced by the confinement transition, which is governed by the dilaton field. 
In the scenario of a 2-step phase transition, the SM particles in the plasma remain massless after the confinement transition. The coupling between the SM particles and the dilaton is suppressed by $T^2/v_\chi^2$, making the friction to the bubble walls negligible. 
In the scenario of a 1-step phase transition, typically supercooled, the nucleation temperature is so low that the friction from the plasma is negligible. The production of gravitational waves in such a supercooled phase transition has been extensively studied~\cite{wang_phase_2020,jinno_gravitational_2019,lewicki_gravitational_2020,ellis_gravitational_2019,sagunski_supercool_2023}.
Given that the friction from the plasma is negligible in both phase transition patterns, the bubble walls are runaway~\cite{bodeker_can_2009}, allowing us to set $\xi_\mathrm{w} \simeq 1$ in our calculation. 
Consequently, the energy density of GWs is primarily contributed by the collision of bubble walls, as given by~\cite{Huber:2008hg,Konstandin:2017sat,Cutting:2018tjt},
\begin{eqnarray}\label{omega_col}
\Omega_\mathrm{col} h^2 = 1.67\times10^{-5} \left(\frac{100}{g_*}\right)^{1/3} \frac{0.44~\xi_\mathrm{w}^3}{1+8.28~\xi_\mathrm{w}^3}~\frac{1}{\tilde{\beta}^2} \left( \frac{\kappa_\phi\alpha}{1+\alpha}\right)^2 C_\mathrm{col}\left({f}/f_\mathrm{col}\right).
\end{eqnarray}
In this equation, we will assume that the energy fraction of the bubble collision, denoted by $\kappa_\phi$, is approximately 1.
The spectral shape function, $C_\mathrm{col}$, is defined as\cite{Konstandin:2017sat}:
\begin{eqnarray}
    C_\mathrm{col}\left(f/f_\mathrm{col}\right) = \frac{3.8 (f/f_\mathrm{col})^{2.9}}{0.9 + 2.9(f/f_\mathrm{col})^{3.8}},
\end{eqnarray}
where the peak frequency is given by:
\begin{eqnarray}
f_\mathrm{col}  = 1.65\times10^{-5} \mathrm{Hz} \left(\frac{T_*}{100~\mathrm{GeV}}\right) \left(\frac{g_*}{100}\right)^{1/6} \frac{1.96~\tilde{\beta}}{1 - 0.051~\xi_\mathrm{w} + 0.88~ \xi_\mathrm{w}^2}.
\end{eqnarray}

\begin{figure}[htp]
	\centering
	\subfigure{
		\begin{minipage}[t]{0.5\linewidth}
			\centering
			\includegraphics[width=\linewidth]{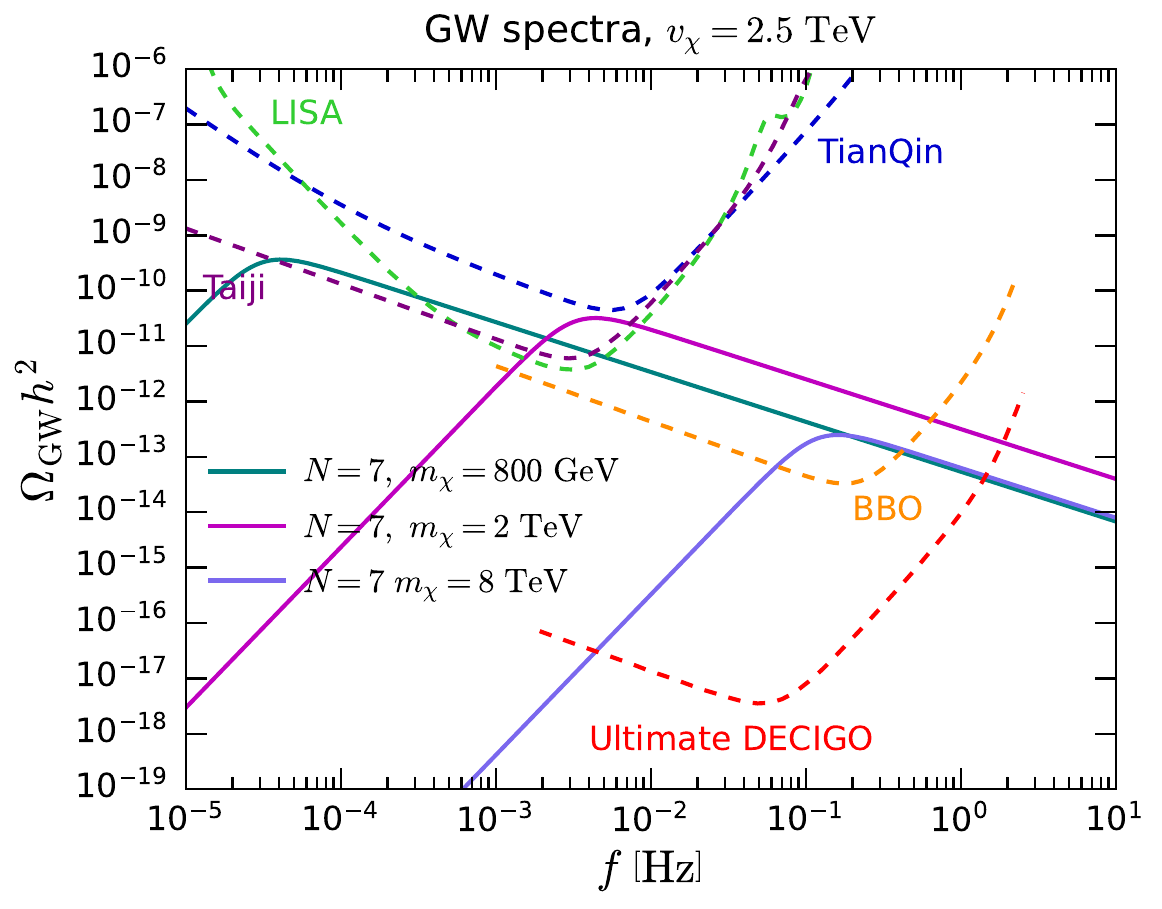}
	\end{minipage}}%
	\subfigure{
		\begin{minipage}[t]{0.5\linewidth}
			\centering
			\includegraphics[width=\linewidth]{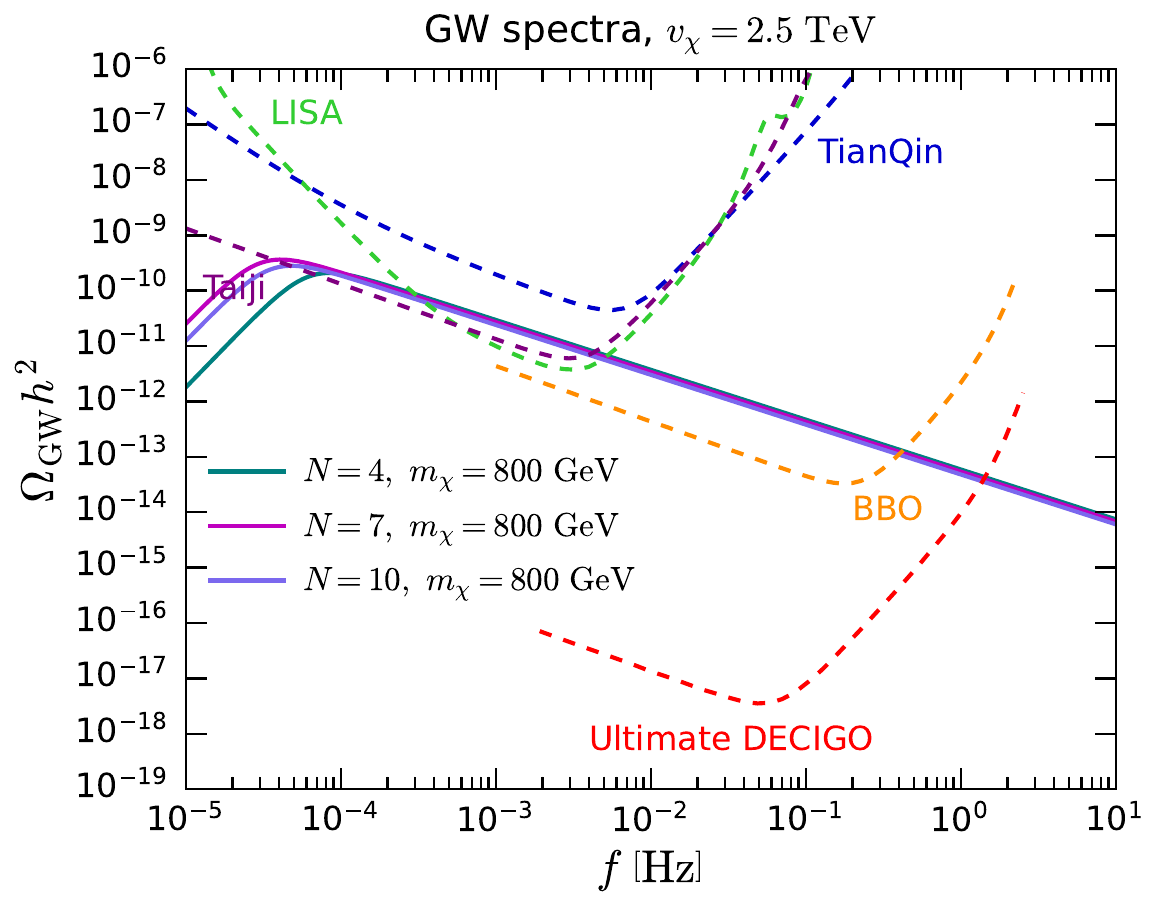}
	\end{minipage}}	
	\caption{ The GW spectra (solid lines) of the benchmark points in Tab.\ref{tab:GW_BP}.
		The sensitivity curves (dashed lines) of LISA~\cite{Audley:2017drz}, TianQin~\cite{Mei:2020lrl}, Taiji~\cite{Guo:2018npi}, BBO~\cite{Cutler:2005qq}, and DECIGO~\cite{Kudoh:2005as} are presented for comparison.
        }\label{fig:GW_BP}
\end{figure}

As previously discussed, a smaller dilaton mass results in a more supercooled phase transition, which in turn leads to a larger value for $\alpha$ and a smaller value for $\tilde{\beta}$. Consequently, according to Eq.\eqref{omega_col}, the peak amplitude of the GW energy density is enhanced.
The GW spectra for the benchmark points listed in Tab.~\ref{tab:GW_BP} are presented in Fig.~\ref{fig:GW_BP}.
In the left panel, we fix $N=7$ and display the spectra for different dilaton masses $m_\chi$. Conversely, in the right panel, we fix the dilaton mass $m_\chi=800~\mathrm{GeV}$ and illustrate the spectra for different values of $N$.
The GW frequency band is chosen to be within the range of $(10^{-5}~\mathrm{Hz},1~\mathrm{Hz})$, which will be detectable in the future space-based GW interferometers experiments, such as LISA, TianQin, Taiji, BBO, and Ultimate DECIGO.
As observed in the left panel, if $m_\chi\lesssim1$~TeV, the GW amplitude can meet the sensitivity of LISA, Taiji, BBO, and Ultimate DECIGO. However, if $m_\chi>2$~TeV, the amplitude can only meet the sensitivity requirements of BBO and Ultimate DECIGO. From the right panel, it can be seen that different values of $N$ only slightly affect the peak, and all spectra with $m_\chi=800$~GeV are expected to be detectable by LISA and Taiji.

\begin{table}[!t]
	\centering
        \setlength\tabcolsep{0.6em}
	\renewcommand{\arraystretch}{1}
	\begin{tabular}{ccccccc}
		\hline\hline
        & \text{pattern}& $N$
		& $m_\chi~\text{[GeV]}$  & $T_\mathrm{n}~\text{[GeV]}$ & $\alpha $ &$\tilde{\beta}$ \\
		\hline
		$\bm{\mathrm{BP}_4}\quad$ & 1-step & 7 & 800 & 5.00 & $1.31\times 10^7$ &  $4.71\times 10^1$  \\
		$\bm{\mathrm{BP}_5}\quad$ & 2-step & 7 & 2000 & 160 & $6.23\times 10^1$  & $1.51\times 10^2$    \\
		$\bm{\mathrm{BP}_6}\quad$ &2-step & 7 & 8000 & 685 & $2.96\times 10^0$  & $1.30\times 10^3$    \\
		$\bm{\mathrm{BP}_1}\quad$ & 1-step & 4  & 800 & 7.78 & $2.21\times 10^6$  & $6.22\times 10^1$   \\
        $\bm{\mathrm{BP}_7}\quad$ & 1-step & 10 & 800 & 5.14 & $1.17\times 10^7$  & $5.36\times 10^1$   \\
        \hline
	\end{tabular}
	\caption{The benchmark points for plotting the GW spectra.}\label{tab:GW_BP}
\end{table}

\section{Conclusion}\label{sec:conclusion}
In this paper, we have conducted a detailed examination of the NMCHM extended with a dilaton $\chi$, including aspects of DM phenomenology, cosmological phase transition dynamics, and gravitational waves (GW).
Due to the mixing between the dilaton and Higgs fields, an accidental cancellation of the DM-nucleon scattering amplitude can occur, allowing some parameter space to evade the stringent constraints of DM direct detection.
We have also investigated the impact of DM indirect detection and relic density on the parameter space. A key distinction of our model from the standard NMCHM is the existence of the $\eta\eta\to\chi\chi$ channel, which can significantly alter the DM annihilation cross section when $m_\eta>m_\chi$.

Our model can easily achieve strong FOPTs within our parameter space of interest. 
This is not surprising, as it is a common characteristic of composite models with near conformal symmetry. 
We have identified two patterns of phase transition: 1-step and 2-step phase transitions. 
The type and strength of the phase transition are primarily influenced by the dilaton mass, the VEV of the dilaton field, and the color number $N$ of SU($N$) strong interaction in the composite sector.
For a given set of $N$ and $v_\chi$, we can determine the critical dilaton mass that separates the two phase transition patterns using an approximation. 
Supercooling is a significant feature in this model. As $N$ increases and $m_\chi$ decreases, supercooling progressively intensifies. This could potentially lead to late-time inflation, as the vacuum energy will dominate the universe at low temperatures. Following the supercooled phase transition, the universe will reheat, which could modify the production mechanism of DM. However, our calculations show that within the range of $m_\eta\lesssim10$~TeV, the traditional thermal freeze-out mechanism still holds. By combining the results from DM phenomenology and phase transition studies, we find that wihin a range of $200~\textrm{GeV}<m_\eta<1$~TeV, our model is consistent with all DM constraints and can simultaneously achieve a strong first-order phase transition.

Finally, we have also explored the stochastic background of gravitational waves produced by the FOPT, which are anticipated to be stronger than those produced by the traditional electroweak phase transition. The spectra of these GWs are expected to be detected by future space-based GW interferometer experiments. Our research can also be applied to electroweak baryogenesis, a topic we plan to investigate in our future studies.

\begin{acknowledgments}
The Large Language Models (LLMs), specifically ChatGPT and Gemini, were solely utilized for language refinement in this paper. The authors retain full responsibility for the content and conclusions presented in this work. This work is supported by the National Natural Science Foundation of China (NSFC) under Grants No.~11905300 and No.~12275367, the Fundamental Research Funds for the Central Universities, the Guangzhou Science and Technology Planning Project under Grant No.~2023A04J0008, and the Sun Yat-Sen University Science Foundation.

\end{acknowledgments}

\appendix
	
\section{Effective Lagrangian of $\text{NMCHM}_{\chi}$}\label{A}
The full Lagrangian of the NMCHM$_{\chi}$ model can be derived using spurion techniques~\cite{blum_wimp_2014,kim_wimp_2016}. By expanding the Higgs field and the dilaton field around their respective VEVs, we can obtain
\begin{align}\label{eq:fullEffL}
	\mathcal{L}_{pheno}&=\frac{1}{2}\partial_{\mu}\hat{h}\partial^{\mu}\hat{h}+\frac{1}{2}\partial_{\mu}\eta\partial^{\mu}\eta+\frac{1}{2}\partial_{\mu}\hat{\chi}\partial^{\mu}\hat{\chi}\notag\\
	&+\left(\frac{\xi}{\sqrt{1-\xi}}\frac{\hat{h}}{v}+\frac{(3\xi+1)\xi}{2(1-\xi)}\frac{\hat{h}^2}{v^2}+\frac{\xi^2}{2(1-\xi)}\frac{\eta^2}{v^2}+\frac{\xi}{2}\frac{\hat{\chi}^2}{v^2}+\sqrt{\xi}\frac{\hat{\chi}}{v}+\frac{2\xi^{\frac{3}{2}}}{\sqrt{1-\xi}}\frac{\hat{\chi}\hat{h}}{v^2}\right)(\partial_{\mu}\hat{h})^2\notag\\
	&+\left(\frac{\xi}{2}\frac{\hat{\chi}^2}{v^2}+\sqrt{\xi}\frac{\hat{\chi}}{v}+\frac{\xi}{2(1-\xi)}\frac{\eta^2}{v^2}\right)(\partial_{\mu}\eta)^2\notag\\
	&+\left(\frac{\xi}{\sqrt{1-\xi}}\frac{\eta}{v}+\frac{\xi(1+\xi)}{1-\xi}\frac{\hat{h}\eta}{v^2}+\frac{2\xi^{\frac{3}{2}}}{\sqrt{1-\xi}}\frac{\hat{\chi}\eta}{v^2}\right)\partial_\mu \hat{h}\partial^{\mu}\eta\notag\\
	&-\left(\lambda_h v^2(1-\xi)\hat{h}^2+\lambda_h v(1-\xi)^{\frac{3}{2}}\hat{h}^3+\frac{\lambda_h}{4}(1-\xi)^2\hat{h}^4\right)\notag\\
	&-\left(\frac{1}{2}m_{\eta}^2\eta^2+\frac{\lambda_\eta}{4}\eta^4\right)\notag\\
	&-\left(\frac{m_\chi^2}{2}\hat{\chi}^2+\lambda_{\chi}^{(3)}\hat{\chi}^3+\lambda_{\chi}^{(4)}\hat{\chi}^4\right)\notag\\
	&-m^2_{h\chi}\hat{h}\hat{\chi}\notag\\
	&-\left(\frac{\lambda_{h\eta}}{2}(1-\xi)\eta^2\hat{h}^2+\lambda_{h\eta}v\sqrt{1-\xi}\eta^2h\right)\notag\\
	&-\left(3\frac{m^2_{\eta}}{v_\chi^2}\eta^2\hat{\chi}^2+2\frac{m^2_{\eta}}{v_\chi}\eta^2\hat{\chi}\right)\notag\\
	&-\left(6\lambda_h\xi(1-\xi)\hat{h}^2\hat{\chi}^2+4\lambda_hv\sqrt{\xi}(1-\xi)\hat{h}^2\hat{\chi}+4\lambda_h\sqrt{\xi}(1-\xi)^{\frac{3}{2}}\hat{h}^3\hat{\chi}\right)\notag\\
	&-\lambda_{h\eta}\sqrt{(1-\xi)\xi}\eta^2\hat{h}\hat{\chi}\notag\\
	&+\left(1+2\sqrt{1-\xi}\frac{h}{v}+(1-\xi)\frac{h^2}{v^2}+2\sqrt{\xi(1-\xi)}(2+\gamma_{V^2})\frac{\hat{h}\hat{\chi}}{v^2}+(2+\gamma_{V^2})\frac{\hat{\chi}}{v_\chi}+(1+\frac{3}{2}\gamma_{V^2})\frac{\hat{\chi}^2}{v_\chi^2}\right)\notag\\
	&\cdot\left(M^2_WW^+_\mu W^{-\mu}+\frac{M^2_Z}{2}Z_\mu Z^{\mu}\right)\notag\\
	&-{\lambda^{(SM)}_\psi}v\left({1+\frac{1-2\xi}{\sqrt{1-\xi}}\frac{\hat{h}}{v}-\frac{\xi(3-2\xi)}{2(1-\xi)}\frac{\hat{h}^2}{v^2}-\frac{\xi}{2(1-\xi)}\frac{\eta^2}{v^2}}\right.\notag\\
	&\left.{+
	(1+\gamma_\psi)\sqrt{\xi}\frac{\hat{\chi}}{v}+\sqrt{\frac{\xi}{1-\xi}}(1+\gamma_\psi-2\xi)\frac{\hat{\chi}\hat{h}}{v^2}+\xi\gamma_\psi\frac{\hat{\chi}^2}{v^2}}\right)\bar{\psi}_f\psi_f\notag\\
    &+\frac{\alpha_s}{8\pi}\left(b^3_{IR}-b^3_{UV}\right)\left(1+\frac{\hat{\chi}}{v_\chi}-\frac{1}{2}\frac{\hat{\chi}^2}{v_\chi^2}\right)G^a_{\mu\nu}G^{a\mu\nu}+\frac{\alpha_{em}}{8\pi}\left(b^{em}_{IR}-b^{em}_{UV}\right)\left(1+\frac{\hat{\chi}}{v_\chi}-\frac{1}{2}\frac{\hat{\chi}^2}{v_\chi^2}\right)F_{\mu\nu}F^{\mu\nu}\notag\\
    &+\text{effective interaction terms for $hgg$, $hhgg$, $h\gamma\gamma$, $\eta\eta gg$, $\eta\eta \gamma\gamma$, $\chi gg$, $\chi\gamma\gamma$...}
     \end{align} 
where $\lambda_\psi^{(SM)}$ is the Yukawa coupling of the Standard Model, $\gamma_\psi$ and $\gamma_{V^2}$ parametrise the breaking of conformal invariance of the fermion sector and the gauge boson sector respectively. The contributions from gauge coupling are small compared to ones from fermions, and thus become negligible, that is, $\gamma_{V^2}\simeq 0$. The loop effects might be crucial for phenomenology of Higgs and dilaton, and thus we present the effective interactions with loop corrections explicitly as follows~\cite{chacko_effective_2013,spira_higgs_2017}:
\begin{align}
\mathcal{L}_{eff.}^{\text{(loop)}}
	&=\frac{\alpha_s}{12\pi}c^{h}_{f(1)}\frac{h}{v}G^a_{\mu\nu}G^{a\mu\nu}+\frac{\alpha_s}{24\pi}(2c^{h}_{f(2)}-(c^{h}_{f(1)})^2)\frac{h^2}{v^2}G^a_{\mu\nu}G^{a\mu\nu}\notag\\
	&+\frac{\alpha_{em}}{8\pi}\left(4Q_t^2c^{h}_{f(1)}+c^h_{W(1)}A^H_W\left(\frac{4M_W^2}{m_h^2}\right)\right)\frac{h}{v}F_{\mu\nu}F^{\mu\nu}\notag\\
	&+\frac{\alpha_s}{8\pi}\left(\frac{c^\chi_{f(1)}}{2}A_Q\left(\frac{4m_t^2}{m_\chi^2}\right)\right)\frac{\hat{\chi}}{v_\chi}G^a_{\mu\nu}G^{a\mu\nu}\notag\\
	&+\frac{\alpha_{em}}{8\pi}\left(\frac{4c^\chi_{f(1)}}{3}A_Q\left(\frac{4m_t^2}{m_\chi^2}\right)-c^\chi_{W(1)}A_W\left(\frac{4M_W^2}{m_\chi^2}\right)\right)\frac{\hat{\chi}}{v_\chi}F_{\mu\nu}F^{\mu\nu}\notag\\
	&+\frac{\alpha_s}{32\pi}\frac{c^\eta_{f(2)}}{2}A_Q\left(\frac{4m_t^2}{m_\eta^2}\right)\frac{\eta^2}{v^2}G^a_{\mu\nu}G^{a\mu\nu}\notag\\
	&+\frac{3\alpha_em}{16\pi}\frac{c^\eta_{f(2)}}{2}Q_t^2A_Q\left(\frac{4m_t^2}{m_\eta^2}\right)\frac{\eta^2}{v^2}F_{\mu\nu}F^{\mu\nu}
\end{align}
Where $A_Q$, $A_W$ and $A_W^H$ are loop functions which can be found in Ref.~\cite{chacko_effective_2013,spira_higgs_2017}.
By expanding the potential of dilaton, we can extract the mass and dominant self-interaction couplings as follows,
\begin{align}
	m^2_\chi&=V^{''}(v_\chi)-3\lambda_h\xi v^2\notag\\
	\lambda_{\chi}^{(3)}&=\frac{1}{6}V^{'''}(v_\chi)-\lambda_h v\xi^{\frac{3}{2}}\notag\\
	\lambda_{\chi}^{(4)}&=\frac{1}{24}V^{''''}(v_\chi)-\frac{\lambda_h}{4} \xi^{2}
\end{align}	
where some approximations have been applied and 
\begin{align}
	V^{''}(v_\chi)&=-\left(4 {c_\chi} {v_\chi}^2 {\gamma_\epsilon} {g_\chi}^2\right)\notag\\
	V^{'''}(v_\chi)&=\frac{2 {g_\chi}^2 {v_\chi} }{{c_\epsilon}}\left(-48 {c_\chi }^2 {c_\epsilon }^2-2 {c_\chi } {c_\epsilon} (\gamma_\epsilon  (\gamma_\epsilon +24)+104)+30 {c_\chi } {c_\epsilon} \sqrt{16 {c_\chi} {c_\epsilon}+(\gamma_\epsilon +4)^2}\right.\notag\\
	&\left.+(\gamma_\epsilon +4) (\gamma_\epsilon +11) \left(\sqrt{16 {c_\chi} {c_\epsilon}+(\gamma_\epsilon +4)^2}-\gamma_\epsilon -4\right)\right)\notag\\
	V^{''''}(v_\chi)&=-\frac{2 g_{\chi }^2}{c_{\epsilon }} \left(96 c_{\chi }^2 c_{\epsilon }^2 \left(\sqrt{16 c_{\chi } c_{\epsilon }+\left(\gamma _{\epsilon }+4\right){}^2}-15\right)\right.\notag\\
	&\left.+2 c_{\chi } c_{\epsilon } \left(\gamma _{\epsilon }^2 \left(\sqrt{16 c_{\chi } c_{\epsilon }+\left(\gamma _{\epsilon }+4\right){}^2}-50\right)+48 \gamma _{\epsilon } \left(\sqrt{16 c_{\chi } c_{\epsilon }+\left(\gamma _{\epsilon }+4\right){}^2}-15\right)\right.\right.\notag\\
	&\left.\left.+7 \left(53 \sqrt{16 c_{\chi } c_{\epsilon }+\left(\gamma _{\epsilon }+4\right){}^2}-328\right)\right)\right.\notag\\
	&\left.+\left(\gamma _{\epsilon }+4\right) \left(\gamma _{\epsilon } \left(\gamma _{\epsilon }+42\right)+203\right) \left(\sqrt{16 c_{\chi } c_{\epsilon }+\left(\gamma _{\epsilon }+4\right){}^2}-\gamma _{\epsilon }-4\right)\right)
\end{align}
For the annihilation of $\eta$, the derivative coupling of $\eta$ can be determined by $p_1^\mu\approx p_2^\mu\approx (m_\eta,0)$. The interactions between $\eta$ and $h$ are given by 
\begin{align}
	\mathcal{F}^{(d)}_{\eta\eta\hat{h}\hat{h}}&=-i\left(\frac{\xi(\xi+1)}{1-\xi}(p_1\cdot p_3+p_1\cdot p_4+p_2\cdot p_3+p_2\cdot p_4)\frac{\hat{h}^2\eta^2}{v^2}-\frac{2\xi^2}{1-\xi}p_3\cdot p_4\frac{\hat{h}^2\eta^2}{v^2}\right)\notag\\
	&\overset{\text{s-channel}}{\simeq} -i\left( \frac{\xi(\xi+1)}{1-\xi}\frac{4m_{\eta}^2}{v^2}\hat{h}^2\eta^2-\frac{2\xi^2}{1-\xi}\frac{2m_\eta^2-m_h^2}{v^2}\hat{h}^2\eta^2\right)\\
	\mathcal{F}^{(d)}_{\eta\eta\hat{h}}&=-i\cdot\frac{\xi}{\sqrt{1-\xi}}(p_1\cdot p_3+p_2\cdot p_3)\frac{\eta^2\hat{h}}{v}\overset{\text{s-channel}}{\simeq} -i\cdot\frac{\xi}{\sqrt{1-\xi}}\frac{4m_\eta^2}{v}\eta^2\hat{h}\\
	\mathcal{F}^{(d)}_{\eta\eta\hat{\chi}\hat{\chi}}&=-i\cdot4\cdot\left(\frac{\xi}{2}\right)(-p_1\cdot p_2)\frac{\eta^2\hat{\chi}^2}{v^2}\overset{\text{s-channel}}{\simeq} -i\cdot\left(-2\xi\frac{m^2_\eta}{v^2}\eta^2\hat{\chi}^2\right)\\
	\mathcal{F}^{(d)}_{\eta\eta\hat{\chi}}&=-i\cdot2\cdot\sqrt{\xi}(-p_1\cdot p_2)\frac{\eta^2\hat{\chi}}{v}\overset{\text{s-channel}}{\simeq} -i\cdot\left(-2\sqrt{\xi}\frac{m^2_\eta}{v}\eta^2\hat{\chi}\right)\\
	\mathcal{F}^{(d)}_{\eta\eta\hat{h}\hat{\chi}}&=-i\cdot\frac{2\xi^{\frac{3}{2}}}{\sqrt{1-\xi}}(p_2\cdot p_4+p_1\cdot p_4)\frac{\eta^2\hat{h}\hat{\chi}}{v^2}\overset{\text{s-channel}}{\simeq}-i\cdot\frac{2\xi^{\frac{3}{2}}}{\sqrt{1-\xi}}\left(\frac{4m^2_\eta-m_\chi^2+m_h^2}{2v^2}\right)\eta^2\hat{h}\hat{\chi}
\end{align}
Finally we obtain the effective Lagrangian modified by the derivative couplings for s-channel annihilation:
\begin{align}\label{eq:deriv_coup}
	\mathcal{L}_{\eta\eta\hat{h}\hat{h}}&=\left[-2\lambda_{h\eta}(1-\xi)+\frac{\xi(\xi+1)}{1-\xi}\frac{4m_{\eta}^2}{v^2}-\frac{2\xi^2}{1-\xi}\frac{2m_\eta^2-m_h^2}{v^2}\right]\frac{\hat{h}^2\eta^2}{4}\notag\\
	\mathcal{L}_{\eta\eta\hat{h}}&=\left[-2\lambda_{h\eta}v\sqrt{1-\xi}+\frac{\xi}{\sqrt{1-\xi}}\frac{4m_\eta^2}{v}\right]\frac{\eta^2\hat{h}}{2}\notag\\
	\mathcal{L}_{\eta\eta\hat{\chi}\hat{\chi}}&=\left[-12\frac{m_\eta^2}{v_{\chi}^2}-2\xi\frac{m_\eta^2}{v^2}\right]\frac{\eta^2\hat{\chi}^2}{4}=-14\frac{m_\eta^2}{v_{\chi}^2}\frac{\eta^2\hat{\chi}^2}{4}\notag\\
	\mathcal{L}_{\eta\eta\hat{\chi}}&=\left[-4\frac{m_\eta^2}{v_{\chi}}-2\sqrt{\xi}\frac{m_\eta^2}{v}\right]\frac{\eta^2\hat{\chi}}{2}=-6\frac{m_\eta^2}{v_{\chi}}\frac{\eta^2\hat{\chi}}{2}\notag\\
	\mathcal{L}_{\eta\eta\hat{h}\hat{\chi}}&=\left[-2\lambda_{h\eta}\sqrt{(1-\xi)\xi}+\frac{2\xi^{\frac{3}{2}}}{\sqrt{1-\xi}}\left(\frac{4m^2_\eta-m_\chi^2+m_h^2}{2v^2}\right)\right]\frac{\eta^2\hat{h}\hat{\chi}}{2}
\end{align}	
The calculation of t,u channel is similar and only requires the replacement of the corresponding momenta.
\section{Field dependent mass matrix elements }\label{appdB}
The specific form of effective masses with Landau gauge is
 \begin{equation}
     m_{W}^{2}(\chi_1,\chi_2,\chi_3)=\frac{m^2_{W_0}}{v^2}\chi^2,\quad m_{Z}^{2}(\chi_1,\chi_2,\chi_3)=\frac{m^2_{Z_0}}{v^2}\chi^2,\quad m_{CFT}=g_\chi^2\chi^2
     \end{equation}
     \begin{align}
        m^2_{11}(\chi_1,\chi_2,\chi_3)&=12(\frac{\mu_{h}^{2}}{2v_{\chi}^2}+\frac{\lambda_{h}}{4}+c_{\chi}g_{\chi}^{2})\chi_{1}^{2}+2(\frac{\mu^{2}_{\eta}+\mu^{2}_{h}}{2v_{\chi}^{2}}+\frac{\lambda_{h\eta}}{2}+2c_{\chi}g_{\chi}^{2})\chi_{2}^{2}\notag\\
        & +2(\frac{\mu_{h}^{2}}{2v_{\chi}^{2}}+2c_{\chi}g_{\chi}^{2})\chi_{3}^{2}-\frac{\partial^2}{\partial\chi_{1}^2}(\epsilon(\chi)\chi^4)
        \end{align}
        \begin{align}
        m^2_{22}(\chi_1,\chi_2,\chi_3)&=12(\frac{\mu_{\eta}^{2}}{2v_{\chi}^2}+\frac{\lambda_{\eta}}{4}+c_{\chi}g_{\chi}^{2})\chi_{2}^{2}+2(\frac{\mu^{2}_{\eta}+\mu^{2}_{h}}{2v_{\chi}^{2}}+\frac{\lambda_{h\eta}}{2}+2c_{\chi}g_{\chi}^{2})\chi_{1}^{2}\notag\\
        & +2(\frac{\mu_{\eta}^{2}}{2v_{\chi}^{2}}+2c_{\chi}g_{\chi}^{2})\chi_{3}^{2}-\frac{\partial^2}{\partial\chi_{2}^2}(\epsilon(\chi)\chi^4)
        \end{align}
        \begin{align}
        m^2_{33}(\chi_1,\chi_2,\chi_3)&=12c_{\chi}g_{\chi}^{2}\chi_{3}^{2}+2(\frac{\mu_{h}^{2}}{2v_{\chi}^{2}}+2c_{\chi}g_{\chi}^{2})\chi_{1}^{2}+2(\frac{\mu_{\eta}^{2}}{2v_{\chi}^{2}}+2c_{\chi}g_{\chi}^{2})\chi_{2}^{2}\notag\\
        &-\frac{\partial^2}{\partial\chi_{3}^2}(\epsilon(\chi)\chi^4)
        \end{align}
        \begin{align}
        m^2_{12}(\chi_1,\chi_2,\chi_3)=4(\frac{\mu^{2}_{\eta}+\mu^{2}_{h}}{2v_{\chi}^{2}}+\frac{\lambda_{h\eta}}{2}+2c_{\chi}g_{\chi}^{2})\chi_{1}\chi_{2}-\frac{\partial^2}{\partial\chi_{1}\partial\chi_{2}}(\epsilon(\chi)\chi^4)
        \end{align}
        \begin{align}
        m^2_{13}(\chi_1,\chi_2,\chi_3)=4(\frac{\mu^{2}_{h}}{2v_{\chi}^{2}}+2c_{\chi}g_{\chi}^{2})\chi_{1}\chi_{3}-\frac{\partial^2}{\partial\chi_{1}\partial\chi_{3}}(\epsilon(\chi)\chi^4)
        \end{align}
        \begin{align}
        m^2_{23}(\chi_1,\chi_2,\chi_3)=4(\frac{\mu^{2}_{\eta}}{2v_{\chi}^{2}}+2c_{\chi}g_{\chi}^{2})\chi_{2}\chi_{3}-\frac{\partial^2}{\partial\chi_{2}\partial\chi_{3}}(\epsilon(\chi)\chi^4)
        \end{align}
        \begin{align}
    m^2_{\Pi \chi_1}=m^2_{\Pi \chi_2}=m^2_{\Pi \chi_3}=0
    \end{align}
    \begin{align}
    m^2_{\Pi}=2[(\frac{\mu^2_{h}}{2v_{\chi}^2}+\frac{\lambda_h}{2})\chi^2_1+(\frac{\mu^2_{h}}{2v_{\chi}^2}+\frac{\lambda_{h\eta}}{2})\chi^2_2+\frac{\mu^2_h}{2v_\chi^2}\chi^2_3]
\end{align}
The mass matrix for $\chi_{(1,2,3)}$ is given by
\begin{equation}
M_S^2 = \left(\begin{array}{ccc}
m_{11}^2&m_{12}^2&m_{13}^2\\
m_{12}^2&m_{22}^2&m_{23}^2\\
m_{13}^2&m_{23}^2&m_{33}^2
\end{array}\right).
\end{equation}
where
    \begin{align}
    \frac{\partial^2}{\partial\chi^2_i}(\epsilon(\chi)\chi^4)&=\frac{\chi^2-\chi_i^2}{\chi^3}(\frac{\partial}{\partial\chi}(\epsilon(\chi)\chi^4))+\frac{\chi_i^2}{\chi^2}(\frac{\partial^2}{\partial\chi^2}(\epsilon(\chi)\chi^4))\\
    \frac{\partial^2}{\partial\chi_i\partial\chi_j}(\epsilon(\chi)\chi^4)&=-\frac{\chi_i\chi_j}{\chi^3}(\frac{\partial}{\partial\chi}(\epsilon(\chi)\chi^4))+\frac{\chi_i\chi_j}{\chi^2}(\frac{\partial^2}{\partial\chi^2}(\epsilon(\chi)\chi^4)),(i\ne j)
  \end{align}
    Solving the RGE with the initial condition $\epsilon(\chi)=\epsilon(v_\chi)$, the CFT deformation coefficient yields
 \begin{align}
    \epsilon(\chi)=\frac{8 c_\chi g_\chi^2 \gamma_\epsilon\left(\chi / v_\chi\right)^{\gamma_\epsilon}}{\gamma_\epsilon\left(4+\gamma_\epsilon+\sqrt{16 c_\epsilon c_\chi+\left(4+\gamma_\epsilon\right)^2}\right)+8 c_\epsilon c_\chi\left(1-\left(\chi / v_\chi\right)^{\gamma_\epsilon}\right)}
    \end{align}
    We can use the RGE to simplify the calculation:
    \begin{align}
    \chi\frac{\partial \epsilon}{\partial\chi}\simeq \gamma_{\epsilon}\epsilon+\frac{c_{\epsilon}}{g_{\chi}^2}\epsilon^2.
    \end{align}
    The derivatives are written as
\begin{align}
    \frac{\partial}{\partial\chi}(\epsilon(\chi)\chi^4)&=\chi^3[(4+\gamma_{\epsilon})\epsilon(\chi)+\frac{c_{\epsilon}}{g_{\chi}^2}\epsilon^2[\chi]]\notag\\
    \frac{\partial^2}{\partial\chi^2}(\epsilon(\chi)\chi^4)&=\chi^2[(3+\gamma_{\epsilon})(4+\gamma_{\epsilon})\epsilon(\chi)+\frac{c_{\epsilon}}{g_{\chi}^2}(7+3\gamma_{\epsilon})\epsilon^2[\chi]+2(\frac{c_{\epsilon}}{g_{\chi}^2})^2\epsilon^3[\chi]]
 \end{align}
    The effective mass of top quark is given by
\begin{align}
    m_t(\chi_1,\chi_2,\chi_3)=\frac{m_{t0}\chi_1}{v\sqrt{1-(\frac{v}{v_\chi})^2}}\sqrt{1-(\frac{\chi_1}{\chi})^2-(\frac{\chi_2}{\chi})^2}
\end{align}
The VEV at $T=0$ yields
\begin{align}
	\left \langle \chi_1 \right \rangle&=\left \langle h \right \rangle=v\approx 246\text{GeV}\notag\\
	\left \langle \chi_2 \right \rangle&=\left \langle \eta \right \rangle=0\notag\\
	\left \langle \chi_3 \right \rangle&=\sqrt{ \left \langle \chi \right \rangle^2- \left \langle \chi_1 \right \rangle^2- \left \langle \chi_2 \right \rangle^2}=\sqrt{v_\chi^2-v^2}
\end{align}
and
\begin{align}
	&g_{\chi}=\frac{4\pi}{\sqrt{N}},\quad\gamma_{\epsilon}=-\frac{m_{\chi}^2}{4c_{\chi}g_{\chi}^2v_\chi^2},\quad m_{\eta}^2=\frac{\partial^2V}{\partial\eta^2}|_{\eta=0,h=v}=\mu^2_{\eta}+\lambda_{h\eta}v^2,\\
	&\left\{\begin{array}{ll}
		\mu_h^2&=-\frac{1}{2}m^2_h
		\\
		\lambda_h&=-\frac{\mu_h^2}{v^2}=\frac{m_h^2}{2v^2}
	\end{array}\right.
\end{align}

\section{High temperature expansion of the effective potential}\label{appdC}
We consider the high temperature corrections by assuming $T^2>>m_i^2$ and the thermal function can be expanded as
\begin{align}
	J_B\left[\frac{m_b^2}{T^2}\right]&=-\frac{\pi^4}{45}+\frac{\pi^2}{12}\frac{m^2_i}{T^2}+\mathcal{O}\left(\frac{1}{T^3}\right)\notag\\
	J_F\left[\frac{m_f^2}{T^2}\right]&=\frac{7\pi^4}{360}-\frac{\pi^2}{24}\frac{m^2_i}{T^2}+\mathcal{O}\left(\frac{1}{T^4}\right)
\end{align}
which yields
\begin{align}
	V_T\approx\frac{1}{24}\sum_{b}n_bm^2_b(\chi_1,\chi_2,\chi)T^2+\frac{1}{48}\sum_{f}n_fm^2_f(\chi_1,\chi_2,\chi)T^2
\end{align}
where $b$ sums over $\chi_i,\Pi_j,W^{\pm},Z,\phi_{CFT}\quad(i,j=1,2,3)$ and $f=t$. In order to simplify the expressions, we turn to the basis with $\chi_1$, $\chi_2$ and $\chi$. Plugging the expression of effective masses into the formula, we obtain
\begin{align}
	V_T&=\frac{1}{24}\left[6\frac{m_W^2}{v^2}\chi_1^2+3\frac{m_Z^2}{v^2}\chi_1^2+\mu_h^2\left(\frac{\chi}{v_\chi}\right)^2+3\lambda_h\chi_1^2+\lambda_{h\eta}\chi_2^2\right.\notag\\
	&\left.+\mu^2_\eta\left(\frac{\chi}{v_\chi}\right)^2+3\lambda_h\chi_1^2+\lambda_{h\eta}\chi_1^2\right.\notag\\
	&\left.+3\left(\mu^2_h\left(\frac{\chi}{v_\chi}\right)^2+\lambda_h\chi_1^2+\lambda_{h\eta}\chi_2^2\right)\right.\notag\\
	&\left.+\frac{\mu_h^2}{v_\chi^2}\chi_1^2+\frac{\mu_\eta^2}{v_\chi^2}\chi_2^2-\frac{\partial^2(\epsilon(\chi)\chi^4)}{\partial\chi^2}
	\right]T^2\notag\\
	&=\frac{1}{24}\left(\frac{6m_W^2}{v^2}+\frac{3m_Z^2}{v^2}+6\lambda_h+\lambda_{h\eta}+\frac{\mu^2_h}{v_\chi^2}+\frac{6m_t^2}{v^2(1-\frac{v^2}{v_\chi^2})}\right)\chi_1^2T^2\notag\\
	&+\frac{1}{24}\left(4\lambda_{h\eta}+3\lambda_{\eta}+\frac{\mu^2_\eta}{v_\chi^2}\right)\chi_2^2T^2\notag\\
	&-\frac{1}{4}\frac{m_t^2}{v^2(1-\frac{v^2}{v_\chi^2})}\frac{1}{\chi^2}\chi_1^2T^2-\frac{1}{4}\frac{m_t^2}{v^2(1-\frac{v^2}{v_\chi^2})}\frac{1}{\chi^2}\chi_1^2\chi_2^2T^2\notag\\
	&+\frac{1}{24}\left[\left(\frac{4\mu^2_h+\mu^2_\eta}{v_\chi^2}+\frac{45}{4}N^2g_\chi^2\right)\chi^2-\frac{\partial^2(\epsilon(\chi)\chi^4)}{\partial\chi^2}\right]T^2
\end{align}
Finally, the effective potential is given by
\begin{align}\label{eq:chithermalpotential}
	&V_{eff}(\chi_1,\chi_2,\chi)=V_{0}(\chi_1,\chi_2,\chi)+V_{1}^T(\chi_1,\chi_2,\chi)\notag\\
	&=\frac{1}{2}\left(\frac{\chi^2}{v_\chi^2}\mu^2_h+c_hT^2\right)\chi_1^2+\frac{1}{2}\left(\frac{\chi^2}{v_\chi^2}\mu^2_\eta+c_\eta T^2\right)\chi_2^2\notag\\
	&+\frac{1}{4}\left(\lambda_h-\frac{4d_h}{\chi^2}T^2\right)\chi_1^4+\frac{\lambda_{h\eta}}{4}\chi_2^4+\frac{1}{2}\left(\lambda_{h\eta}-\frac{2d_{h\eta}}{\chi^2}T^2\right)\chi_1^2\chi_2^2\notag\\
	&+\left(c_\chi g^2_{\chi}-\epsilon(\chi)-\kappa(\chi,T)\right)\chi^4
\end{align}
The coefficients are shown explicitly as follows,
\begin{align}\label{eq:chithermalpotentialcoe}
	c_h&=\frac{1}{12}\left(\frac{6m_W^2}{v^2}+\frac{3m_Z^2}{v^2}+6\lambda_h+\lambda_{h\eta}+\frac{\mu^2_h}{v_\chi^2}+\frac{6m_t^2}{v^2(1-\frac{v^2}{v_\chi^2})}\right)\notag\\
	c_\eta&=\frac{1}{12}\left(4\lambda_{h\eta}+3\lambda_{\eta}+\frac{\mu^2_\eta}{v_\chi^2}\right)\notag\\
	d_h&=d_{h\eta}=\frac{1}{4}\frac{m_t^2}{v^2(1-\frac{v^2}{v_\chi^2})}\notag\\
	\kappa(\chi,T)&=-\frac{1}{24}\left[\left(\frac{4\mu^2_h+\mu^2_\eta}{v_\chi^2}+\frac{45}{4}N^2g_\chi^2\right)\frac{1}{\chi^2}-\frac{\partial^2(\epsilon(\chi)\chi^4)}{\partial\chi^2}\frac{1}{\chi^4}\right]T^2
\end{align}
In the case of 2-step phase transitions, we can approximately calculate the PT in the Higgs direction by fixing the dilaton field with its VEV in the second step of phase transitions. 

In the canonical basis which parametrizing the scalar fields as $\chi_1$, $\chi_2$ and $\chi_3$, the effective potential is given by
\begin{align}
	&V_{eff}(\chi_1,\chi_2,\chi_3)=V_{0}(\chi_1,\chi_2,\chi_3)+V_{T}^{(1)}(\chi_1,\chi_2,\chi_3)\notag\\
	&=\frac{T^2}{24}\chi_1^2\left[22\frac{\mu_h^2}{2v_\chi^2}+2\frac{\mu_\eta^2}{2v_\chi^2}+6\lambda_{h}+\lambda_{h\eta}+\frac{6m_{W}^2}{v^2}+\frac{3m_{Z}^2}{v^2}+\frac{6m_t^2}{v^2\left(1-\frac{v^2}{v_\chi^2}\right)}+20c_\chi g_\chi^2+\frac{45}{4}N^2g_\chi^2\right]\notag\\
	&+\frac{T^2}{24}\chi_2^2\left[8\frac{\mu_h^2}{2v_\chi^2}+16\frac{\mu_\eta^2}{2v_\chi^2}+3\lambda_{\eta}+4\lambda_{h\eta}+20c_\chi g_\chi^2+\frac{45}{4}N^2g_\chi^2\right]\notag\\
	&+\frac{T^2}{24}\chi_3^2\left[8\frac{\mu_h^2}{2v_\chi^2}+2\frac{\mu_\eta^2}{2v_\chi^2}+20c_\chi g_\chi^2+\frac{45}{4}N^2g_\chi^2\right]\notag\\
	&-\frac{T^2}{4}\frac{m_t^2}{v^2\left(1-\frac{v^2}{v_\chi^2}\right)}\frac{\chi_1^4}{\chi_1^2+\chi_2^2+\chi_3^2}-\frac{T^2}{4}\frac{m_t^2}{v^2\left(1-\frac{v^2}{v_\chi^2}\right)}\frac{\chi_1^2\chi_2^2}{\chi_1^2+\chi_2^2+\chi_3^2}\notag\\
	&+\left(\frac{\mu_{h}^{2}}{2v_\chi^2}+\frac{\lambda_{h}}{4}+c_{\chi}g_{\chi}^{2}\right)\chi_{1}^{4}+\left(\frac{\mu_{\eta}^{2}}{2v_{\chi}^2}+\frac{\lambda_{\eta}}{4}+c_{\chi}g_{\chi}^{2}\right)\chi_{2}^{4}+c_{\chi}g_{\chi}^{2}\chi_{3}^{4}\notag\\
	&+\left(\frac{\mu^{2}_{\eta}+\mu^{2}_{h}}{2v_{\chi}^{2}}+\frac{\lambda_{h\eta}}{2}+2c_{\chi}g_{\chi}^{2}\right)\chi_{1}^{2}\chi_{2}^{2}+\left(\frac{\mu_{\eta}^{2}}{2v_{\chi}^{2}}+2c_{\chi}g_{\chi}^{2}\right)\chi_{2}^{2}\chi_{3}^{2}+\left(\frac{\mu_{h}^{2}}{2v_{\chi}^{2}}+2c_{\chi}g_{\chi}^{2}\right)\chi_{1}^{2}\chi_{3}^{2}\notag\\
	& -\epsilon(\chi)(\chi_{1}^{2}+\chi_{2}^{2}+\chi_{3}^{2})^{2}\notag\\
	&-\frac{T^2}{24}\chi^2\left[(4+\gamma_\epsilon)(5+\gamma_\epsilon)\epsilon(\chi)+\frac{c_\epsilon}{g_\chi^2}(9+3\gamma_\epsilon)\epsilon^2(x)+2\left(\frac{c_\epsilon}{g_\chi^2}\right)^2\epsilon^3(\chi)\right]
\end{align}
The expression of the effective potential in canonical basis can be used to calculate the thermal masses of $\chi_1$, $\chi_2$ and $\chi_3$. This kind of parametrization is valid for the origin $\chi=0$ as well since the coordinate singularity is eliminated. 

\section{Triangular potential approximation solution of tunneling rate}\label{appdD}
\begin{figure}[htb!!]
		\centering
		\includegraphics[width=0.5\linewidth]{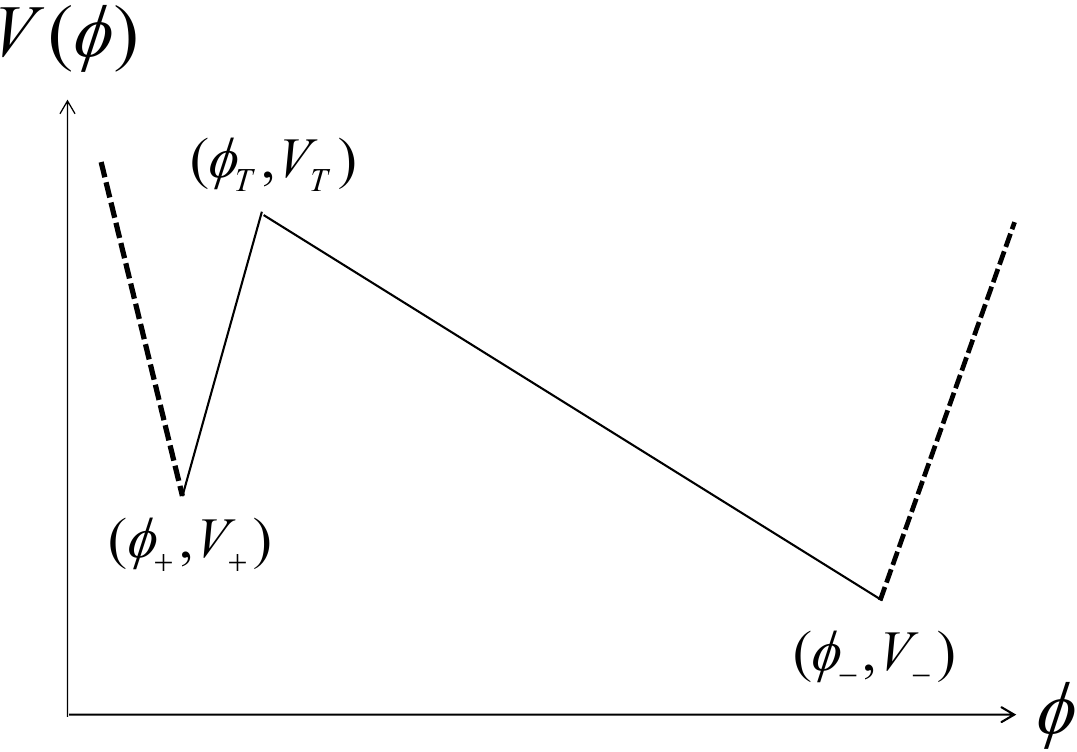}
		\caption{Schematic diagram of the triangular potential approximation and the related parameters}\label{figestimatepoten}
	\end{figure}

In this section, we follow the method given by Ref.~\cite{Duncan:1992ai} to derive the thermal tunneling rate for an approximate single-field solution. According to the shape of dilaton potential, we approximate the barrier with a triangular potential (see FIG.\ref{figestimatepoten}). 
The thermal tunneling rate is determined by the $O(3)\text{-symmetric}$ action $S_3$ defined by
\begin{align}
    \frac{S_3}{T}&=\frac{1}{T}\int d^3x\left[\frac{1}{2}(\nabla\phi)^2+V(\phi,T)\right]=\frac{4\pi}{T}\int r^2dr\left[\frac{1}{2}\left(\frac{\partial\phi}{\partial r}\right)^2+V(\phi(r,T),T)\right]
\end{align}
The equation of motion yields
\begin{align}
\frac{\partial^2\phi}{\partial r^2}+\frac{2}{r}\frac{\partial\phi}{d\partial r}=V^\prime(\phi(r,T),T)
\end{align}
where $V^\prime\equiv \partial V/\partial phi$. In the calculation, we assume the starting point $\phi_0$ is smaller than the true vacuum $\phi_{-}$ and in this case the boundary conditions are
\begin{equation}
\left\{\begin{aligned}
    &\phi(r)=\phi_+,\quad \dot{\phi}(r)=0,\quad r=R_+\\
    &\phi(r)=\phi_0,\quad \dot{\phi}(r)=0,\quad r=0
    \end{aligned}\right.
\end{equation}
where $\phi_+$ is the false vacuum.  To simplify the calculation, we define the magnitudes of the gradients of the potential as
\begin{align}
    \lambda_{\pm}=\frac{\Delta V_{\pm}}{\Delta\phi_\pm},\quad \Delta\phi_\pm=\pm(\phi_T),\quad \Delta V_\pm=(V_T-V_\pm)
\end{align}
where $\phi_T$ and $V_T$ are the maximum points and maximum values of the effective potential. We can view the effective potential as a piecewise function which is the benefit of using a line segment approximation and solve for the expected value of the field on either side of the maximum. That is
\begin{equation}
    \left\{\begin{aligned}
        \frac{1}{r^2}\partial_r\left(r^2\frac{\partial\phi}{\partial r}\right)=-\lambda_-\Rightarrow r^2\frac{\partial\phi}{\partial r}\bigg|^r_0=-\int^r_0\lambda_-r^2 dr,\quad 0\leq r\leq R_T\\
        \frac{1}{r^2}\partial_r\left(r^2\frac{\partial\phi}{\partial r}\right)=\lambda_+\Rightarrow r^2\frac{\partial\phi}{\partial r}\bigg|^{R_+}_r=\int^{R_+}_r\lambda_+r^2 dr,\quad R_T\leq r\leq R_+
    \end{aligned}
    \right.
\end{equation}
The solutions become
\begin{equation}\label{phisolutions}
    \left\{\begin{aligned}
        \phi_R(r)&=\phi_0-\frac{\lambda_-}{6},\quad &\quad 0\leq r\leq R_T\\
        \phi_L(r)&=\phi_++\frac{\lambda_+}{6r}\left[r^3+2R_+^3-3R^2_+r\right],\quad \quad &R_T\leq r\leq R_+
    \end{aligned}
    \right.
\end{equation}
Plugging into the boundary conditions and the continuity conditions
\begin{align}
   {\phi}_L(R_T)={\phi}_R(R_T),\quad  \dot{\phi}_L(R_T)= \dot{\phi}_R(R_T)
\end{align}
and all parameters can be expressed in terms of $R_T$
\begin{equation}
    \left\{\begin{aligned}
        R^3_+&=(1+c)R^3_T\\
        \Delta\phi_+&=\frac{\lambda_+}{6}\left(1+2(1+c)-3(1+c)^{\frac{2}{3}}\right)R_T^2
    \end{aligned}\right.
\end{equation}
where $c=\frac{\lambda_c}{\lambda_+}$, which totally depends on the shape of effective potential. In order to calculate the action, we also need the expression of the effective potential. Similar to the calculation of field value, we divide the effective potential into the left and right sides and treat them respectively
\begin{equation}\label{Vsolutions}
    \left\{\begin{aligned}
       V(\phi(r,T),T)_R&=V_-+\lambda_-\left(\phi_--\phi_L(r)\right),\quad \quad 0\leq r\leq R_T\\
       V(\phi(r,T),T)_L&=V_++\lambda_+\left(\phi_L(r)-\phi_+\right),\quad R_T\leq r\leq R_+
    \end{aligned}\right.
\end{equation}
The action yields
\begin{align}\label{S3solution}
    \frac{S_3}{T}&=\frac{4\pi}{T}\int^{R_T}_0r^2dr\left[\frac{1}{2}\left(\frac{\partial \phi_R}{\partial r}\right)^2+V(\phi(r,T),T)_R\right]\notag\\
    &+\frac{4\pi}{T}\int^{R_+}_{R_T}r^2dr\left[\frac{1}{2}\left(\frac{\partial \phi_L}{\partial r}\right)^2+V(\phi(r,T),T)_L\right]
\end{align}
Substitute Eq.(\ref{phisolutions}) and Eq.(\ref{Vsolutions}) into the above formula, after some tedious calculation we finally obtain
\begin{align}
     \frac{S_3}{T}&=\frac{4\pi}{T}\left[\left(\mathcal{F}^{(1)}+\mathcal{F}^{(2)}\right)R^5_T+\left(\mathcal{T}^{(1)}+\mathcal{T}^{(2)}\right)R^3_T\right]
\end{align}
where
\begin{align}
    \mathcal{F}^{(1)}&=-\frac{1}{90}\lambda^2_-,\quad \mathcal{T}^{(1)}=\frac{1}{3}(V_-+\lambda_-\Delta \phi_-),\quad\mathcal{T}^{(2)}=-\frac{\Delta V_+}{3}c \notag\\
     \mathcal{F}^{(2)}&=\frac{\lambda_+^2}{3}\left[-\frac{2}{15}+\frac{1}{2}(1+c)^{\frac{2}{3}}-\frac{1}{3}(1+c)-\frac{1}{5}(1+c)^{\frac{5}{3}}+\frac{1}{6}(1+c)^2\right]
\end{align}
The tunneling rate can be estimated by
\begin{align}
    \Gamma\propto e^{-S_E}
\end{align}
where
\begin{align}
    S_E=\frac{S_3}{T}\left(\phi(r)\right)-\frac{S_3}{T}\left(\phi_+\right)
\end{align}
In our model, when the tunneling occur, the absolute value of the maximum potential $V_T$ is relatively small and can be approximated by $\sim 0$. The coordinate is chosen such that $\phi_+=0$. The parameters in the approximate calculation can be concretely represented by the parameters in the dilaton model
\begin{align}
   & \Delta V_+\approx-V_+=\frac{\pi^2}{8}N^2T^4,\quad \Delta V_-=-V_-=\frac{1}{16}m^2_\chi v_\chi^2\notag\\
    &\phi_+=0,\quad \phi_-=v_\chi,\quad\phi_T=\Delta\phi_+\sim a\cdot T/g_\chi,\quad \Delta\phi_-=v_\chi-a\cdot T/g_\chi\notag\\
    &R_T=\left[\frac{6\Delta\phi_+}{\lambda_+\left(1+2(1+c)-3(1+c)^\frac{2}{3}\right)}\right]^{\frac{1}{2}}
\end{align}
where we have made a linear approximation for $\Delta{\phi_+}$ with a proportional coefficient $a$. By experimenting with different parameter points, we set $a$ as $10$.

\section{Calculation of nucleation condition in supercooled phase transition }\label{appdE}
The nucleation rate is given by Eq.~\eqref{eq:nucleationrate}
\begin{equation}
  \Gamma \approx A T^4 e^{- S (T)}.
\end{equation}
The nucleation criterion requires
\begin{equation}
  \int_{t_{\text{c}}}^{t_{\text{n}}} d t \frac{\Gamma}{H^3} = 1.
\end{equation}
which means at least one bubble is generated per Hubble volume per Hubble time.
For simplicity, We expand the action with $T-T_0$ to the linear order,
\begin{equation}
  S (T) \approx S \left( T_{\text{0}} \right) + \left. \frac{d S}{d T}
  \right|_{T = T_0} (T - T_0) = S \left( T_{\text{0}} \right) + \left. \frac{d
  S}{d T} \right|_{T = T_0} (T - T_0) = S \left( T_{\text{0}} \right) + S'
  (T_0) (T - T_0) .
\end{equation}
In the case of vacuum energy domination, by assuming a constant vacuum energy, we have
\begin{equation}
  H_{\mathrm{vac}} = \sqrt{\frac{\rho_{\mathrm{vac}}}{3 M^2_{\text{Pl}}}}, \qquad
  a (t) = a_0 e^{H_{\mathrm{vac}} \left( t - t_{\text{c}} \right)} .
\end{equation}
Since $s a^3 = \mathrm{const}.$, we can find
\begin{equation}
  d t = - (H_{\mathrm{vac}} T)^{- 1} d T, \qquad T = T_0 e^{- H_{\mathrm{vac}}
  \left( t - t_{\text{c}} \right)} .
\end{equation}
The nucleation condition now becomes
\begin{eqnarray}
 1= \int_{t_{\text{c}}}^{t_{\text{n}}} d t \frac{\Gamma}{H_{\mathrm{vac}}^3} & = &
  \int_{T_{\text{n}}}^{T_{\text{c}}} d T \frac{\Gamma}{H^4_{\mathrm{vac}} T}
  \nonumber\\
  & = & \frac{A}{H^4_{\mathrm{vac}}} \int_{T_{\text{n}}}^{T_{\text{c}}} d T
  e^{- S (T)} T^3 \nonumber\\
  & \approx & \frac{A}{H^4_{\mathrm{vac}}} \int_{T_{\text{n}}}^{\infty} d T
  e^{- S (T)} T^3 \nonumber\\
  & \approx & \frac{A}{H^4_{\mathrm{vac}}} \int_{T_{\text{n}}}^{\infty} d T
  e^{- \left[ S \left( T_{\text{n}} \right) + S' \left( T_{\text{n}} \right)
  \left( T - T_{\text{n}} \right) \right]} T^3 \\
  & =& \frac{A}{H^4_{\mathrm{vac}}} e^{- \left[ S \left( T_{\text{n}}
  \right) - S' \left( T_{\text{n}} \right) T_{\text{n}} \right]}
  \int_{T_{\text{n}}}^{\infty} d T e^{- S' \left( T_{\text{n}} \right) T} T^3
  \nonumber\\
  & = & \frac{A T_{\text{n}}^4}{H^4_{\mathrm{vac}}} e^{- S \left( T_{\text{n}}
  \right)} \left( 6 \tilde{\beta}_{\text{n}}^{- 4} + 6
  \tilde{\beta}_{\text{n}}^{- 3} + 3 \tilde{\beta}_{\text{n}}^{- 2} +
  \tilde{\beta}_{\text{n}}^{- 1} \right), \nonumber
\end{eqnarray}
thus
\begin{equation}
  S \left( T_{\text{n}} \right) \simeq 4 \ln \left(
  \frac{T_{\text{n}}}{H_{\mathrm{vac}}} \right) + \ln \left( 6
  \tilde{\beta}_{\text{n}}^{- 4} + 6 \tilde{\beta}_{\text{n}}^{- 3} + 3
  \tilde{\beta}_{\text{n}}^{- 2} + \tilde{\beta}_{\text{n}}^{- 1} \right),
\end{equation}
where $\beta \equiv - d S / d t = H T d S / d T, \tilde{\beta} \equiv \beta /
H$.

Since the supercooled FOPTs are mainly determined by the dilaton, we can simplify the discussion by approximating the full potential with the dilaton potential,
\begin{equation}
  V_0 (\chi) = [c_{\chi} g_{\chi}^2 - \epsilon (\chi)] \chi^4,
\end{equation}
with $g_{\chi} = 4 \pi / \sqrt{N}$, and
\begin{equation}
  \epsilon (\chi) = \frac{8 c_{\chi} g_{\chi}^2 \gamma_{\epsilon} (\chi /
  v_{\chi})^{\gamma_{\epsilon}}}{\gamma_{\epsilon} \left( 4 +
  \gamma_{\epsilon} + \sqrt{16 c_{\epsilon} c_{\chi} + (4 +
  \gamma_{\epsilon})^2} \right) + 8 c_{\epsilon} c_{\chi} (1 - (\chi /
  v_{\chi})^{\gamma_{\epsilon}})}
\end{equation}
The renormalization group equation for $\epsilon(\chi)$ is given by
\begin{equation}
  \frac{\partial \epsilon (\chi)}{\partial \ln \chi} = \gamma_{\epsilon}
  \epsilon^2 (\chi) + c^{(1)} \epsilon^2 (\chi) .
\end{equation}

There are two minima of the potential. One is the high temperature vacuum $\chi = 0$, while the other one is the physical vacuum $\chi \simeq v_{\chi}$.
The mass and the potential corresponding to the latter are  
\begin{equation}
  m^2_{\chi} \simeq - 4 \gamma_{\epsilon} c_{\chi} g_{\chi}^2 v_{\chi}^2,
  \qquad V_0 (v_{\chi}) \simeq \frac{\gamma_{\epsilon} c_{\chi} g_{\chi}^2}{4}
  v_{\chi}^4 = - \frac{1}{16} m^2_{\chi} v_{\chi}^2 .
\end{equation}
Once the finite temperature corrections are included, the potential for these two vacuum are given by
\begin{equation}
  V (0, T) = - \frac{\pi^2}{8} N^2 T^4, \qquad V (v_{\chi}, T) \simeq -
  \frac{1}{16} m^2_{\chi} v_{\chi}^2,
\end{equation}
Below the critical temperature, the energy density difference of these two
vacuum is
\begin{eqnarray}
  \rho_{\mathrm{vac}} & = & V (0, T) - \left. T \frac{\partial V}{\partial T}
  \right|_{\chi = 0} - \left( V (v_{\chi}, T) - \left. T \frac{\partial
  V}{\partial T} \right|_{\chi = v_{\chi}} \right) \nonumber\\
  & = & - \frac{\pi^2}{8} N^2 T^4 - T \left( - \frac{\pi^2}{2} N^2 T^3
  \right) + \frac{1}{16} m^2_{\chi} v_{\chi}^2 \nonumber\\
  & = & \frac{3 \pi^2}{8} N^2 T^4 + \frac{1}{16} m^2_{\chi} v_{\chi}^2 
\end{eqnarray}
The Friedmann equation is given by
\begin{eqnarray}
  H & = & \frac{d a}{a d t} = \sqrt{\frac{\rho_{\mathrm{vac}}}{3
  M^2_{\text{Pl}}}}\nonumber\\
  & = & \frac{1}{\sqrt{3} M_{\text{Pl}}} \sqrt{\frac{1}{16} m^2_{\chi}
  v_{\chi}^2 + \frac{3 \pi^2}{8} N^2 T^4}\nonumber\\
  & = & \frac{m_{\chi} v_{\chi}}{4 \sqrt{3} M_{\text{Pl}}} \sqrt{1 + \frac{6
  \pi^2 N^2 T^4}{m^2_{\chi} v_{\chi}^2}}\nonumber\\
  & \equiv & p \sqrt{1 + q a^{- 4}}
\end{eqnarray}
It can be solved with initial condition $a (0) = 1$, and the result is
\begin{equation}
    a (t) = q^{1 / 4} \sinh^{1 / 2} \left[ 2 p t + \mathrm{arcsinh} \left(
   \frac{1}{\sqrt{q}} \right) \right] \simeq e^{p t}
\end{equation}
Therefore, we can treat $H_{\mathrm{vac}} $ as a constant with a value of
\begin{equation}
H_{\mathrm{vac}} = \frac{m_{\chi} v_{\chi}}{4 \sqrt{3} M_{\text{Pl}}}
\end{equation}
Finally, we obtain the nucleation condition for NMCHM$_\chi$
\begin{eqnarray}
  S \left( T_{\text{n}} \right) & = & 4 \ln \left( \frac{4 \sqrt{3}
  M_{\text{Pl}} T_{\text{n}}}{m_{\chi} v_{\chi}} \right) + \ln \left( 6
  \tilde{\beta}_{\text{n}}^{- 4} + 6 \tilde{\beta}_{\text{n}}^{- 3} + 3
  \tilde{\beta}_{\text{n}}^{- 2} + \tilde{\beta}_{\text{n}}^{- 1} \right)\notag\\
  & \simeq & 4 \ln \left( \frac{4 \sqrt{3} M_{\text{Pl}}
  T_{\text{n}}}{m_{\chi} v_{\chi}} \right) - \ln \left(
  \tilde{\beta}_{\text{n}} \right)\notag\\
  & \simeq & 131.98 - 4 \ln \left( \frac{m_{\chi}}{1~{\mathrm{TeV}}} \right) - 4
  \ln \left( \frac{v_{\chi}}{2.5~\mathrm{TeV}} \right) + 4 \ln \left(
  \frac{T_{\text{n}}}{100~\mathrm{GeV}} \right) - \ln \left(
  \frac{\tilde{\beta}_{\text{n}}}{100} \right).
\end{eqnarray}

\bibliographystyle{apsrev4-1-JHEPfix}
\bibliography{ref}
\end{document}